\pdfoutput=1
\documentclass[11pt]{article}
\usepackage[margin=1in]{geometry}
\usepackage{amsmath,amssymb,amsthm,mathtools,bm}
\usepackage{booktabs,array,longtable}
\usepackage{graphicx}
\usepackage{natbib}
\usepackage[hidelinks]{hyperref}
\usepackage{xcolor}
\usepackage{enumitem}
\usepackage{microtype}
\usepackage{setspace}
\newtheorem{theorem}{Theorem}[section]
\newtheorem{lemma}[theorem]{Lemma}
\newtheorem{corollary}[theorem]{Corollary}
\newtheorem{proposition}[theorem]{Proposition}
\newtheorem{assumption}[theorem]{Assumption}
\theoremstyle{remark}
\newtheorem{remark}[theorem]{Remark}

\newcommand{\E}{\mathbb E}
\newcommand{\Var}{\operatorname{Var}}
\newcommand{\Cov}{\operatorname{Cov}}
\newcommand{\pto}{\overset{p}{\to}}
\newcommand{\dto}{\overset{d}{\to}}
\newcommand{\op}{o_p}

\title{Bootstrap Inference with a Randomly Assigned Regressor:\\ Covariance Filtering and the Limits of Marginal Resampling}
\author{
Ulrich Hounyo\thanks{Department of Economics, University at Albany, State University of New York, 1400 Washington Avenue, Albany, NY 12222, USA. Email: \texttt{khounyo@albany.edu}.}
\and
Jungbin Hwang\thanks{Department of Economics, University of Connecticut, 309 Susan V. Herbst Hall Storrs, CT, USA. Email: \texttt{jungbin.hwang@uconn.edu}.}
}
\date{October 2, 2026}

\begin{document}
\maketitle

\begin{abstract}
Random assignment can make ordinary least squares (OLS) inference insensitive to outcome dependence, yet iid resampling can still fail because assignment and resampling need not remove the same covariance terms. With binary treatment, the iid variance target differs from the sampling variance by exactly minus aggregate cross-unit covariance of treatment effects. Under a Gaussian first-order limit, positive covariance leads to over-rejection and negative covariance to under-rejection. Two designs can generate the same distribution for each observation but require variance corrections of opposite signs, so no marginal-only variance correction is first-order exact for both. For first-order Gaussian inference, only one covariance component---the covariance carried by the randomized-regressor score---must be recovered. Standard dependence estimators on that score restore validity under suitable ordered or grouped dependence conditions. In simulations with ordered data, fixed-bandwidth calibration reduces several large-bandwidth size distortions.
\end{abstract}


\section{Introduction}

Random assignment can make inference much simpler than the dependence structure of the observed data suggests. \citet{CHLS2026} (hereafter CHLS) show that, for the ordinary least squares (OLS) coefficient on a randomly assigned regressor, residualizing the regressor against fixed-dimensional controls is asymptotically absorbed by the assignment mechanism: under their conditions, the leading score is the centered randomized regressor multiplied by the regression error. Broad dependence in outcomes, controls, and regression errors can therefore disappear from the first-order variance or survive only through a restricted covariance component. That simplification might suggest that iid resampling is automatically valid. Our question is sharper: \emph{does independent resampling reproduce exactly the covariance cancellations induced by random assignment in the sampling law?}

The answer is no in general. An iid bootstrap deletes cross-observation covariance from the resampled score by construction. Random assignment also filters covariance, but it need not delete the same terms. Random assignment may therefore simplify the sampling variance even though a seemingly natural iid bootstrap still targets the wrong first-order variance. This problem is distinct from the bootstrap-moment pathology of \citet{HahnLiao2021}: the variance target can already be wrong before conditional second moments become an issue.

\paragraph{Scope of the assignment design.}
Throughout the formal results, assignment is iid at the unit level. Fixed-count complete randomization, stratified or covariate-adaptive assignment, matched designs, and cluster-level assignment alter the assignment covariance structure and can change the relevant filter at first order; Remark~\ref{rem:designboundary} gives an explicit fixed-count counterexample. The formulas below do not carry over automatically to those designs. For such experiments, one must first derive the design-specific score filter and then choose an appropriate resampling or variance correction; see, for example, \citet{BugniCanayShaikh2018} for covariate-adaptive randomization.

The mechanism is most transparent with binary treatment. Suppose $D_i$ is independently randomized and
\[
Y_i=Y_i(0)+D_i\tau_i,
\qquad
\tau_i=Y_i(1)-Y_i(0).
\]
Let $V_n$ denote the first-order sampling variance of the leading OLS statistic and $V_n^\#$ the diagonal variance targeted by iid score resampling. Define
\[
\Delta_{\tau,n}=\frac1n\sum_{i\ne j}\Cov(\tau_i,\tau_j).
\]
We prove the exact leading-score identity
\[
\boxed{V_n^\#-V_n=-\Delta_{\tau,n}.}
\]
Under the standardized Gaussian first-order limit used for inference below, positive aggregate covariance makes iid inference anti-conservative; negative covariance makes it conservative; covariance cancellation can leave it first-order correct despite dependence. The variance discrepancy itself does not depend on the Bernoulli assignment probability $p$, even though both $V_n$ and $V_n^\#$ do.

Correlated treatment effects under individual assignment are economically natural when units share \emph{effect modifiers}, even without interference. Returns to an individually randomized training program can co-move within local labor markets because local demand conditions affect the payoff to training. Returns to a household subsidy can co-move across nearby cohorts because macroeconomic conditions affect take-up returns. Treatment effects can also co-move within schools because students face common teachers, curricula, or institutional environments. In each example assignment can remain iid at the unit level while the response to treatment is correlated across units.

The general covariance filter nests the binary identity. When the regression error takes the assignment-dependent form
\[
\varepsilon_i=\lambda(D_i)'e_i,
\]
with a mean-zero latent vector $e_i$, let
\[
\sigma_D^2=\Var(D_i),
\qquad
\mu_A=E[(D_i-E D_i)\lambda(D_i)].
\]
Random assignment eliminates every off-diagonal covariance direction except the scalar projection $\mu_A'e_i$. If
\[
\Delta_{A,n}=\frac1n\sum_{i\ne j}\Cov(\mu_A'e_i,\mu_A'e_j),
\]
then
\[
\boxed{V_n^\#-V_n=-\sigma_D^{-4}\Delta_{A,n}.}
\]
The bootstrap problem is therefore lower-dimensional than the observed-data dependence problem: only the covariance surviving the assignment-induced projection matters for the coefficient of interest.

The information-limit result makes this restriction explicit. We construct two Gaussian moving-average (MA(1)) designs satisfying the maintained assumptions with exactly the same one-observation law of $(D_i,Y_i)$ and the same diagonal bootstrap target, but opposite treatment-effect covariance and different first-order sampling variances. Any variance rule asymptotically determined only by the one-observation marginal law must fail for at least one design. For centered-Gaussian bootstraps this rules out universal first-order validity over the pair; under the conditional moment condition of Theorem~\ref{thm:ui}, the same information limit applies to bootstrap second moments more generally. Ordered or grouped observations can distinguish the designs. The missing information is therefore genuinely joint rather than marginal.

A complementary minimality result shows that full dependence estimation is unnecessary. Conditional on consistently estimating the observation-by-observation variance component, a centered-Gaussian procedure is first-order valid if and only if it recovers the single surviving covariance contribution at the first-order variance scale. The associated covariance quotient is one-dimensional whenever $\mu_A\ne0$. The information lower bound and the dimensionality result therefore line up: marginals are insufficient, but one projected covariance contribution is enough.

\paragraph{Relation to design-based clustering and causal bootstrap.}
\citet{AbadieEtAl2023} show in a design-based framework that clustering can matter when assignment is clustered or heterogeneous treatment effects co-move within clusters. Our binary identity is related to that heterogeneity channel; our question is different: whether independent resampling removes covariance that random assignment leaves in the sampling law. Our comparison leads to the general covariance-filter identity, the marginal-information limit, and the one-dimensional minimality result. \citet{ZhangEtAl2025} provide a sampling-side counterpart by showing that least-squares inference with random regressors can be robust to correlated errors. Our strong-exogeneity benchmark shares that intuition; our contribution begins when the regression error responds to assignment.

\citet{AbadieAtheyImbensWooldridge2020} distinguish sampling- and design-based uncertainty, while \citet{ImbensMenzel2021} develop a finite-population causal bootstrap. Bootstrap inference under matched-pair or covariate-adaptive assignment is studied by \citet{JiangEtAl2024,ZhangZheng2020}, with general covariate-adaptive inference in \citet{BugniCanayShaikh2018}. Those designs make assignment itself dependent; here assignment is iid, so surviving covariance comes from potential outcomes or errors. Our constructive step then uses standard heteroskedasticity-and-autocorrelation-consistent (HAC), block, cluster, or dependent-bootstrap methods \citep{Kunsch1989,PolitisRomano1994,Shao2010,Hounyo2023,CameronEtAl2008}; related cluster guidance includes \citet{DjogbenouEtAl2019,CanaySantosShaikh2021,MacKinnonNielsenWebb2023}. Group-based $t$ and network-HAC alternatives include \citet{IbragimovMuller2010,KojevnikovEtAl2021}.

Ordinary pairs resampling does not evade the problem. Its linearized target is the usual heteroskedasticity-robust HC0 variance for the residualized randomized-regressor score and therefore inherits the same diagonal discrepancy. Under their respective consistency conditions, HAC, cluster, or spatial variance estimators applied to that score can recover $V_n$. The problem is dependence-ignorant diagonal inference, whether analytic or based on iid row resampling.

The simulations illustrate these implications. Time-series designs cover positive, negative, and cancellation cases. In a cluster design with iid individual assignment and correlated treatment effects, iid score inference rejects $18.5\%$ at a nominal $5\%$ level when cluster size is eight, while the correctly grouped projected correction rejects about $5.1\%$. Deliberate grouping misspecification shows that the correction is only as good as the dependence structure supplied to it. For ordered data we document bandwidth sensitivity and compare small-bandwidth normal with fixed-bandwidth calibration. Because the relevance ratio is noisier, we treat it only as a descriptive summary rather than a pretest.

The remainder of the paper first defines the sampling and bootstrap targets, establishes the strong-exogeneity benchmark, and derives the covariance filter. It then proves the marginal information limit and one-dimensional minimality result, specializes to binary treatment and ordinary pairs resampling, develops the projected-score correction and diagnostic, and reports Monte Carlo evidence. We also show that once the iid and projected tests are calibrated to the same first-order size, their Gaussian local-power curves coincide. Proofs, primitive sufficient conditions, and additional simulations are collected in the numbered Online Appendix.

\section{Setup and inferential target}\label{sec:setup}
We use a triangular array where needed and suppress $n$ subscripts when no ambiguity arises. Within each row, the assignment variables $\{D_{i,n}\}_{i=1}^n$ are iid with a common law and the full assignment vector is independent of the latent/control array $\{(e_{i,n}',W_{i,n}')'\}_{i=1}^n$. Dependence across observations in the latent variables and controls is otherwise allowed subject to the conditions of each result. The dimensions of $W_{i,n}$ and $e_{i,n}\in\mathbb R^L$ are fixed as $n$ grows. The first component of $W_i$ is a nonzero constant. We assume $\sigma_{D,n}^2>0$ for all sufficiently large $n$; stronger lower bounds are imposed only where needed. Thus $\mu_D$ and $\sigma_D^2$ below denote the row-$n$ quantities $\mu_{D,n}$ and $\sigma_{D,n}^2$ whenever a triangular-array statement is in force. Sample-matrix inverses are understood on their nonsingularity events. On exceptional samples the associated statistic may be assigned an arbitrary fixed value; the relevant theorems ensure that such events have probability tending to zero. Define
\[
\mu_D=\E[D_i],\qquad \sigma_D^2=\Var(D_i),\qquad D_i^c=D_i-\mu_D.
\]
The regression model is
\[
Y_i=D_i\beta+W_i'\gamma+\varepsilon_i.
\]
Stack the observations as $Y=(Y_1,\ldots,Y_n)'$, $D=(D_1,\ldots,D_n)'$, and $\varepsilon=(\varepsilon_1,\ldots,\varepsilon_n)'$, and let $W$ be the $n\times K$ matrix with $i$th row $W_i'$, where the fixed control dimension is $K$. Let $I_n$ denote the $n\times n$ identity matrix and define the residual-maker
\[
M_W=I_n-W(W'W)^{-1}W'.
\]
The scalar $\widehat\beta$ denotes the OLS coefficient on $D$. By the Frisch--Waugh--Lovell theorem,
\[
\widehat\beta-\beta=(D'M_WD)^{-1}D'M_W\varepsilon.
\]
The Frisch--Waugh--Lovell formula above is an exact finite-sample identity for OLS. The next step is asymptotic. We take from CHLS the representation
\begin{equation}
\sqrt n(\widehat\beta-\beta)=\sigma_{D,n}^{-2}\frac1{\sqrt n}\sum_{i=1}^n D_{i,n}^c\varepsilon_{i,n}+\op(1).
\label{eq:chlsrep}
\end{equation}
Equation~\eqref{eq:chlsrep} requires one clarification about the controls. In the exact Frisch--Waugh--Lovell formula, $D$ is residualized against $W$. Under the CHLS conditions, iid assignment makes the normalized sample projection of the centered assignment draw on the fixed-dimensional controls asymptotically negligible, while $n^{-1}D'M_WD$ converges to the assignment variance. Consequently, the control-projection terms contribute only $o_p(1)$ to the normalized numerator and the first-order score reduces to $(D_{i,n}-\mu_{D,n})\varepsilon_{i,n}$. This is a first-order statement, not a claim that the controls are absent from the finite-sample OLS formula or from the joint distribution of the data.

Our target is the unconditional CHLS superpopulation sampling law, not the finite-population randomization distribution; this distinction follows the sampling-versus-design perspective emphasized by \citet{AbadieAtheyImbensWooldridge2020}. Accordingly, the variance objects below are deterministic variance sequences for the \emph{leading score} in \eqref{eq:chlsrep}; they are not asserted to equal the exact finite-sample variance of the nonlinear OLS estimator.

Define the first-order score
\[
\psi_i=D_i^c\varepsilon_i,
\]
its sampling quadratic scale
\[
\sigma_{D\varepsilon,n}^2
=
\Var\!\left(
n^{-1/2}\sum_{i=1}^n\psi_i
\right),
\]
the diagonal score scale
\[
s_{\#,n}^2
=
\frac1n\sum_{i=1}^nE[\psi_i^2],
\]
the corresponding diagonal bootstrap target
\[
V_n^\#
=
\sigma_{D,n}^{-4}s_{\#,n}^2,
\]
and the first-order sampling-variance sequence
\[
V_n=\sigma_{D,n}^{-4}\sigma_{D\varepsilon,n}^2.
\]
Thus $V_n$ is exactly the variance of the leading linearized statistic
\[
\sigma_{D,n}^{-2}n^{-1/2}\sum_{i=1}^n\psi_{i,n},
\]
while its interpretation as the first-order variance of $\sqrt n(\widehat\beta-\beta)$ uses \eqref{eq:chlsrep}. If $V_n$ can shrink, the unscaled remainder $o_p(1)$ in \eqref{eq:chlsrep} is not by itself enough to obtain a standardized limit: one needs the stronger remainder $o_p(V_n^{1/2})$ or, equivalently for our purposes, a standardized CHLS conclusion stated directly. Every theorem below that permits a shrinking variance scale therefore assumes the standardized sampling law explicitly. These objects are defined before specializing the error structure, so the strong-exogeneity and assignment-dependent-heteroskedasticity branches share the same normalization. Throughout, $P^*$, $E^*$, and $\Var^*$ denote probability, expectation, and variance conditional on the observed sample. We write $\Phi$ for the standard normal distribution function and $z_q=\Phi^{-1}(q)$ for its $q$th quantile. To avoid collisions, $L$ is reserved for the latent-vector dimension, $m_n$ for a theoretical kernel denominator, $\ell$ for the largest included positive lag reported in the Bartlett simulations, $\varrho_{\psi,n}(h)$ and $\varrho_{\tau,n}(h)$ for score and treatment-effect autocovariances, and $h_{\mathrm{loc}}$ for a local-alternative shift.

\begin{assumption}[Stable scale]\label{ass:stable}
There exist $0<c<C<\infty$ such that for all sufficiently large $n$,
\[
c\le \sigma_{D,n}^2\le C,
\qquad
c\le \sigma_{D\varepsilon,n}^2\le C,
\qquad
c\le s_{\#,n}^2\le C.
\]
\end{assumption}

\begin{lemma}[Randomized-design LLN]\label{lem:design}
Let $\{D_{i,n}\}_{i\le n}$ be row-wise iid with mean $\mu_{D,n}$ and variance $\sigma_{D,n}^2$, and define
\[
\bar D_n=n^{-1}\sum_{i=1}^nD_{i,n},
\qquad
\widehat Q_n=n^{-1}\sum_{i=1}^n(D_{i,n}-\bar D_n)^2.
\]
Suppose for some $\delta\in(0,2]$,
\[
\sup_n E|D_{1,n}-\mu_{D,n}|^{2+\delta}<\infty.
\]
Then
\[
\bar D_n-\mu_{D,n}=O_p(n^{-1/2}),
\qquad
\widehat Q_n-\sigma_{D,n}^2=o_p(1).
\]
If, in addition, $\inf_n\sigma_{D,n}^2>0$, then $\widehat Q_n/\sigma_{D,n}^2\to_p1$.
\end{lemma}

\subsection{Assignment-dependent heteroskedasticity and heterogeneous effects}
Let $\lambda(d)\in\mathbb R^L$ denote the assignment-dependent loading vector and write
\[
\varepsilon_i=\lambda(D_i)'e_i,
\qquad A_i=(D_i-\mu_D)\lambda(D_i),
\qquad \mu_A=\E[A_i],
\]
with a mean-zero latent vector $e_i\in\mathbb R^L$, as in CHLS, so that $\E e_i=0$. Put $\widetilde A_i=A_i-\mu_A$. Then
\[
\sigma_{D\varepsilon,n}^2
=
\sigma_{e,1,n}^2+\sigma_{e,2,n}^2,
\]
where
\[
\sigma_{e,1,n}^2=\frac1n\sum_{i=1}^n\E[(\widetilde A_i'e_i)^2],
\qquad
\sigma_{e,2,n}^2=\Var\left(n^{-1/2}\sum_{i=1}^n\mu_A'e_i\right).
\]
Thus the CHLS conditional-heteroskedastic representation decomposes the generic score variance rather than defining a separate normalization.

\subsection{Gaussian multiplier bootstrap target and null-imposed score}
Using the score $\psi_i$ defined above, let $\xi_i\sim N(0,1)$ be iid and independent of the data. Define
\[
T_{n,o}^\#=\sigma_{D,n}^{-2}n^{-1/2}\sum_{i=1}^n\xi_i\psi_{i,n},
\qquad
\widehat V_{n,o}^\#=\sigma_{D,n}^{-4}n^{-1}\sum_{i=1}^n\psi_{i,n}^2.
\]
This is an infeasible device that isolates the covariance effect.

For testing the null hypothesis $H_0:\beta=\beta_0$, define the sample assignment mean $\bar D=n^{-1}\sum_{i=1}^nD_i$ and the null-imposed nuisance estimator
\[
\widetilde\gamma(\beta_0)=(W'W)^{-1}W'(Y-D\beta_0),
\]
the restricted residual
\[
\widetilde\varepsilon_i(\beta_0)=Y_i-D_i\beta_0-W_i'\widetilde\gamma(\beta_0),
\]
and the restricted score
\[
\widetilde\psi_i(\beta_0)=(D_i-\bar D)\widetilde\varepsilon_i(\beta_0).
\]
Let
\[
\widehat Q_n=n^{-1}\sum_{i=1}^n(D_i-\bar D)^2,
\]
and
\[
T_n^\#(\beta_0)=\widehat Q_n^{-1}n^{-1/2}\sum_{i=1}^n\xi_i\widetilde\psi_i(\beta_0).
\]
Conditional on the sample, this statistic is exactly Gaussian with variance
\[
\widetilde V_n^\#(\beta_0)=\widehat Q_n^{-2}n^{-1}\sum_{i=1}^n\widetilde\psi_i(\beta_0)^2.
\]

\paragraph{Test inversion and confidence sets.}
The feasible score and its variance are null imposed: both are recomputed at the hypothesized value $\beta_0$. Accordingly, the formal results below justify tests of $H_0:\beta=\beta_0$. A $(1-\alpha)$ confidence set is obtained by inversion,
\[
\mathcal C_{1-\alpha}
=
\{\beta_0:\text{the level-}\alpha\text{ null-imposed test does not reject}\}.
\]
In general this set need not equal a Wald interval of the form $\widehat\beta\pm z_{1-\alpha/2}\widehat{\mathrm{se}}$ because the variance estimate may vary with $\beta_0$. A conventional null-independent Wald interval would require a separate consistency argument for an unrestricted-score variance estimator; we do not infer that result from the null-imposed theory.

The following branch-neutral lemma records the only nuisance-replacement result needed by both the strong-exogeneity benchmark and the feasible conditional-heteroskedastic analysis.

\begin{lemma}[Restricted-score replacement]\label{lem:restricted}
Under $H_0:\beta=\beta_0$, suppose
\[
\|(n^{-1}W'W)^{-1}\|=O_p(1),
\qquad
n^{-1}W'\varepsilon=o_p(1),
\]
\[
\frac1n\sum_i\varepsilon_i^2=O_p(1),
\qquad
\frac1n\sum_i\|W_i\|^2=O_p(1),
\qquad
\sup_n\Var(D_{1,n})<\infty.
\]
Under the assignment structure in Section~\ref{sec:setup},
\[
\frac1n\sum_{i=1}^n
\{\widetilde\psi_i(\beta_0)-\psi_i\}^2=o_p(1).
\]
Under $H_0$,
\[
\widetilde\gamma(\beta_0)-\gamma
=(n^{-1}W'W)^{-1}(n^{-1}W'\varepsilon)=o_p(1)
\]
under the stated conditions. More precisely,
\[
\frac1n\sum_i(\widetilde\psi_i-\psi_i)^2
\le
2(\bar D-\mu_D)^2\frac1n\sum_i\varepsilon_i^2
+
2\|\widetilde\gamma-\gamma\|^2 B_n,
\]
where
\[
B_n=\frac1n\sum_i(D_i-\bar D)^2\|W_i\|^2=O_p(1).
\]
\end{lemma}

\section{Benchmark: when random assignment justifies iid bootstrap inference}

For row $n$, let
\[
\mathcal D_n=\sigma(D_{1,n},\ldots,D_{n,n}),
\qquad
\mathcal E_n=\sigma(\varepsilon_{1,n},\ldots,\varepsilon_{n,n}).
\]
We say that \emph{strong exogeneity} holds when the sigma-fields $\mathcal D_n$ and $\mathcal E_n$ are independent. Thus no regression error depends on any assignment draw. This is stronger than the assignment-dependent representation used later: in Section~\ref{sec:filter}, assignment is independent of the latent array $e$, but $\varepsilon_i=\lambda(D_i)'e_i$ may itself depend on $D_i$. Controls may remain dependent across units and correlated with the errors; assignment--control independence is already imposed in Section~\ref{sec:setup}.

\begin{lemma}[Scalar randomized-score LLN]\label{lem:scalar-diag}
Suppose the assignment structure in Section~\ref{sec:setup} and strong exogeneity hold and, for some $\delta\in(0,2]$,
\[
\sup_n E|D_{1,n}-\mu_{D,n}|^{2+\delta}<\infty,
\qquad
\frac1n\sum_{i=1}^n|\varepsilon_{i,n}|^{2+\delta}=O_p(1),
\]
and
\[
\frac1n\sum_{i=1}^n
\{\varepsilon_{i,n}^2-E(\varepsilon_{i,n}^2)\}
=o_p(1).
\]
Then
\[
\frac1n\sum_{i=1}^n
(D_{i,n}-\mu_{D,n})^2\varepsilon_{i,n}^2
-
\frac1n\sum_{i=1}^n
E[(D_{i,n}-\mu_{D,n})^2\varepsilon_{i,n}^2]
=o_p(1).
\]
\end{lemma}

Under strong exogeneity, for $i\ne j$,
\[
\E[(D_{i,n}-\mu_{D,n})(D_{j,n}-\mu_{D,n})\varepsilon_{i,n}\varepsilon_{j,n}]=0
\]
even if $\E[\varepsilon_i\varepsilon_j]\ne0$.

\begin{theorem}[Feasible iid multiplier score bootstrap under strong exogeneity]\label{thm:strong}
Suppose strong exogeneity holds, the conditions of Lemmas~\ref{lem:design}, \ref{lem:scalar-diag}, and \ref{lem:restricted}, and Assumption~\ref{ass:stable} hold. Let the standardized CHLS conclusion hold:
\[
\frac{\sqrt n(\widehat\beta-\beta)}{\sqrt{V_n}}\dto N(0,1).
\]
Under strong exogeneity,
\[
V_n^\#=V_n.
\]
Under $H_0:\beta=\beta_0$,
\[
\widetilde V_n^\#(\beta_0)-V_n^\#=\op(1),
\]
and hence
\[
\widetilde V_n^\#(\beta_0)-V_n=\op(1),
\]
and
\[
\sup_{t\in\mathbb R}\left|
\Pr^*\{T_n^\#(\beta_0)\le t\}
-\Pr\{\sqrt n(\widehat\beta-\beta_0)\le t\}
\right|=\op(1).
\]
The distributional conclusion does not require $V_n$ to converge to a fixed constant. Under Assumption~\ref{ass:stable}, however, $V_n=V_n^\#$ is bounded away from zero, so the absolute variance consistency displayed above is also relative consistency.
\end{theorem}

Because $\E^*T_n^\#=0$ exactly,
\[
\E^*[(T_n^\#)^2]=\widetilde V_n^\#.
\]
Thus, if $V_n\to V$ for some fixed $V\in(0,\infty)$,
\[
\E^*[(T_n^\#)^2]\pto V.
\]
This does not follow from weak convergence alone; here it follows directly from variance consistency.

\section{When dependence survives random assignment: the covariance filter}\label{sec:filter}
Define
\[
\sigma_{\mu,0,n}^2=\frac1n\sum_i\Var(\mu_A'e_i),
\]
\[
\Delta_{A,n}=\frac1n\sum_{i\ne j}\Cov(\mu_A'e_i,\mu_A'e_j).
\]
Equivalently, with
\[
\Omega_{e,n}=\Var\left(n^{-1/2}\sum_i e_i\right),\qquad
\Omega_{e,0,n}=n^{-1}\sum_i\Var(e_i),
\]
we have
\[
\Delta_{A,n}=\mu_A'(\Omega_{e,n}-\Omega_{e,0,n})\mu_A.
\]

\begin{lemma}[Randomized diagonal-score LLN]\label{lem:diag}
Suppose the assignment structure in Section~\ref{sec:setup} holds and, for some $\delta\in(0,2]$,
\[
\sup_n E|D_{1,n}-\mu_{D,n}|^{2+\delta}<\infty,
\qquad
\frac1n\sum_{i=1}^n \|e_{i,n}\|^{2+\delta}=O_p(1),
\]
the assignment loading is uniformly bounded over the support of the randomized regressor,
\[
\sup_n\sup_{d\in\operatorname{supp}(D_{1,n})}\|\lambda_n(d)\|<\infty,
\]
and
\[
\left\|
\frac1n\sum_{i=1}^n
\{e_{i,n}e_{i,n}'-E[e_{i,n}e_{i,n}']\}
\right\|=o_p(1).
\]
Then
\[
\frac1n\sum_{i=1}^n\psi_{i,n}^2
-
\frac1n\sum_{i=1}^nE[\psi_{i,n}^2]
=o_p(1).
\]
\end{lemma}

\begin{theorem}[Oracle covariance-filter identity for the linearized score]\label{thm:filter}
Suppose the conditional-heteroskedastic representation in Section~\ref{sec:setup} holds, the iid assignment vector is independent of the full latent array $\{e_{i,n}\}_{i=1}^n$, $\sigma_{D,n}^2>0$ for all sufficiently large $n$, and, for every sufficiently large $n$ and $i\le n$,
\[
E\|A_{i,n}\|^2<\infty,
\qquad
E\|e_{i,n}\|^2<\infty.
\]
Because $\varepsilon_{i,n}=\lambda_n(D_{i,n})'e_{i,n}$, this assumption does \emph{not} require assignment to be independent of the regression-error array: the error may respond to $D_{i,n}$ through the loading $\lambda_n(D_{i,n})$. Independence is imposed on assignment and the latent array $e$, which is the condition used by the covariance decomposition below.
Define
\[
\sigma_{e,1,n}^2
=
\frac1n\sum_{i=1}^n
E[((A_{i,n}-\mu_{A,n})'e_{i,n})^2],
\]
\[
\sigma_{\mu,0,n}^2
=
\frac1n\sum_{i=1}^n
\Var(\mu_{A,n}'e_{i,n}).
\]
Then the generic diagonal target in Section~\ref{sec:setup} satisfies the following exact finite-$n$ identity at the level of the leading-score variance:
\[
V_n^\#
=
\sigma_{D,n}^{-4}
\left(
\sigma_{e,1,n}^2+\sigma_{\mu,0,n}^2
\right).
\]
Moreover, independence of iid assignment from the full latent array implies the exact orthogonal decomposition
\[
\sigma_{D\varepsilon,n}^2
=
\sigma_{e,1,n}^2
+
\Var\!\left(
n^{-1/2}\sum_{i=1}^n\mu_{A,n}'e_{i,n}
\right).
\]
Hence
\[
V_n
=
\sigma_{D,n}^{-4}
\left[
\sigma_{e,1,n}^2
+
\Var\!\left(
n^{-1/2}\sum_{i=1}^n\mu_{A,n}'e_{i,n}
\right)
\right],
\]
and therefore
\[
\boxed{
V_n^\#-V_n
=
-\sigma_{D,n}^{-4}\Delta_{A,n}
}
\]
with
\[
\Delta_{A,n}
=
\frac1n\sum_{i\neq j}
\Cov(\mu_{A,n}'e_{i,n},\mu_{A,n}'e_{j,n}).
\]
Equivalently,
\[
\Delta_{A,n}
=
\mu_{A,n}'(\Omega_{e,n}-\Omega_{e,0,n})\mu_{A,n}.
\]

These equalities are exact algebra for the deterministic variance sequences of the leading score; no claim is made here that $V_n$ equals $\Var\{\sqrt n(\widehat\beta-\beta)\}$ at finite $n$. That link is asymptotic and comes from the CHLS linearization \eqref{eq:chlsrep} together with the scale conditions used in the distributional results below.

If, in addition, the conditions of Lemma~\ref{lem:diag} hold, then the oracle diagonal estimator satisfies
\[
\boxed{
\sigma_{D,n}^{4}\{\widehat V_{n,o}^\#-V_n^\#\}=o_p(1).
}
\]
If also $\inf_n\sigma_{D,n}^2>0$, then
\[
\widehat V_{n,o}^\#-V_n^\#=o_p(1).
\]
\end{theorem}

\begin{remark}[From the score identity to OLS inference]\label{rem:score-vs-ols}
Theorem~\ref{thm:filter} is exact for the variance of the leading linearized score. Its implication for standardized OLS inference also requires the remainder in \eqref{eq:chlsrep} to be negligible at the relevant variance scale. If $V_n$ is bounded away from zero, the displayed $o_p(1)$ remainder is sufficient. If $V_n\to0$, one instead needs
\[
\sqrt n(\widehat\beta-\beta)
-
\sigma_{D,n}^{-2}n^{-1/2}\sum_i\psi_{i,n}
=
o_p(V_n^{1/2}),
\]
or a standardized CHLS conclusion stated directly. The scale-free distributional results below impose that standardized conclusion explicitly.
\end{remark}

The covariance identities in Theorem~\ref{thm:filter} are finite-$n$ consequences of the assignment structure and second moments. The additional conditions of Lemma~\ref{lem:diag} enter only for consistency of the oracle diagonal estimator $\widehat V_{n,o}^\#$.

\begin{remark}[Why only one covariance direction survives]\label{rem:filterorthogonality}
The identity in Theorem~\ref{thm:filter} does not require the latent array $\{e_{i,n}\}$ to be independent across $i$. What matters is that the iid assignment vector is independent of the full latent array. Writing $A_{i,n}=\mu_{A,n}+\widetilde A_{i,n}$, centering and iid assignment eliminate every cross-index covariance term containing $\widetilde A_{i,n}$. Dependence survives only through $\mu_{A,n}'e_{i,n}$, which is why the relevant off-diagonal covariance correction is the scalar
\[
\mu_{A,n}'(\Omega_{e,n}-\Omega_{e,0,n})\mu_{A,n}
\]
rather than the full latent covariance matrix.
\end{remark}

\begin{remark}[Why non-iid assignment requires a new filter]\label{rem:designboundary}
The iid assignment assumption matters. Under complete randomization with exactly $n_{1,n}$ treated units, put $p_n=n_{1,n}/n$. Then $\sum_i(D_{i,n}-p_n)=0$ identically, so the Bernoulli covariance filter need not survive as an $O(n^{-1})$ perturbation. For example, if $Y_i(0)=U$, $EU=0$, $E[U^2]>0$, and the treatment effect is zero, then
\[
n^{-1/2}\sum_i(D_{i,n}-p_n)U
\]
is identically zero under fixed-count assignment but has variance $p_n(1-p_n)E[U^2]$ under iid Bernoulli assignment with probability $p_n\in(0,1)$. Matched, covariate-adaptive, and cluster-level assignment likewise change the assignment covariance structure. Such designs require their own score filter rather than a generic correction. Regression adjustment can reduce residual heterogeneity when covariates explain common effect modifiers \citep{Lin2013}, but its adjusted score and surviving covariance projection must also be re-derived.
\end{remark}

Define the relevance index
\[
R_{A,n}=\frac{\Delta_{A,n}}{\sigma_{D\varepsilon,n}^2}.
\]
The exact oracle identity implies
\[
\frac{V_n^\#}{V_n}=1-R_{A,n}.
\]
Hence $R_{A,n}$ is the signed off-diagonal covariance contribution relative to the first-order sampling variance in the direction that survives random assignment. It is this dimensionless quantity---rather than dependence in the full outcome process---that governs dependence-ignorant bootstrap validity.

\begin{remark}[The phase function is well defined, and the failure is one-sided]\label{prop:rbound}
Suppose the covariance-filter conditions of Theorem~\ref{thm:filter} hold and $V_n>0$. Then $s_{\#,n}^2>0$ and therefore $R_{A,n}<1$ for every $n$ satisfying these conditions. Without a uniform lower bound on the diagonal score scale, however, a sequence $R_{A,n}$ may still approach one. If Assumption~\ref{ass:stable} also holds, then for all sufficiently large $n$,
\[
R_{A,n}\le1-\frac{c}{C}<1,
\]
where $c$ and $C$ are the constants in Assumption~\ref{ass:stable}.
\end{remark}

\begin{theorem}[Marginal indistinguishability and an information limit]\label{thm:marginalimpossibility}
Fix $p\in(0,1)$ and $\theta\in(0,1)$. Let
\[
D_i\stackrel{iid}{\sim}\operatorname{Bernoulli}(p),\qquad
u_i\stackrel{iid}{\sim}N(0,\sigma_u^2),\qquad
\eta_i\stackrel{iid}{\sim}N(0,1),
\]
with the three sequences mutually independent, and consider the following two binary-treatment designs
\[
Y_i^{\pm}(0)=u_i,
\qquad
\tau_i^{\pm}=\eta_i\pm\theta\eta_{i-1},
\qquad
Y_i^{\pm}=Y_i^{\pm}(0)+D_i\tau_i^{\pm}.
\]
Then:
\begin{enumerate}[label=(\roman*)]
\item the one-observation law of $(D_i,Y_i^+)$ is identical to that of $(D_i,Y_i^-)$;
\item the diagonal bootstrap target is identical under the two designs and equals
\[
V^{\#}
=
\frac{\sigma_u^2}{p(1-p)}+
\frac{1+\theta^2}{p};
\]
\item their aggregate treatment-effect covariance contributions satisfy
\[
\Delta_{\tau,n}^{\pm}
=
\pm 2\theta\left(1-\frac1n\right),
\]
so their leading-score sampling variances satisfy
\[
V_n^{\pm}
=
V^{\#}
\pm2\theta\left(1-\frac1n\right),
\qquad
V_n^+-V_n^-
=
4\theta\left(1-\frac1n\right).
\]
Under both designs the population treatment coefficient is zero. The leading score is stationary and one-dependent, and, in this intercept-only construction,
\[
\frac{\sqrt n\,\widehat\beta_n^{\pm}}{(V_n^{\pm})^{1/2}}\dto N(0,1).
\]
\end{enumerate}
Call a variance estimator $\widehat V_n^{m}$ \emph{asymptotically marginal} on these two designs if there exists a deterministic sequence of functionals $\{\nu_n(\cdot)\}$ of the one-observation law such that, under either design,
\[
\widehat V_n^{m}
-
\nu_n\!\left(\mathcal L(D_i,Y_i)\right)
\pto0.
\]
Allowing the functional to vary with $n$ makes the class broader than one with a fixed probability-limit functional.
No asymptotically marginal variance estimator can satisfy
\[
\widehat V_n^{m}/V_n^{+}\pto1
\quad\text{under the $+$ design}
\qquad\text{and}\qquad
\widehat V_n^{m}/V_n^{-}\pto1
\quad\text{under the $-$ design}.
\]
Consequently, no conditionally centered-Gaussian bootstrap whose conditional variance is asymptotically marginal can be first-order valid under both designs. More generally, if a bootstrap law is first-order valid and conditionally uniformly integrable in squared standardized units as in Theorem~\ref{thm:ui}, then its conditional second moment cannot be asymptotically marginal over this pair.
\end{theorem}

Theorem~\ref{thm:marginalimpossibility} concerns \emph{marginal-only} procedures; it does not rule out data-driven estimators that use joint dependence information. The two designs are indistinguishable in the one-observation law, but not in the joint law of the sample: their cross-unit covariance structures differ in sign. Joint-dependence information is therefore indispensable, but it need not arrive through a known ordering; a grouping, spatial metric, network, or other defensible dependence model can distinguish the designs. What no refinement of the one-observation resampling law can supply is the sign and magnitude of the surviving covariance.

The construction is intercept-only by design. If the analyst observes covariates that proxy the latent effect modifiers, the one-observation law of $(D_i,Y_i,W_i)$ need not remain common across the two designs, and the impossibility theorem does not apply to procedures that exploit those covariates. Observed effect modifiers therefore provide an important qualification. Treatment interactions or regression adjustment based on such covariates, in the spirit of \citet{Lin2013}, can reduce the residual heterogeneity whose covariance generates $\Delta_{\tau,n}$; they complement rather than replace a dependence correction for the remaining score.

The information limit also has a useful finite-population analogue. Classical Neyman variance calculations depend on treatment-effect heterogeneity that is not point identified from the two marginal potential-outcome distributions; \citet{AronowGreenLee2014} develop sharp variance bounds for that problem. Here the unidentified-from-marginals object is cross-unit treatment-effect covariance. Because it can have either sign, marginals alone do not tell the researcher whether to raise or lower the diagonal target: exact first-order correction requires joint-dependence information. This does not preclude conservative bounds under additional restrictions.

\begin{theorem}[Covariance quotient and scale-free minimal dependence correction]\label{thm:minimality}
Let $\mathbb S^L$ denote the ambient space of symmetric $L\times L$ matrices and define
\[
H_n=\Omega_{e,n}-\Omega_{e,0,n},
\qquad
\mathcal L_n(B)=\mu_{A,n}'B\mu_{A,n},
\]
with null space
\[
\mathcal N_n=\{B\in\mathbb S^L:\mathcal L_n(B)=0\}.
\]
Under the exact leading-score covariance-filter conditions of Theorem~\ref{thm:filter}, define the first-order variance correction
\[
c_n:=V_n-V_n^\#
=
\sigma_{D,n}^{-4}\mathcal L_n(H_n).
\]
Hence:
\begin{enumerate}[label=(\roman*)]
\item holding the assignment law, $\mu_{A,n}$, $\sigma_{D,n}^2$, and the diagonal bootstrap target $V_n^\#$ fixed, two \emph{model-compatible off-diagonal covariance operators} $H_n$ and $\widetilde H_n$ generate the same deterministic first-order variance target $V_n$ whenever
\[
H_n-\widetilde H_n\in\mathcal N_n;
\]
\item if $\mu_{A,n}\neq0$, then the ambient quotient $\mathbb S^L/\mathcal N_n$ is one-dimensional; model-compatible covariance operators inherit the equivalence relation by intersection with its cosets. If $\mu_{A,n}=0$, every model-compatible off-diagonal covariance direction is irrelevant for the first-order variance target $V_n$;
\item suppose $V_n>0$ and $V_n^\#>0$ for all sufficiently large $n$,
\[
\frac{\widetilde V_n^\#-V_n^\#}{V_n}\pto0,
\]
and define
\[
\widehat V_n^{\,*}
=
\widetilde V_n^\#+\widehat c_n,
\qquad
P(\widehat V_n^{\,*}>0)\to1.
\]
On the event $\{\widehat V_n^{\,*}>0\}$, let $T_n^*$ be conditionally centered Gaussian with conditional variance $\widehat V_n^{\,*}$; on its complement, define $T_n^*$ arbitrarily. If
\[
\frac{\sqrt n(\widehat\beta-\beta_0)}{V_n^{1/2}}\dto N(0,1),
\]
then
\[
\sup_x\left|
P^*\{T_n^*\le x\}
-
P\{\sqrt n(\widehat\beta-\beta_0)\le x\}
\right|\pto0
\]
if and only if
\[
\boxed{
\frac{\widehat c_n-c_n}{V_n}
\pto0.
}
\]
\end{enumerate}
In particular, if
\[
\widehat c_n
=
\sigma_{D,n}^{-4}\mathcal L_n(\widehat H_n),
\]
then first-order validity requires and, within the stated centered-Gaussian class, is implied by
\[
\boxed{
\frac{
\sigma_{D,n}^{-4}\mathcal L_n(\widehat H_n-H_n)
}{V_n}
\pto0.
}
\]
Thus full-matrix covariance consistency is generally stronger than necessary. Errors in directions belonging to $\mathcal N_n$ are exactly irrelevant, while errors outside $\mathcal N_n$ need only be negligible at the first-order sampling-variance scale.
\end{theorem}

\begin{remark}[Representation invariance of the filtered covariance]\label{rem:representation}
The latent factorization $\varepsilon_i=\lambda(D_i)'e_i$ is a convenient representation, not an identified coordinate system. The filtered covariance correction is invariant to nonsingular changes of latent coordinates. In particular, for any nonsingular $L\times L$ matrix $C_n$, define
\[
e_{i,n}^{\star}=C_ne_{i,n},
\qquad
\lambda_n^{\star}(d)=C_n^{-\prime}\lambda_n(d).
\]
Then $\varepsilon_{i,n}=\lambda_n^{\star}(D_{i,n})'e_{i,n}^{\star}$,
\[
\mu_{A,n}^{\star}=C_n^{-\prime}\mu_{A,n},
\qquad
H_n^{\star}=C_nH_nC_n',
\]
and therefore
\[
(\mu_{A,n}^{\star})'H_n^{\star}\mu_{A,n}^{\star}
=
\mu_{A,n}'H_n\mu_{A,n}.
\]
The scalar correction $c_n$ and the one-dimensional quotient are therefore invariant under nonsingular reparameterization. The ambient null space itself changes coordinates, but the induced equivalence classes are mapped isomorphically. What matters is the scalar filtered covariance, not a particular choice of latent basis.
\end{remark}

\begin{lemma}[The covariance quotient is attained by model-compatible processes]\label{lem:sharpquotient}
Consider the binary specialization of Section~\ref{sec:binary} with fixed $p\in(0,1)$ and $q=p(1-p)$. Let $Y_i(0)=u_i$, where $\{u_i\}$ is iid with finite variance, and let
\[
\tau_i=\eta_i+\theta\eta_{i-1},\qquad \theta\in\mathbb R,
\]
with $\{\eta_i\}$ iid, mean zero, variance one, and independent of $\{u_i\}$ and assignment. Then the latent process is stationary and satisfies the maintained second-moment and assignment conditions, and
\[
\mathcal L_n(H_n)=q^2\Delta_{\tau,n}=2q^2\theta\left(1-\frac1n\right).
\]
Hence, for every $n\ge2$, varying $\theta$ moves the model-compatible covariance class through a continuum of distinct cosets of $\mathcal N_n$, including both signs. The one-dimensional quotient in Theorem~\ref{thm:minimality}(ii) is therefore attained rather than merely an ambient algebraic possibility.
\end{lemma}

\begin{remark}[Geometry versus asymptotic relevance]\label{rem:minimality}
Applying HAC or dependent multipliers to an asymptotically linear score is standard. Theorem~\ref{thm:minimality} identifies the additional content supplied by random assignment: before any HAC device is chosen, random assignment quotients out, in the ambient symmetric-matrix representation, all off-diagonal covariance directions in $\mathcal N_n$. When $\mu_{A,n}\neq0$ this quotient is algebraically one-dimensional, but its statistical relevance can still vanish along a triangular array. In particular,
\[
\frac{c_n}{V_n}\to0
\]
makes the iid correction asymptotically unnecessary even if $\mu_{A,n}\neq0$ for every $n$. Conversely, $\|\mu_{A,n}\|\to0$ alone does not imply $c_n/V_n\to0$ unless the latent covariance operator is controlled. The scale-free object is the surviving covariance contribution relative to $V_n$, not the norm of $\mu_{A,n}$ by itself.
\end{remark}

\begin{theorem}[Hahn--Liao moment bridge]\label{thm:ui}
Let $Z_n^*=T_n^*/V_n^{1/2}$, where $V_n>0$. Suppose
\[
\sup_x\left|P^*\{Z_n^*\le x\}-\Phi(x)\right|\pto0
\]
and $\{(Z_n^*)^2\}$ is conditionally uniformly integrable in probability, meaning that for every $\varepsilon>0$,
\[
\lim_{M\to\infty}\limsup_{n\to\infty}
P\!\left(
E^*\!\left[(Z_n^*)^2\mathbf 1\{|Z_n^*|>M\}\right]>\varepsilon
\right)=0.
\]
Then
\[
E^*[(Z_n^*)^2]\pto1,\qquad E^*[Z_n^*]\pto0,
\qquad \frac{\Var^*(T_n^*)}{V_n}\pto1.
\]
The second conclusion is needed because $T_n^*$ is not assumed conditionally centered in this theorem. Conditional uniform integrability of $(Z_n^*)^2$ implies conditional uniform integrability of $Z_n^*$; together with conditional weak convergence this yields $E^*[Z_n^*]\to_p0$. The variance conclusion therefore follows from the first two moments rather than from the second moment alone. If, in addition,
\[
E^*[(T_n^*)^2]=\widetilde V_n^\#+\widehat c_n,
\qquad
\frac{\widetilde V_n^\#-V_n^\#}{V_n}\pto0,
\]
then weak validity necessarily implies
\[
\boxed{\frac{\widehat c_n-c_n}{V_n}\pto0},
\qquad
c_n:=V_n-V_n^\#.
\]
Under the covariance-filter conditions of Theorem~\ref{thm:filter}, this correction is
\[
c_n=\sigma_{D,n}^{-4}\mathcal L_n(H_n).
\]
Thus, under moment control, an arbitrary bootstrap law must recover the same filtered second moment identified by Theorem~\ref{thm:minimality}. The converse is not asserted outside the Gaussian class.
\end{theorem}

A primitive sufficient condition for the uniform-integrability requirement is
\[
E^*|Z_n^*|^{2+\eta}=O_p(1)
\quad\text{for some }\eta>0.
\]
This condition is automatic for the Gaussian projected-dependence bootstrap whenever its conditional variance ratio is tight. Online Appendix Section~\ref{app:ui} also verifies it under the existing fourth-moment conditions for the regularized pairs bootstrap and, after conditioning on nondegenerate resamples, for the binary-treatment pairs bootstrap. It is \emph{not} automatic for unrestricted ordinary pairs OLS with controls: rare nearly singular resampled Hessians can preserve weak validity while destroying bootstrap moments. This is the failure mode emphasized by Hahn--Liao.

\begin{assumption}[Scale-free design and diagonal-score consistency]\label{ass:scalefree}
Under $H_0:\beta=\beta_0$, suppose $\sigma_{D,n}^2>0$ and $s_{\#,n}^2>0$. Restoring the row index in the design scale, assume
\[
\frac{\widehat Q_n}{\sigma_{D,n}^2}\pto1,
\qquad
\frac{n^{-1}\sum_{i=1}^n\{\widetilde\psi_i(\beta_0)-\psi_i\}^2}{s_{\#,n}^2}=o_p(1),
\qquad
\frac{n^{-1}\sum_{i=1}^n\psi_{i,n}^2-s_{\#,n}^2}{s_{\#,n}^2}=o_p(1).
\]
\end{assumption}

Assumption~\ref{ass:scalefree} collects the three relative requirements needed when the treatment or diagonal-score scale can shrink: treatment-design concentration, feasible-score replacement, and oracle diagonal-score concentration. Under Assumption~\ref{ass:stable}, Lemma~\ref{lem:design} and Lemmas~\ref{lem:restricted}--\ref{lem:diag} imply these requirements from the primitive conditions used below. If $s_{\#,n}$ can shrink, the absolute score-replacement bound
\[
n^{-1}\sum_i
\{\widetilde\psi_i(\beta_0)-\psi_i\}^2=o_p(1)
\]
is not enough by itself; a rate relative to $s_{\#,n}^2$ is genuinely needed. Primitive sufficient rates for that case are given in Online Appendix Section~\ref{app:proofs-main}.

The relative formulation matters in triangular arrays with diminishing treatment variation, for example Bernoulli designs with $p_n\to0$ or $p_n\to1$, and in sequences where the diagonal score scale itself becomes small. In sparse-treatment designs $\sigma_{D,n}^2=p_n(1-p_n)$ vanishes and the OLS variance need not shrink---it can instead diverge. This is not weak identification in the instrumental-variables sense. Rather, an absolute $o_p(1)$ error may still be large relative to a moving design or score scale, which is why the scale-free conditions are stated in relative terms.

\begin{theorem}[Feasible variance target]\label{thm:feasible}
Under $H_0:\beta=\beta_0$, suppose the exact covariance-filter conditions of Theorem~\ref{thm:filter} hold. No lower bound on $V_n$ is needed for this variance-target result.
\begin{enumerate}[label=(\alph*)]
\item If Assumption~\ref{ass:scalefree} holds, then
\[
\frac{\widehat Q_n}{\sigma_{D,n}^2}\pto1,
\qquad
\frac{\widetilde V_n^\#(\beta_0)}{V_n^\#}\pto1.
\]
\item Suppose, in addition, the conditions of Lemmas~\ref{lem:diag} and \ref{lem:restricted} and Assumption~\ref{ass:stable} hold. Then Lemma~\ref{lem:design} and Lemmas~\ref{lem:diag} and \ref{lem:restricted} imply all three clauses of Assumption~\ref{ass:scalefree}, and
\[
\widetilde V_n^\#(\beta_0)-V_n^\#=\op(1).
\]
Consequently,
\[
\widetilde V_n^\#(\beta_0)-V_n
=
-\sigma_{D,n}^{-4}\Delta_{A,n}+\op(1).
\]
\end{enumerate}
\end{theorem}

\begin{theorem}[Bootstrap validity frontier and size phase diagram]\label{thm:phase}
Suppose the covariance-filter conditions of Theorem~\ref{thm:filter} hold, $V_n>0$ and $V_n^\#>0$ for all sufficiently large $n$, and
\[
\frac{\widetilde V_n^\#(\beta_0)}{V_n^\#}\pto1,
\]
as guaranteed by Theorem~\ref{thm:feasible} under either its scale-free or stable-scale route, and suppose the CHLS standardized sampling law satisfies
\[
\frac{\sqrt n(\widehat\beta-\beta_0)}{V_n^{1/2}}\dto N(0,1)
\]
under $H_0$. If
\[
R_{A,n}\to r<1,
\]
then
\[
\frac{\sqrt n(\widehat\beta-\beta_0)}
{\{\widetilde V_n^\#(\beta_0)\}^{1/2}}
\dto
N\!\left(0,\frac1{1-r}\right).
\]
Consequently, the two-sided nominal level-$\alpha$ test based on the iid multiplier score-bootstrap standard error has limiting rejection probability
\[
\boxed{
\pi_\alpha(r)
=
2\Phi\!\left(
-z_{1-\alpha/2}\sqrt{1-r}
\right).
}
\]
Thus $r=0$ is the first-order validity frontier for the iid multiplier score-bootstrap standard error, $r>0$ yields anti-conservative inference, and $r<0$ yields conservative inference. Remark~\ref{prop:rbound} gives $R_{A,n}<1$ at each finite $n$ satisfying the stated conditions, but sequences may approach one. The maintained condition $r<1$ excludes that scale-free boundary, where $V_n^\#/V_n\to0$; Assumption~\ref{ass:stable} ensures uniform separation from one.
\end{theorem}

\begin{corollary}[Oracle equal-size benchmark and feasible local power]\label{cor:localpower}
Suppose the covariance-filter conditions of Theorem~\ref{thm:filter} hold, $V_n\to V\in(0,\infty)$, $V_n^\#>0$ for all sufficiently large $n$, and $R_{A,n}\to r<1$. Along the local alternatives
\[
\beta_n=\beta_0+\frac{c}{\sqrt n},
\]
suppose
\[
\frac{\sqrt n(\widehat\beta-\beta_n)}{V_n^{1/2}}\dto N(0,1),
\qquad
\frac{\widetilde V_n^\#(\beta_0)}{V_n^\#}\pto1.
\]
Let $\delta=c/V^{1/2}$ and $z=z_{1-\alpha/2}$. Then the unadjusted iid-variance test has limiting local rejection probability
\[
\Phi\!\left(-z\sqrt{1-r}-\delta\right)
+1-\Phi\!\left(z\sqrt{1-r}-\delta\right).
\]
Any first-order valid projected test whose variance estimator remains relatively consistent for $V_n$ along the same local alternatives has limiting local power
\[
\Phi(-z-\delta)+1-\Phi(z-\delta).
\]
As a theoretical equal-size benchmark, if the iid statistic is given the oracle critical value $z/\sqrt{1-r}$ so that its asymptotic null rejection probability is exactly $\alpha$, its local-power curve is also
\[
\Phi(-z-\delta)+1-\Phi(z-\delta).
\]
Thus the raw first-order power difference generated by the wrong iid variance is a calibration effect. The oracle iid statistic can attain the same first-order curve only by using the unknown relevance-index calibration $r$. A first-order valid projected procedure attains that curve feasibly by estimating the surviving covariance contribution. The correction restores calibration and feasibility; it does not create a first-order efficiency gain in this Gaussian local experiment.
\end{corollary}

Corollary~\ref{cor:localpower} separates calibration from efficiency. Positive surviving covariance can make the raw iid test appear more powerful only because it over-rejects under the null, while negative covariance can make it appear less powerful because it under-rejects. Once the two procedures are calibrated to the same asymptotic size, the first-order local-power difference disappears. Feasibility is what remains: the equal-size iid benchmark requires the unknown limiting relevance index, whereas the projected procedure estimates the covariance contribution that determines it. The projected correction should therefore be viewed as a feasible size correction, not as a source of first-order power gains.

\begin{figure}[!htbp]
\centering
\includegraphics[width=.72\textwidth]{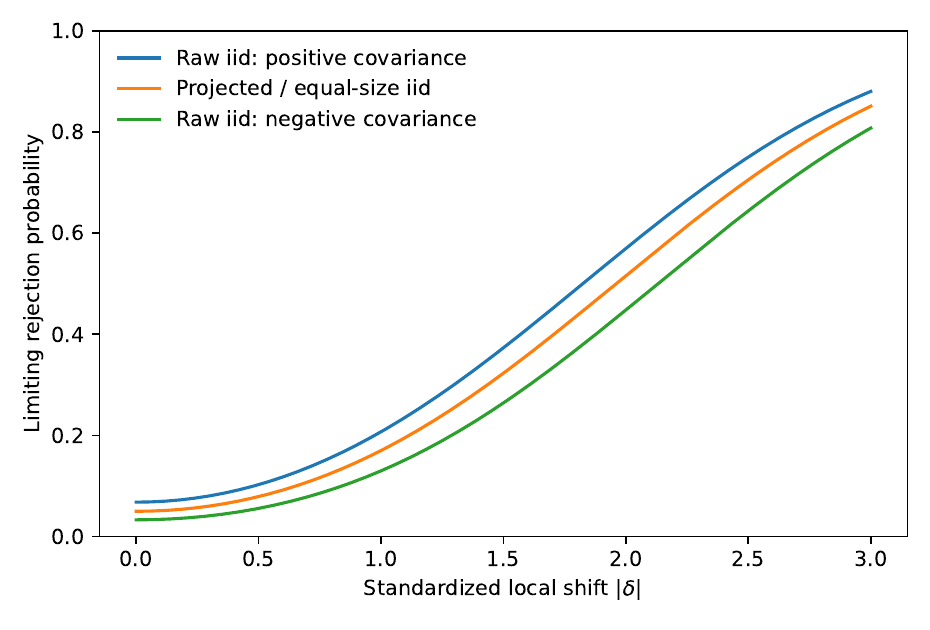}
\caption{Asymptotic local rejection probabilities as a function of the standardized local shift $|\delta|$. The raw iid curves use the limiting relevance indices of the positive and negative MA(1) benchmark designs. The projected curve coincides exactly with the iid curve after equal-size calibration, so the raw power differences are calibration effects rather than first-order efficiency gains.}
\label{fig:localpower}
\end{figure}

\begin{remark}[Separation from CHLS]\label{rem:chlsnovelty}
CHLS characterize the first-order OLS variance represented here by $V_n$. We take their sampling linearization as given; the corresponding multivariate expansion is developed in their setup and formalized in their main theory. The present paper characterizes the bootstrap target $V_n^\#$ induced by dependence-ignorant weighting and compares it with $V_n$. The covariance-filter identity, validity frontier, minimality characterization, and projected repair are therefore bootstrap results rather than alternative derivations of the CHLS variance formula.
\end{remark}

\section{Binary treatment and heterogeneous effects}\label{sec:binary}

For the cleanest potential-outcomes specialization, take $W_{i,n}=1$ and let
\[
D_{i,n}\stackrel{iid}{\sim}\operatorname{Bernoulli}(p_n),
\qquad
0<p_n<1,
\]
independently of the full potential-outcome array. Write
\[
Y_{i,n}
=
Y_{i,n}(0)+D_{i,n}\tau_{i,n},
\qquad
\tau_{i,n}=Y_{i,n}(1)-Y_{i,n}(0).
\]
In this binary representation, constant treatment effects recover the strong-exogeneity benchmark; heterogeneous $\tau_{i,n}$ is the economically important route through which assignment-dependent regression errors arise.
Assume that, within each row,
\[
E Y_{i,n}(0)=\alpha_n,
\qquad
E\tau_{i,n}=\beta_n
\]
for every $i$. Then the population coefficient in the regression of $Y_{i,n}$ on an intercept and $D_{i,n}$ is exactly $\beta_n$. Indeed, with
\[
q_n=p_n(1-p_n),
\]
independence of assignment and potential outcomes gives
\[
E[(D_{i,n}-p_n)Y_{i,n}]
=
q_n\beta_n.
\]
The associated regression error is
\[
\varepsilon_{i,n}
=
Y_{i,n}(0)-\alpha_n
+
D_{i,n}(\tau_{i,n}-\beta_n).
\]
Thus the CHLS representation can be written as
\[
e_{i,n}
=
\begin{pmatrix}
Y_{i,n}(0)-\alpha_n\\
\tau_{i,n}-\beta_n
\end{pmatrix},
\qquad
\lambda_n(D_{i,n})
=
\begin{pmatrix}
1\\D_{i,n}
\end{pmatrix}.
\]
Since
\[
A_{i,n}
=
(D_{i,n}-p_n)
\begin{pmatrix}
1\\D_{i,n}
\end{pmatrix},
\]
we have exactly
\[
\mu_{A,n}
=
\begin{pmatrix}
0\\q_n
\end{pmatrix},
\qquad
\mu_{A,n}'e_{i,n}
=
q_n(\tau_{i,n}-\beta_n).
\]

\begin{corollary}[Treatment-effect covariance discrepancy]\label{cor:te}
Under the exact covariance-filter conditions of Theorem~\ref{thm:filter} and the binary-treatment specialization above, define
\[
\Delta_{\tau,n}
=
\frac1n\sum_{i\ne j}
\Cov(\tau_{i,n},\tau_{j,n}).
\]
Then, exactly for every $n$ satisfying the stated conditions at the level of the leading-score variance sequences,
\[
\Delta_{A,n}=q_n^2\Delta_{\tau,n}
=\sigma_{D,n}^4\Delta_{\tau,n},
\]
and therefore
\[
\boxed{
V_n^\#-V_n=-\Delta_{\tau,n}.
}
\]
The displayed equality is not an exact finite-sample identity for $\Var\{\sqrt n(\widehat\beta-\beta)\}$; its implication for OLS inference uses the CHLS linearization.
Whenever $V_n>0$,
\[
\boxed{
R_{A,n}=\frac{\Delta_{\tau,n}}{V_n}.
}
\]
\end{corollary}

The sign classification concerns a non-negligible asymptotic contribution. Under the standardized sampling central limit theorem (CLT) in Theorem~\ref{thm:phase}, for example, if
\[
V_n\to V\in(0,\infty),
\qquad
\Delta_{\tau,n}\to\Delta_\tau,
\qquad
V-\Delta_\tau>0,
\]
then $\Delta_\tau>0$ implies anti-conservative dependence-ignorant inference, $\Delta_\tau=0$ implies first-order correct inference, and $\Delta_\tau<0$ implies conservative inference. More generally the relevant object is the relative discrepancy $\Delta_{\tau,n}/V_n$. Treatment-effect independence is sufficient but not necessary for first-order validity.

\begin{corollary}[Binary-treatment size distortion]\label{cor:size}
Under the conditions of Theorem~\ref{thm:phase}, in the binary-treatment specialization suppose
\[
\frac{\Delta_{\tau,n}}{V_n}\to\rho<1.
\]
Finite-$n$ strictness from Remark~\ref{prop:rbound} does not rule out the boundary limit $\rho=1$; Assumption~\ref{ass:stable} does. Under the maintained $\rho<1$, the two-sided nominal level-$\alpha$ iid multiplier score-bootstrap test has limiting rejection probability
\[
2\Phi\!\left(
-z_{1-\alpha/2}\sqrt{1-\rho}
\right).
\]
Thus positive aggregate treatment-effect covariance is anti-conservative, negative aggregate covariance is conservative, and asymptotically negligible aggregate covariance is first-order correct.
\end{corollary}

\subsection{Relation to design-based clustering and practical guidance}\label{sec:practice}
The binary identity is closely related to the treatment-effect-heterogeneity channel emphasized by \citet{AbadieEtAl2023} and to the randomization/clustering perspective of \citet{BarriosEtAl2012}, but it answers a different inferential question. Their design-based analysis asks when clustered variance estimation is warranted by the sampling and assignment design. Here assignment remains iid at the unit level and the object of interest is whether an independent resampling scheme reproduces the covariance that survives random assignment. The exact identity adds two features that are especially useful for practice: the correction is signed, and only the treatment-effect covariance contribution relevant to the scalar randomized-regressor score must be recovered.

Three practical implications follow for the randomized-regressor coefficient under the maintained iid unit-level assignment. First, outcome dependence alone is not a sufficient reason to cluster this coefficient: under strong exogeneity, Theorem~\ref{thm:strong} gives $V_n^\#=V_n$ even with broad cross-unit dependence in outcomes and controls. Second, when heterogeneous treatment effects co-move, the relevant dependence geometry is the geometry of the effect modifiers---shared teachers, local labor markets, institutions, spatial exposure, or another defensible grouping---rather than automatically the level at which treatment was assigned. Third, the correction is not a one-way conservative adjustment: negative treatment-effect covariance makes dependence-ignorant inference conservative, so a valid correction can narrow rather than widen confidence intervals.

In practice, we recommend reporting the diagonal score variance together with projected variances under substantively defensible, preferably prespecified, dependence specifications. If several are plausible, report all as sensitivity analysis rather than select among them ex post. Their gap is the primary object; the ratio $\widehat r_n$ is a descriptive normalization, not a sign pretest. A useful empirical placebo, when pre-treatment outcomes are available, is to apply the same projected-score calculation to a pre-treatment outcome: under the maintained mechanism there is then no treatment-effect covariance to survive random assignment, even if the outcome itself is strongly correlated across groups. This is a diagnostic implication, not a substitute for the iid-assignment requirement stated in the Introduction.

For a covariance-stationary treatment-effect process with autocovariance
\[
\varrho_{\tau,n}(h)=\Cov(\tau_{i,n},\tau_{i-h,n})
\]
that does not depend on $i$ within row $n$,
\[
\Delta_{\tau,n}
=
2\sum_{h=1}^{n-1}
\left(1-\frac hn\right)\varrho_{\tau,n}(h).
\]
If $\varrho_{\tau,n}(h)=\varrho_\tau(h)$ and
$\sum_{h\ge1}|\varrho_\tau(h)|<\infty$, then
\[
\Delta_{\tau,n}
\to
2\sum_{h\ge1}\varrho_\tau(h).
\]

\section{Implication for ordinary pairs bootstrap}\label{sec:pairs}
Let $\widehat\gamma$ and $\widehat\varepsilon_i=Y_i-D_i\widehat\beta-W_i'\widehat\gamma$ denote the unrestricted OLS nuisance estimate and residuals. Let
\[
\widehat v=M_WD,\qquad \widehat Q_D=n^{-1}\sum_i\widehat v_i^{\,2}=n^{-1}D'M_WD.
\]
The ordinary pairs bootstrap draws $n$ rows with replacement. Equivalently,
\[
(N_1^*,\ldots,N_n^*)\sim\operatorname{Multinomial}(n;n^{-1},\ldots,n^{-1})
\]
counts how often each observation is drawn. Its conditional moments are $E^*N_i^*=1$ and $\Var^*(N_i^*)=1-n^{-1}$, while $\Cov^*(N_i^*,N_j^*)=-n^{-1}$ for $i\ne j$.
Ordinary pairs resampling does not evade the covariance filter. Classical analyses of regression resampling include \citet{Freedman1981} and \citet{GoncalvesWhite2005}. Because row resampling perturbs both the score and the design matrix, its linearization requires leverage and empirical-Lindeberg regularity beyond the baseline CHLS conditions. Under those conditions, the first-order pairs statistic is
\[
\frac1{\sqrt n}\sum_{i=1}^n(N_i^*-1)\widehat\phi_i,
\qquad
\widehat\phi_i
=
\widehat Q_D^{-1}\widehat v_i\widehat\varepsilon_i,
\]
and its conditional variance is exactly the usual HC0 heteroskedasticity-robust coefficient variance,
\[
\widehat V_{\mathrm{HC0},\beta}
=
\widehat Q_D^{-2}\frac1n\sum_{i=1}^n
\widehat v_i^2\widehat\varepsilon_i^2.
\]
The equality is exact: the multinomial cross-terms contribute a term proportional to
\[
-\left(\sum_i\widehat v_i\widehat\varepsilon_i\right)^2,
\]
which vanishes identically because $\sum_i\widehat v_i\widehat\varepsilon_i=D'M_W\widehat\varepsilon=0$ by the OLS first-order conditions.
Theorem~\ref{thm:pairs} in the Online Appendix, under Assumption~\ref{ass:pairs}, proves the corresponding conditional central limit theorem and shows
\[
\widehat V_{\mathrm{HC0},\beta}
=
V_n^\#+o_p(1).
\]
Hence ordinary pairs resampling has the same first-order diagonal target as the iid multiplier score bootstrap. Analytical variance estimation is not inherently invalid here. HC0 and iid row resampling fail because both are diagonal and therefore discard surviving cross-unit score covariance. A dependence-aware HAC, cluster, or spatial variance estimator for the relevant residualized score can instead target $V_n$. Combining this result with Theorem~\ref{thm:filter},
\[
\widehat V_{\mathrm{HC0},\beta}-V_n
=
-\sigma_{D,n}^{-4}\Delta_{A,n}+o_p(1).
\]
For binary randomized treatment this reduces to
\[
\widehat V_{\mathrm{HC0},\beta}-V_n
=
-\Delta_{\tau,n}+o_p(1).
\]
Thus the pairs bootstrap inherits the same variance-target phase. Online Appendix Corollary~\ref{cor:pairsphase} formalizes the corresponding size result: under the sampling-law and relevance-index conditions of Theorem~\ref{thm:phase} with $R_{A,n}\to r<1$, a two-sided test using ordinary-pairs quantiles of $\sqrt n(\widehat\beta^*-\widehat\beta)$ has the same limiting rejection probability $\pi_\alpha(r)$ as the iid multiplier score-bootstrap test. It is therefore anti-conservative for $r>0$, first-order correct for $r=0$, and conservative for $r<0$. Without those sampling-law and relative-gap conditions, the displays above are variance-target statements rather than a rejection-probability theorem.

This first-order statement does not by itself imply convergence of the conditional variance of the nonlinear pairs-OLS estimator; that requires conditional uniform integrability or an equivalent tail condition. Online Appendix Section~\ref{app:pairs-full} treats this Hahn--Liao distinction, a regularized pairs construction with controls, and an unregularized binary-treatment procedure conditional on nondegenerate resamples whose bootstrap variance satisfies
\[
V_{\mathrm{pairs},H,n}^*-V_n
=
-\Delta_{\tau,n}+o_p(1).
\]
These results concern iid row resampling and do not cover fixed-treatment-count complete randomization, matched-pair, covariate-adaptive, or cluster-level assignment.

\section{Dependence correction on the projected score}\label{sec:repair}

The covariance-filter and minimality theorems suggest correcting only the dependence that survives random assignment. Theorem~\ref{thm:minimality} shows that, given a consistent diagonal component, the surviving covariance contribution measured relative to $V_n$ is necessary and sufficient for first-order conditionally centered-Gaussian bootstrap validity; the relevant first-order object is therefore the scalar randomized-regressor score, not the full vector of outcomes and controls.

Let $\widetilde\psi_i(\beta_0)$ be the feasible null-imposed score. Let
\[
\xi_n^*=(\xi_{1,n}^*,\ldots,\xi_{n,n}^*)'
\]
be conditionally mean-zero multipliers with
\[
E^*(\xi_n^*\xi_n^{*'})=K_n=(\kappa_{ij,n})_{i,j\le n},
\]
where $K_n$ is symmetric positive semidefinite. For Gaussian multipliers this condition is exactly what guarantees the existence of the stated conditional covariance law.
Define the \emph{projected-dependence multiplier implementation}
\[
T_{\mathrm{PD},n}^*(\beta_0)
=
\widehat Q_n^{-1}\frac1{\sqrt n}
\sum_{i=1}^n\xi_{i,n}^*\widetilde\psi_i(\beta_0),
\]
with conditional variance
\[
\widetilde V_{\mathrm{PD},n}(\beta_0)
=
\widehat Q_n^{-2}\frac1n
\sum_{i,j=1}^n
\kappa_{ij,n}\widetilde\psi_i(\beta_0)\widetilde\psi_j(\beta_0).
\]
The iid multiplier score bootstrap is the special case $K_n=I_n$. Block, cluster, spatial, or dependent-multiplier choices of $K_n$ alter only the covariance of the scalar score multipliers; they do not resample the full data vector or recompute OLS. The projection is a dimension reduction, not a model-free dependence estimator: $K_n$ must still be chosen so that the resulting scalar quadratic form consistently targets the first-order score variance. When Gaussian multipliers are used, the conditional uniform-integrability requirement behind the Hahn--Liao moment issue is automatic as soon as the projected variance ratio is tight.

\begin{theorem}[Projected-dependence bootstrap principle]\label{thm:pd}
Under $H_0$, suppose $V_n>0$ for all sufficiently large $n$,
\[
P\{\widetilde V_{\mathrm{PD},n}(\beta_0)>0\}\to1,
\]
and
\[
\frac{\sqrt n(\widehat\beta-\beta_0)}{V_n^{1/2}}\dto N(0,1),
\qquad
\frac{\widetilde V_{\mathrm{PD},n}(\beta_0)}{V_n}\pto1,
\]
and, on the event $\{\widetilde V_{\mathrm{PD},n}(\beta_0)>0\}$, conditionally on the data,
\[
\sup_x\left|
P^*\!\left(
\frac{T_{\mathrm{PD},n}^*(\beta_0)}
{\{\widetilde V_{\mathrm{PD},n}(\beta_0)\}^{1/2}}\le x
\right)-\Phi(x)
\right|\pto0.
\]
On the complementary event the standardized statistic may be assigned any fixed bookkeeping value.
Then
\[
\sup_x\left|
P^*\{T_{\mathrm{PD},n}^*(\beta_0)\le x\}
-
P\{\sqrt n(\widehat\beta-\beta_0)\le x\}
\right|\pto0.
\]
Thus, once the standardized conditional bootstrap law is asymptotically Gaussian, first-order validity can be achieved by consistently reproducing the covariance of the projected score only; reproducing the full joint dependence of $(Y_i,W_i)$ is unnecessary. Necessity within the centered-Gaussian class is supplied separately by Theorem~\ref{thm:minimality}.
\end{theorem}

\paragraph{Recentering the null-imposed covariance score.}
One may replace $\widetilde\psi_i(\beta_0)$ inside the covariance estimator by
\[
\widetilde\psi_i^{c}(\beta_0)
=
\widetilde\psi_i(\beta_0)-\overline{\widetilde\psi}(\beta_0),
\qquad
\overline{\widetilde\psi}(\beta_0)=n^{-1}\sum_{i=1}^n\widetilde\psi_i(\beta_0).
\]
This is harmless only under an appropriate bandwidth regime. If $K_n$ is the covariance kernel, $\overline{\widetilde\psi}=O_p(n^{-1/2})$, $\mathbf 1'K_n\widetilde\psi=O_p(m_n\sqrt n)$, and $\mathbf 1'K_n\mathbf 1=O(nm_n)$, then the centered and uncentered quadratic forms differ by $O_p(m_n/n)$. Hence they are first-order equivalent under the small-bandwidth routes when $m_n=o(n)$ and the target scale is bounded away from zero. At fixed bandwidth, $m_n/n\to b>0$, the centering term is generally first order and can change the Brownian-motion reference functional to a bridge-type functional. The fixed-bandwidth theorem below therefore uses the nonrecentered null-imposed score; a recentered fixed-bandwidth implementation requires its own joint limit. The exact algebra and the stated sufficient orders are recorded in Online Appendix Remark~\ref{rem:recenter}.

Primitive implementations are given in the Online Appendix. Theorem~\ref{thm:pdmixing} treats a stationary strong-mixing projected score, a positive-semidefinite lag kernel, and kernel denominator $m_n=o(n^{1/3})$; its proof separates kernel bias, quadratic-form concentration, feasible-score replacement, and the finite-sample/long-run-variance edge correction. Theorem~\ref{thm:pdcumulant} gives a complementary fourth-order-cumulant route with $m_n=o(n)$, while Proposition~\ref{prop:pdcluster} gives an independent-cluster route under a cluster-sum fourth-moment diffuseness condition. These are sufficient implementations of Theorem~\ref{thm:pd}; their role is to show how the one-dimensional covariance target can be estimated under familiar dependence structures. A useful feature of the cumulant route is that the nuisance-estimation error is low rank: exploiting that structure sharpens feasible replacement from a crude bandwidth-times-estimation-error bound to $O_p(n^{-1/2})+O_p(m_n/n)$, which is what permits $m_n=o(n)$. Online Appendix Proposition~\ref{prop:pduniform} records the corresponding uniform-size transfer when the sampling CLT, positivity, and projected-variance consistency hold uniformly over a class of data-generating processes. The HAC theorems use deterministic $m_n$; plug-in bandwidths require an additional validity argument.

\subsection{Fixed-bandwidth calibration for ordered data}\label{sec:fixedb}
The small-bandwidth theory above uses a kernel bandwidth that is negligible relative to $n$ and therefore a standard-normal reference law. When the kernel bandwidth is a non-negligible fraction of the sample size, that approximation can be inaccurate even when the projected covariance target is correct. Fixed-bandwidth asymptotics instead let the kernel denominator $m_n$ satisfy $m_n/n\to b\in(0,1]$ and let the reference law depend on the kernel and $b$ \citep{KieferVogelsang2005,SunPhillipsJin2008}.

Let $K_{n,b}$ have entries $k((i-j)/m_n)$, where $m_n/n\to b$, and let $\widetilde V_{\mathrm{PD},n}^{(b)}$ denote the corresponding projected quadratic form. In the Bartlett implementation used below, $\ell$ denotes the largest included positive lag and the weights are $1-h/(\ell+1)$, so $m_n=\ell+1$ and the reported finite-sample bandwidth fraction is $b_n=(\ell+1)/n$. For a continuous positive-semidefinite kernel, define the fixed-bandwidth Brownian functional
\[
Q_{k,b}(W)=\int_0^1\!\int_0^1 k\!\left(\frac{r-s}{b}\right)dW(r)dW(s),
\]
interpreted as the standard $L^2$ limit of the associated quadratic forms.

\begin{theorem}[Fixed-bandwidth calibration of the projected statistic]\label{thm:fixedb}
Suppose $b\in(0,1]$ and $P\{\widetilde V_{\mathrm{PD},n}^{(b)}(\beta_0)>0\}\to1$. For some positive score scale $\Lambda_n$, assume
\[
\left(
\frac{n^{-1/2}\sum_{i=1}^n\psi_{i,n}}{\Lambda_n^{1/2}},
\frac{n^{-1}\psi_n'K_{n,b}\psi_n}{\Lambda_n}
\right)
\dto
\{W(1),Q_{k,b}(W)\},
\]
with $Q_{k,b}(W)>0$ almost surely. Suppose also that the OLS linearization and feasible-score replacement are negligible at this fixed-bandwidth scale:
\[
\frac{\sqrt n(\widehat\beta-\beta_0)-\sigma_{D,n}^{-2}n^{-1/2}\sum_i\psi_{i,n}}
{\sigma_{D,n}^{-2}\Lambda_n^{1/2}}\pto0,
\qquad
\frac{\widehat Q_n}{\sigma_{D,n}^2}\pto1,
\]
\[
\frac{n^{-1}\{\widetilde\psi'K_{n,b}\widetilde\psi-\psi'K_{n,b}\psi\}}{\Lambda_n}\pto0.
\]
Then
\[
\frac{\sqrt n(\widehat\beta-\beta_0)}
{\{\widetilde V_{\mathrm{PD},n}^{(b)}(\beta_0)\}^{1/2}}
\dto
\frac{W(1)}{\{Q_{k,b}(W)\}^{1/2}}.
\]
Hence, if $c_{1-\alpha}(b,k)$ is a continuity-point critical value satisfying
\[
P\!\left\{\frac{|W(1)|}{Q_{k,b}(W)^{1/2}}>c_{1-\alpha}(b,k)\right\}=\alpha,
\]
then the two-sided test that rejects when the absolute studentized statistic exceeds $c_{1-\alpha}(b,k)$ has asymptotic size $\alpha$.
\end{theorem}

The theorem concerns calibration; it does not claim that the fixed-bandwidth quadratic form consistently estimates $V_n$. It implements fixed-bandwidth inference directly through the projected HAC/HAR variance and the corresponding nonstandard critical value. A different route would derive the fixed-bandwidth conditional law of an unstudentized dependent-multiplier or dependent-wild-bootstrap statistic and then construct bootstrap critical values or adjusted $p$-values. That route requires a separate conditional joint-limit theorem and is not implied by the covariance calculation here. In particular, the Gaussian multiplier scheme in Theorem~\ref{thm:pd} does not automatically generate the nonstandard fixed-bandwidth law after re-studentization; that too requires a separate bootstrap-$t$ argument. Fixed-bandwidth calibration changes the reference law, not the covariance direction. A misspecified $K_{n,b}$ that omits relevant score covariance remains misspecified. The Brownian-motion functional is tied to the oracle null-imposed score and the explicit fixed-bandwidth feasible-score replacement condition; if estimation altered the score partial-sum process at first order, the maintained joint limit---and hence the reference functional---would have to be changed accordingly.

\begin{corollary}[Binary randomized treatment]\label{cor:pdbinary}
Under the binary-treatment conditions of Corollary~\ref{cor:te}, if the projected-dependence procedure satisfies Theorem~\ref{thm:pd}, then it is first-order valid. Thus the dependence correction need only recover the scalar treatment-effect covariance contribution identified in Corollary~\ref{cor:te} at the first-order sampling-variance scale; it need not reproduce the full dependence of the potential-outcome vector.
\end{corollary}

For Gaussian multipliers with the projected covariance kernel under the small-bandwidth theory of Theorem~\ref{thm:pd}, the conditional law is exactly normal, so the corresponding test is numerically identical to normal inference using the projected HAC or cluster variance estimator. The fixed-bandwidth calibration in Theorem~\ref{thm:fixedb} changes the reference law when the kernel-bandwidth fraction is not negligible. No bootstrap simulation is required. We use this representation because it makes the variance target transparent, not because Gaussian multipliers constitute new resampling technology. Non-Gaussian dependent multipliers may change higher-order behavior, as in the wild-bootstrap literature \citep{Mammen1993,DavidsonFlachaire2008,Shao2010,Hounyo2023}, but covariance matching alone does not make an arbitrary non-Gaussian multiplier array valid: the conditional standardized CLT in Theorem~\ref{thm:pd} must also hold. Higher-order refinement is outside the present first-order characterization. Online Appendix Sections~\ref{app:repair}--\ref{app:supplementary} give time-series and cluster implementations.
\paragraph{Descriptive relevance summary.}
On $\{\widetilde V_{\mathrm{PD},n}(\beta_0)>0\}$ define
\[
\widehat r_n=1-\frac{\widetilde V_n^{\#}(\beta_0)}{\widetilde V_{\mathrm{PD},n}(\beta_0)},
\]
with an arbitrary bookkeeping value on the complement. Online Appendix Proposition~\ref{prop:rhat} shows that $\widehat r_n$ consistently estimates $R_{A,n}$ when both variance estimators are relatively consistent. We report it only as a normalization of the gap between those two variance estimates. Its finite-sample sign is noisy in moderate samples, and a valid test of $R_{A,n}=0$ would require additional joint limit theory for an estimator of the aggregate off-diagonal covariance. In particular, simply summing all sample off-diagonal score products does not yield a root-$n$ estimator of that object. The diagnostic is therefore descriptive, not a pretest or a stand-alone classifier.

\section{Monte Carlo evidence}\label{sec:mc}

We use iid Bernoulli assignment and $Y_i(0)=u_i$, $Y_i(1)=u_i+\tau+b_i$, where $u_i$ is independent of treatment and of the stationary treatment-effect deviation $b_i$. The benchmark processes are
\[
b_i=\eta_i+\tfrac12\eta_{i-1},\qquad
b_i=\eta_i+\eta_{i-1}-\tfrac12\eta_{i-2},
\]
and $b_i=\eta_i-\tfrac12\eta_{i-1}$. Their limiting aggregate treatment-effect covariance contributions are, respectively, $1$, $0$, and $-1$; the finite-$n$ quantities include the usual edge corrections. We use $p\in\{0.25,0.5,0.75\}$, $n\in\{100,250,500,1000\}$, and 20,000 replications per cell. Monte Carlo standard errors are reported for rejection frequencies.

\begin{table}[!htbp]\centering
\caption{Monte Carlo results at $n=1000$ (20,000 replications per cell).}\label{tab:mc1000}
\scriptsize
\begin{tabular}{lrrrrrr}\toprule
Design & $p$ & $V_n^\#$ & $V_n$ & Avg. null variance & Rejection & MC s.e.\\\midrule
Constant effects & 0.25 & 5.333 & 5.333 & 5.344 & 4.72\% & 0.15 pp \\
Constant effects & 0.50 & 4.000 & 4.000 & 4.002 & 4.98\% & 0.15 pp \\
Constant effects & 0.75 & 5.333 & 5.333 & 5.342 & 5.07\% & 0.16 pp \\
Positive MA(1) & 0.25 & 10.333 & 11.332 & 10.342 & 6.12\% & 0.17 pp \\
Positive MA(1) & 0.50 & 6.500 & 7.499 & 6.498 & 6.76\% & 0.18 pp \\
Positive MA(1) & 0.75 & 7.000 & 7.999 & 7.011 & 6.88\% & 0.18 pp \\
Cancellation MA(2) & 0.25 & 14.333 & 14.334 & 14.342 & 4.81\% & 0.15 pp \\
Cancellation MA(2) & 0.50 & 8.500 & 8.501 & 8.504 & 5.04\% & 0.15 pp \\
Cancellation MA(2) & 0.75 & 8.333 & 8.334 & 8.335 & 4.92\% & 0.15 pp \\
Negative MA(1) & 0.25 & 10.333 & 9.334 & 10.354 & 3.92\% & 0.14 pp \\
Negative MA(1) & 0.50 & 6.500 & 5.501 & 6.500 & 3.27\% & 0.13 pp \\
Negative MA(1) & 0.75 & 7.000 & 6.001 & 7.010 & 3.38\% & 0.13 pp \\
\bottomrule\end{tabular}
\end{table}

At $p=0.5$ and $n=1000$, positive dependence gives $V_n^\#=6.5$ and $V_n=7.499$, so the dependence-ignorant target is about $13.3\%$ below the first-order sampling variance; in Table~\ref{tab:mc1000} the average null-imposed iid score variance is $6.498$ and rejection is $6.76\%$ (Monte Carlo standard error $0.18$ percentage points). With negative dependence, $V_n=5.501$ while the iid target remains $6.5$, and rejection is $3.27\%$ (standard error $0.13$ percentage points). In the cancellation design, the theoretical variance is $8.501$, the average null-imposed variance is $8.504$, and rejection is $5.04\%$ (standard error $0.15$ percentage points). These numbers come from the baseline grid reported in the table; the separate projected experiment below uses independent Monte Carlo streams. The rejection patterns track the variance-target gap, not dependence by itself.

\begin{table}[!htbp]\centering
\caption{Convergence at $p=0.5$ (20,000 replications per cell).}\label{tab:mcp05}
\scriptsize
\begin{tabular}{lrrrrrr}\toprule
Design & $n$ & $V_n$ & Emp. var. & Avg. null variance & Rejection & MC s.e.\\\midrule
Constant effects & 100 & 4.000 & 4.024 & 3.993 & 4.93\% & 0.15 pp \\
Constant effects & 250 & 4.000 & 3.995 & 4.003 & 5.04\% & 0.15 pp \\
Constant effects & 500 & 4.000 & 4.019 & 4.001 & 4.78\% & 0.15 pp \\
Constant effects & 1000 & 4.000 & 3.944 & 4.002 & 4.98\% & 0.15 pp \\
Positive MA(1) & 100 & 7.490 & 7.600 & 6.486 & 7.01\% & 0.18 pp \\
Positive MA(1) & 250 & 7.496 & 7.596 & 6.507 & 7.06\% & 0.18 pp \\
Positive MA(1) & 500 & 7.498 & 7.499 & 6.499 & 6.86\% & 0.18 pp \\
Positive MA(1) & 1000 & 7.499 & 7.495 & 6.498 & 6.76\% & 0.18 pp \\
Cancellation MA(2) & 100 & 8.510 & 8.601 & 8.486 & 5.31\% & 0.16 pp \\
Cancellation MA(2) & 250 & 8.504 & 8.618 & 8.504 & 5.03\% & 0.15 pp \\
Cancellation MA(2) & 500 & 8.502 & 8.657 & 8.495 & 5.10\% & 0.16 pp \\
Cancellation MA(2) & 1000 & 8.501 & 8.480 & 8.504 & 5.04\% & 0.15 pp \\
Negative MA(1) & 100 & 5.510 & 5.546 & 6.518 & 3.26\% & 0.13 pp \\
Negative MA(1) & 250 & 5.504 & 5.470 & 6.503 & 3.23\% & 0.13 pp \\
Negative MA(1) & 500 & 5.502 & 5.654 & 6.505 & 3.68\% & 0.13 pp \\
Negative MA(1) & 1000 & 5.501 & 5.466 & 6.500 & 3.27\% & 0.13 pp \\
\bottomrule\end{tabular}
\end{table}

Table~\ref{tab:mcp05} reports convergence over sample size at $p=0.5$. Additional baseline figures and remaining cells are reported in Online Appendix Section~\ref{app:supplementary}. Null imposition also improves finite-sample targeting in imbalanced-treatment cells, without changing the structural discrepancy $V_n^\#-V_n$.

\subsection{Finite-sample performance of projected dependence correction}\label{sec:mcprojected}

We begin with the cluster geometry because it maps directly to settings in which individually randomized units share a known grouping structure---for example, schools, firms, villages, or local labor markets. Table~\ref{tab:mccluster} keeps iid unit-level assignment but gives treatment-effect heterogeneity an independent cluster component. The groups used by the variance estimator are the true groups in the data-generating process. As cluster size grows from two to eight, iid rejection rises from $7.11\%$ to $18.54\%$ at a nominal $5\%$ level, whereas the projected cluster correction stays near $5\%$. The experiment changes the dependence geometry of treatment effects, not the assignment mechanism.

\begin{table}[!htbp]\centering
\caption{Projected cluster correction under iid unit-level assignment ($n=1000$, $p=0.5$, 20,000 common Monte Carlo samples). Parentheses report Monte Carlo standard errors in percentage points.}\label{tab:mccluster}
\resizebox{\textwidth}{!}{\begin{tabular}{rrrrrrrr}\toprule
Cluster size & Clusters & $V_n^\#$ & $V_n$ & IID reject. & Projected-cluster reject. & Oracle-cluster reject. & Mean $\widehat r_n$\\
\midrule
2 & 500 & 6.000 & 7.000 & 7.11\% (0.18) & 5.14\% (0.16) & 5.18\% (0.16) & 0.141 \\
4 & 250 & 6.000 & 9.000 & 11.10\% (0.22) & 5.04\% (0.15) & 5.02\% (0.15) & 0.329 \\
8 & 125 & 6.000 & 13.000 & 18.54\% (0.27) & 5.08\% (0.16) & 5.03\% (0.15) & 0.533 \\
\bottomrule\end{tabular}}\end{table}

We then return to the ordered time-series designs, where the sign of surviving covariance can be varied transparently. We compare iid multiplier-score inference with projected Bartlett-HAC inference applied only to the null-imposed scalar score and, in the baseline cell, an oracle-HAC benchmark. For this experiment we use $\ell=\lfloor n^{1/4}\rceil$ with Bartlett weights $1-h/(\ell+1)$. This deterministic rule is chosen for transparency, not as an optimal bandwidth prescription. Standard automatic HAC rules such as \citet{Andrews1991HAC} and \citet{NeweyWest1994}, and test-oriented criteria such as \citet{SunPhillipsJin2008}, can be used in applications on the projected score, but our deterministic-bandwidth theory does not itself validate plug-in selection. Because the relevance diagnostic is a ratio, a bandwidth optimized for long-run-variance estimation need not optimize the diagnostic itself. We therefore report explicit bandwidth sensitivity rather than presenting one rule as universally preferred.

\begin{table}[!htbp]\centering
\caption{Projected-HAC correction at $n=1000$, $p=0.5$ (20,000 common Monte Carlo samples within each design). Parentheses report Monte Carlo standard errors in percentage points.}\label{tab:mcprojected}
\scriptsize
\begin{tabular}{lrrrrrrr}\toprule
Design & $L$ & IID reject. & Projected reject. & $E[\widetilde V_n^\#]$ & $E[\widetilde V_{\mathrm{PD},n}]$ & $R_{A,n}$ & Mean $\widehat r_n$\\\midrule
Positive MA(1) & 6 & 6.93\% (0.18) & 5.07\% (0.16) & 6.502 & 7.353 & 0.133 & 0.109 \\
Cancellation MA(2) & 6 & 4.77\% (0.15) & 4.55\% (0.15) & 8.503 & 8.647 & 0.000 & 0.009 \\
Negative MA(1) & 6 & 3.40\% (0.13) & 4.78\% (0.15) & 6.497 & 5.649 & -0.182 & -0.159 \\
\bottomrule\end{tabular}\end{table}

\begin{figure}[!htbp]
\centering
\includegraphics[width=.85\textwidth]{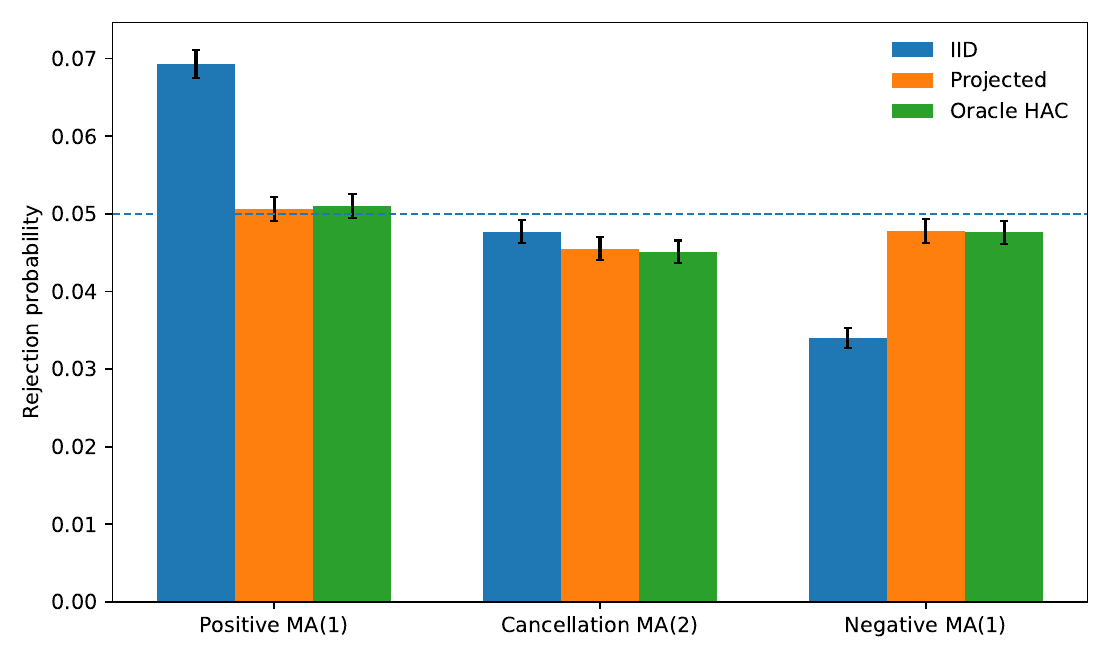}
\caption{Rejection probabilities for iid, projected-HAC, and oracle-HAC score inference at nominal 5\% (20,000 common Monte Carlo samples within each design).}
\label{fig:mcprojected}
\end{figure}

The projected correction behaves in the directions predicted by the phase diagram. Positive surviving covariance makes iid inference anti-conservative and the projected variance moves rejection toward nominal size; negative surviving covariance makes iid inference conservative and the correction moves rejection upward; under covariance cancellation both procedures remain close. The projected test is not exact in finite samples, however. The Bartlett lag window creates the usual bias--variability tradeoff, and the direction of the lag-window bias interacts with the sign of the surviving autocovariance: in Table~\ref{tab:mcprojected} the mean projected variance is below $V_n$ in the positive design and above $V_n$ in the negative design. This helps explain why the projected rejection rates are not symmetric around $5\%$.

Bandwidth sensitivity is economically relevant rather than cosmetic. Online Appendix Table~\ref{tab:stressL} shows that at $n=1000$ the normal-calibrated projected rejection rate falls from $5.45\%$ to $4.49\%$ in the positive design as $\ell$ rises from 2 to 25, with analogous movement elsewhere. A new moderate-sample experiment at $n=250$ shows larger drift. Table~\ref{tab:fixedbmain} therefore compares the same projected statistic under the standard-normal and fixed-bandwidth reference laws at selected bandwidths. Over the grid examined, fixed-bandwidth calibration removes much of the residual under-rejection at large $\ell$, while leaving the covariance target unchanged; it is a calibration device, not a guarantee of uniformly smaller finite-sample size error.

The dependence specification also matters. Online Appendix Table~\ref{tab:groupmisspec} uses split, merged, and scrambled versions of the true cluster structure. Correct grouping restores size; splitting or scrambling the effect-modifier clusters leaves material covariance unaccounted for, while merging independent true clusters preserves the target on average but can increase variance-estimation noise. This is the finite-sample counterpart of the minimality theorem: only error in the surviving covariance direction matters, but a practitioner must still supply a defensible dependence model.

\begin{table}[!htbp]
\centering
\caption{Selected fixed-bandwidth calibration results. Here $\ell$ is the largest included positive Bartlett lag and $b=(\ell+1)/n$ is the kernel-bandwidth fraction. The projected statistic is identical across the last two rejection columns; only the reference critical value changes. Parentheses report Monte Carlo standard errors in percentage points.}
\label{tab:fixedbmain}
\small
\begin{tabular}{lrrrrr}
\toprule
Design & $n$ & $\ell$ & $b$ & Normal reject. & Fixed-$b$ reject.\\
\midrule
Positive MA(1) & 250 & 16 & 0.068 & 4.01\% (0.14) & 5.38\% (0.16) \\
Cancellation MA(2) & 250 & 16 & 0.068 & 3.74\% (0.13) & 5.10\% (0.16) \\
Negative MA(1) & 250 & 16 & 0.068 & 3.77\% (0.13) & 5.11\% (0.16) \\
Positive MA(1) & 1000 & 25 & 0.026 & 4.66\% (0.15) & 5.12\% (0.16) \\
Cancellation MA(2) & 1000 & 25 & 0.026 & 4.39\% (0.14) & 4.86\% (0.15) \\
Negative MA(1) & 1000 & 25 & 0.026 & 4.48\% (0.15) & 4.86\% (0.15) \\
\bottomrule
\end{tabular}
\end{table}

The ratio $\widehat r_n$ is reported in the tables as a descriptive normalization only. Table~\ref{tab:rhatdist} and Online Appendix Tables~\ref{tab:stressn}--\ref{tab:stresstheta} document that its finite-sample sign can be noisy and more bandwidth-sensitive than the projected test.

\begin{table}[!htbp]\centering
\caption{Finite-sample behavior of the relevance diagnostic ($p=0.5$, $L=\lfloor n^{1/4}\rceil$, 20,000 replications per row). ``Wrong sign'' is relative to the sign of $R_{A,n}$.}\label{tab:rhatdist}
\scriptsize
\begin{tabular}{lrrrrrrr}\toprule
Design & $n$ & $R_{A,n}$ & Mean $\widehat r_n$ & 5th pct. & 95th pct. & Wrong sign & Projected reject.\\\midrule
Positive MA(1) & 100 & 0.132 & 0.068 & -0.245 & 0.311 & 30.6\% & 5.00\% \\
Positive MA(1) & 250 & 0.133 & 0.091 & -0.132 & 0.275 & 21.7\% & 5.03\% \\
Positive MA(1) & 500 & 0.133 & 0.102 & -0.072 & 0.249 & 14.8\% & 5.16\% \\
Positive MA(1) & 1000 & 0.133 & 0.110 & -0.020 & 0.227 & 8.0\% & 5.32\% \\
Negative MA(1) & 100 & -0.180 & -0.164 & -0.576 & 0.160 & 24.7\% & 3.88\% \\
Negative MA(1) & 250 & -0.181 & -0.160 & -0.455 & 0.083 & 16.3\% & 4.61\% \\
Negative MA(1) & 500 & -0.181 & -0.160 & -0.391 & 0.037 & 10.1\% & 4.47\% \\
Negative MA(1) & 1000 & -0.182 & -0.160 & -0.332 & -0.002 & 4.7\% & 4.33\% \\
\bottomrule\end{tabular}\end{table}

\section{Conclusion}

Random assignment can simplify bootstrap inference even when the observed data remain dependent. What matters for the randomized-regressor coefficient is whether the bootstrap reproduces the part of that dependence that survives the assignment filter.

Under strong exogeneity, random assignment removes the relevant off-diagonal score covariance, so an iid multiplier bootstrap can be first-order valid despite dependence in outcomes and controls. When the regression error changes with assignment, a scalar covariance component can survive. An iid bootstrap deletes that component. With binary treatment, the resulting first-order variance gap is exactly minus the aggregate cross-observation covariance of heterogeneous treatment effects. Under the standardized Gaussian sampling limit, this gap makes iid inference over-reject, under-reject, or remain first-order correct according to the sign of the relative surviving covariance.

A valid centered-Gaussian procedure need not reproduce the full dependence structure; it needs only the covariance component that survives the randomized-score projection. The information-limit theorem shows why marginal information alone cannot recover that component in general. For first-order Gaussian inference, this means estimating dependence in the scalar randomized-regressor score rather than resampling the full observed-data vector. With a small bandwidth, the Gaussian implementation is a HAC or cluster-normal test on that score. When the bandwidth is non-negligible relative to sample size, the reference law instead requires fixed-bandwidth calibration. The ratio $\widehat r_n$ is best treated as a descriptive normalization of the diagonal--projected variance gap. The grouping experiments also make clear what projection cannot do: it reduces the dimension of the dependence problem but does not identify the appropriate ordering, clusters, or network for the researcher.

The local-power calculation leads to the same conclusion. The apparent power advantage or disadvantage of the raw iid test comes from its size distortion: after equal-size calibration, the iid and projected procedures have the same Gaussian first-order local-power curve. The oracle iid calibration, however, depends on the unknown relevance index, whereas the projected procedure estimates the needed covariance contribution. Its gain is calibration and feasibility, not first-order efficiency. In short, random assignment and resampling need not remove the same covariance. Once the surviving covariance component has been identified, inference can focus on that component instead of modeling the full dependence structure.

\clearpage
\thispagestyle{plain}
\begin{center}
  {\Large\bfseries Online Appendix to\par}
  \vspace{0.7em}
  {\Large ``Bootstrap Inference with a Randomly Assigned Regressor:\\
  Covariance Filtering and the Limits of Marginal Resampling''\par}
  \vspace{1.5em}
  {\large Ulrich Hounyo \qquad Jungbin Hwang\par}
  \vspace{0.8em}
  {\large October 2, 2026\par}
\end{center}
\vspace{2em}

\appendix
\section{Proofs for the setup and main covariance results}\label{app:proofs-main}

\begin{lemma}[Gaussian scale comparison]\label{lem:gaussianscale}
Let $\Phi$ denote the standard normal distribution function. Let $a_n>0$ and $b_n>0$ be deterministic or random scales. If $a_n/b_n\to_p1$, then
\[
\sup_{x\in\mathbb R}\left|\Phi(x/a_n)-\Phi(x/b_n)\right|\to_p0.
\]
Conversely, if the displayed Kolmogorov distance converges to zero in probability, then $a_n/b_n\to_p1$.
\end{lemma}

\paragraph{Proof.}
Put $r_n=a_n/b_n$ and change variables $z=x/b_n$. The displayed distance equals
\[
d(r_n):=\sup_{z\in\mathbb R}|\Phi(z/r_n)-\Phi(z)|.
\]
For $r\in[1/2,2]$, choose $M$ so large that both Gaussian tails outside $[-M,M]$ are uniformly smaller than an arbitrary $\eta>0$. On $[-M,M]\times[1/2,2]$, the map $(z,r)\mapsto\Phi(z/r)-\Phi(z)$ is uniformly continuous. Hence $d(r)\to0$ as $r\to1$, proving the first assertion.

For the converse, the supremum is taken pointwise for each realization, so it may be bounded below by its value at the (possibly random) point $x=b_n$:
\[
|\Phi(1/r_n)-\Phi(1)|\to_p0.
\]
Because $\Phi(1)$ lies in the interior of $(0,1)$ and $\Phi^{-1}$ is continuous there, $1/r_n\to_p1$, hence $r_n\to_p1$. \qed

\subsection{Proof of Lemma~\ref{lem:design}}
Let $Z_{i,n}=(D_{i,n}-\mu_{D,n})^2-\sigma_{D,n}^2$ and $r=1+\delta/2\in(1,2]$. The variables are row-wise iid and mean zero, and the maintained $(2+\delta)$ moment bound implies $\sup_nE|Z_{1,n}|^r<\infty$. Von Bahr--Esseen gives
\[
E\left|n^{-1}\sum_iZ_{i,n}\right|^r\le Cn^{1-r}\to0.
\]
Thus $n^{-1}\sum_i(D_{i,n}-\mu_{D,n})^2-\sigma_{D,n}^2=o_p(1)$. The same moment bound uniformly bounds $\sigma_{D,n}^2$, so Chebyshev gives $\bar D_n-\mu_{D,n}=O_p(n^{-1/2})$. Since
\[
\widehat Q_n=n^{-1}\sum_i(D_{i,n}-\mu_{D,n})^2-(\bar D_n-\mu_{D,n})^2,
\]
the absolute conclusion follows. Division by a positive uniform lower bound on $\sigma_{D,n}^2$ yields the relative conclusion. \qed

\subsection{Proof of Lemma~\ref{lem:scalar-diag}}

Fix a row $n$ and write
\[
X_{i,n}
=
(D_{i,n}-\mu_{D,n})^2\varepsilon_{i,n}^2.
\]
Let
\[
\mathcal F_{\varepsilon,n}
=
\sigma(\varepsilon_{1,n},\ldots,\varepsilon_{n,n}).
\]
By iid assignment and independence of the full assignment vector from the error array, the variables $X_{i,n}$ are conditionally independent given $\mathcal F_{\varepsilon,n}$. Put
\[
Z_{i,n}
=
X_{i,n}
-
E(X_{i,n}\mid\mathcal F_{\varepsilon,n}).
\]
Choose $r=1+\delta/2\in(1,2]$. Conditional von Bahr--Esseen and the same truncation-of-the-random-bound argument used in Lemma~\ref{lem:diag} yield
\[
\frac1n\sum_i Z_{i,n}=o_p(1),
\]
because
\[
E(|Z_{i,n}|^r\mid\mathcal F_{\varepsilon,n})
\le
C|\varepsilon_{i,n}|^{2+\delta}
E|D_{1,n}-\mu_{D,n}|^{2+\delta}
\]
and $n^{-1}\sum_i|\varepsilon_{i,n}|^{2+\delta}=O_p(1)$.

For the conditional mean,
\[
E(X_{i,n}\mid\mathcal F_{\varepsilon,n})
=
\sigma_{D,n}^2\varepsilon_{i,n}^2.
\]
Therefore
\[
\frac1n\sum_i
\left\{
E(X_{i,n}\mid\mathcal F_{\varepsilon,n})
-
E X_{i,n}
\right\}
=
\sigma_{D,n}^2
\frac1n\sum_i
\{\varepsilon_{i,n}^2-E(\varepsilon_{i,n}^2)\}.
\]
The treatment variance is uniformly bounded under the maintained moment condition, and the final factor is $o_p(1)$ by assumption. Combining the conditional-fluctuation and conditional-mean terms proves the lemma. \qed

\subsection{Proof of Theorem~\ref{thm:strong}}
By the definition of strong exogeneity, the full error array is independent of the iid assignment vector. Hence, for $i\neq j$,
\[
\E[(D_{i,n}-\mu_{D,n})(D_{j,n}-\mu_{D,n})\varepsilon_{i,n}\varepsilon_{j,n}\mid \varepsilon_1,\ldots,\varepsilon_n]=0.
\]
Consequently the first-order score variance is diagonal:
\[
\sigma_{D\varepsilon,n}^2
=
\frac1n\sum_i\E[(D_i-\mu_{D,n})^2\varepsilon_i^2]
=
s_{\#,n}^2.
\]
Hence the generic targets in Section~\ref{sec:setup} satisfy
\[
V_n=V_n^\#.
\]
By Lemma~\ref{lem:scalar-diag}, the oracle scalar score satisfies the required unnormalized diagonal-score LLN. Assumption~\ref{ass:stable} bounds $\sigma_{D,n}^2$ away from zero, so
\[
\widehat V_{n,o}^\#-V_n^\#=\op(1).
\]
The restricted-score replacement argument below and $\widehat Q_n\pto\sigma_{D,n}^2$ then imply
\[
\widetilde V_n^\#(\beta_0)-V_n^\#=\op(1),
\]
and therefore, because $V_n^\#=V_n$,
\[
\widetilde V_n^\#(\beta_0)-V_n=\op(1).
\]
Conditional on the data, Gaussian multipliers give the exact identity
\[
T_n^\#(\beta_0)\mid\mathcal F_n\sim N\{0,\widetilde V_n^\#(\beta_0)\}.
\]
Assumption~\ref{ass:stable} bounds both the treatment-design scale and the score scale, and therefore bounds $V_n$ away from zero and infinity. Therefore
\[
\frac{\{\widetilde V_n^\#(\beta_0)\}^{1/2}}{V_n^{1/2}}\pto1,
\]
and Lemma~\ref{lem:gaussianscale} gives
\[
\sup_t\left|\Phi\!\left(t/\sqrt{\widetilde V_n^\#}\right)-
\Phi\!\left(t/\sqrt{V_n}\right)\right|=\op(1).
\]
By assumption,
\[
\frac{\sqrt n(\widehat\beta-\beta_0)}{\sqrt{V_n}}\dto N(0,1).
\]
Because the standard normal cdf is continuous, P\'olya's theorem yields
\[
\sup_x\left|
\Pr\!\left\{\frac{\sqrt n(\widehat\beta-\beta_0)}{\sqrt{V_n}}\le x\right\}
-\Phi(x)
\right|\to0.
\]
The change of variable $x=t/\sqrt{V_n}$ therefore gives
\[
\sup_t\left|\Pr\{\sqrt n(\widehat\beta-\beta_0)\le t\}-
\Phi\!\left(t/\sqrt{V_n}\right)\right|\to0.
\]
The triangle inequality proves the asserted conditional Kolmogorov approximation. Notice that convergence of $V_n$ to a fixed constant is not needed.

\subsection{Proof of Lemma~\ref{lem:diag}}

Fix $n$ and suppress the row subscript when convenient. Let
\[
\mathcal F_{e,n}=\sigma(e_{1,n},\ldots,e_{n,n}).
\]
By iid assignment within row $n$ and independence of the full assignment vector from the latent $e$-array, the variables
\[
X_{i,n}:=(A_{i,n}'e_{i,n})^2
\]
are conditionally independent given $\mathcal F_{e,n}$. Put
\[
Z_{i,n}
=
X_{i,n}
-
E[X_{i,n}\mid\mathcal F_{e,n}].
\]
Then
\[
E[Z_{i,n}\mid\mathcal F_{e,n}]=0.
\]

Choose
\[
r=1+\frac{\delta}{2}\in(1,2].
\]
Conditional von Bahr--Esseen gives
\[
E\left[
\left|
\frac1n\sum_{i=1}^n Z_{i,n}
\right|^r
\Bigm|
\mathcal F_{e,n}
\right]
\le
C_r n^{-r}
\sum_{i=1}^n
E(|Z_{i,n}|^r\mid\mathcal F_{e,n}).
\]
Using
\[
|x-E(x\mid\mathcal F)|^r
\le
2^{r-1}\{|x|^r+E(|x|^r\mid\mathcal F)\},
\]
we have
\[
E(|Z_{i,n}|^r\mid\mathcal F_{e,n})
\le
C
E(|A_{i,n}'e_{i,n}|^{2r}\mid\mathcal F_{e,n}).
\]
Since $2r=2+\delta$ and Lemma~\ref{lem:diag} assumes
$\sup_n\sup_{d\in\operatorname{supp}(D_{1,n})}\|\lambda_n(d)\|<\infty$,
\[
\|A_{i,n}\|
\le
C|D_{i,n}-\mu_{D,n}|,
\]
hence
\[
E(|A_{i,n}'e_{i,n}|^{2+\delta}\mid\mathcal F_{e,n})
\le
C\|e_{i,n}\|^{2+\delta}
E|D_{1,n}-\mu_{D,n}|^{2+\delta}.
\]
Therefore
\[
E\left[
\left|
\frac1n\sum_i Z_{i,n}
\right|^r
\Bigm|
\mathcal F_{e,n}
\right]
\le
C n^{-r}\sum_i\|e_{i,n}\|^{2+\delta}
=
O_p(n^{1-r}).
\]
Because $r>1$, the right-hand side is $o_p(1)$. To make the conditioning step explicit, fix $\eta,\rho>0$ and choose $C_\rho<\infty$ so that
\[
P\!\left\{\frac1n\sum_i\|e_{i,n}\|^{2+\delta}\le C_\rho\right\}\ge1-\rho
\]
for all sufficiently large $n$. On that event, conditional Markov gives
\[
P\!\left(
\left|\frac1n\sum_i Z_{i,n}\right|>\eta
\Bigm|\mathcal F_{e,n}
\right)
\le C_{\rho,\eta}n^{1-r}.
\]
Taking expectations,
\[
P\!\left(\left|\frac1n\sum_iZ_{i,n}\right|>\eta\right)
\le \rho+C_{\rho,\eta}n^{1-r}.
\]
Letting first $n\to\infty$ and then $\rho\downarrow0$ proves
\[
\frac1n\sum_i Z_{i,n}=o_p(1).
\]

For the conditional expectation term, the assignment law and $\lambda_n(\cdot)$ are common across $i$ within row $n$, so
\[
M_{A,n}:=E[A_{i,n}A_{i,n}']
\]
does not depend on $i$. Thus
\[
\frac1n\sum_iE[X_{i,n}\mid\mathcal F_{e,n}]
=
\operatorname{tr}\left(
M_{A,n}\frac1n\sum_i e_{i,n}e_{i,n}'
\right),
\]
while
\[
\frac1n\sum_iE[X_{i,n}]
=
\operatorname{tr}\left(
M_{A,n}\frac1n\sum_i E[e_{i,n}e_{i,n}']
\right).
\]
The latent second-moment LLN assumed in Lemma~\ref{lem:diag} gives
\[
\left\|
\frac1n\sum_i
\{e_{i,n}e_{i,n}'-E[e_{i,n}e_{i,n}']\}
\right\|=o_p(1).
\]
Moreover, the assignment moment bound and boundedness of $\lambda_n(\cdot)$ imply
\[
\sup_n\|M_{A,n}\|<\infty.
\]
Since $L$ is fixed, the trace difference is $o_p(1)$. Combining the two pieces proves the lemma. \qed

\subsection{Proof of Theorem~\ref{thm:filter}}

Within row $n$, write
\[
A_i=\mu_A+\widetilde A_i,
\qquad
E\widetilde A_i=0,
\qquad
\psi_i=A_i'e_i.
\]
Let
\[
\mathcal F_{e,n}
=
\sigma(e_{1,n},\ldots,e_{n,n}).
\]
The full assignment vector is independent of $\mathcal F_{e,n}$, and assignment is iid across $i$. Hence
\[
E(\widetilde A_i\mid\mathcal F_{e,n})=0,
\]
and, for $i\ne j$,
\[
E(\widetilde A_i\widetilde A_j'\mid\mathcal F_{e,n})
=
E(\widetilde A_i)E(\widetilde A_j')
=
0.
\]
The latent vector is centered, so $Ee_i=0$. Therefore every covariance term containing at least one centered assignment loading vanishes. For arbitrary $i,j$,
\[
E[(\widetilde A_i'e_i)(\mu_A'e_j)]
=
E\!\left[
e_i'E(\widetilde A_i\mid\mathcal F_{e,n})\mu_A'e_j
\right]
=
0.
\]
For $i\ne j$,
\[
E[(\widetilde A_i'e_i)(\widetilde A_j'e_j)]
=
E\!\left[
e_i'E(\widetilde A_i\widetilde A_j'\mid\mathcal F_{e,n})e_j
\right]
=
0.
\]
Since $E\psi_i=0$,
\[
\Cov(\psi_i,\psi_j)
=
\begin{cases}
E[(\widetilde A_i'e_i)^2]+\Var(\mu_A'e_i), & i=j,\\[3pt]
\Cov(\mu_A'e_i,\mu_A'e_j), & i\ne j.
\end{cases}
\]
In particular,
\[
E[\psi_i^2]
=
E[(\widetilde A_i'e_i)^2]
+
\Var(\mu_A'e_i).
\]
Summing gives
\[
s_{\#,n}^2
=
\sigma_{e,1,n}^2+\sigma_{\mu,0,n}^2,
\]
hence, by the generic definition of $V_n^\#$,
\[
V_n^\#
=
\sigma_{D,n}^{-4}
\{\sigma_{e,1,n}^2+\sigma_{\mu,0,n}^2\}.
\]

The same cross-index cancellations yield
\[
\begin{aligned}
\sigma_{D\varepsilon,n}^2
&=
\Var\!\left(n^{-1/2}\sum_i\psi_i\right)\\
&=
\frac1n\sum_iE[(\widetilde A_i'e_i)^2]
+
\Var\!\left(n^{-1/2}\sum_i\mu_A'e_i\right)\\
&=
\sigma_{e,1,n}^2
+
\Var\!\left(n^{-1/2}\sum_i\mu_A'e_i\right).
\end{aligned}
\]
Multiplying by $\sigma_{D,n}^{-4}$ gives the stated representation of $V_n$. Moreover,
\[
\begin{aligned}
\Var\!\left(
\frac1{\sqrt n}\sum_i\mu_{A,n}'e_{i,n}
\right)
&=
\frac1n\sum_i\Var(\mu_{A,n}'e_{i,n})\\
&\quad+
\frac1n\sum_{i\ne j}
\Cov(\mu_{A,n}'e_{i,n},\mu_{A,n}'e_{j,n})\\
&=
\sigma_{\mu,0,n}^2+\Delta_{A,n}.
\end{aligned}
\]
Thus
\[
V_n
=
V_n^\#
+
\sigma_{D,n}^{-4}\Delta_{A,n},
\]
which proves the covariance-filter identity.

Finally,
\[
\Omega_{e,n}
=
\Var\!\left(n^{-1/2}\sum_ie_{i,n}\right),
\qquad
\Omega_{e,0,n}
=
\frac1n\sum_i\Var(e_{i,n}),
\]
so
\[
\Omega_{e,n}-\Omega_{e,0,n}
=
\frac1n\sum_{i\ne j}\Cov(e_{i,n},e_{j,n}),
\]
where the sum is over ordered pairs. Premultiplying and postmultiplying by $\mu_{A,n}$ gives
\[
\Delta_{A,n}
=
\mu_{A,n}'(\Omega_{e,n}-\Omega_{e,0,n})\mu_{A,n}.
\]

It remains only to establish the optional oracle-estimator consistency statement. Under the additional conditions of Lemma~\ref{lem:diag},
\[
\frac1n\sum_i\psi_i^2-s_{\#,n}^2=o_p(1).
\]
Since
\[
\widehat V_{n,o}^\#
=
\sigma_{D,n}^{-4}\frac1n\sum_i\psi_i^2,
\]
we obtain
\[
\sigma_{D,n}^{4}
\{\widehat V_{n,o}^\#-V_n^\#\}
=
o_p(1).
\]
If $\inf_n\sigma_{D,n}^2>0$, then $\sigma_{D,n}^{-4}=O(1)$ and therefore
\[
\widehat V_{n,o}^\#-V_n^\#=o_p(1).
\]
\qed

\subsection{Proof of Lemma~\ref{lem:restricted} and Theorem~\ref{thm:feasible}}

Under $H_0:\beta=\beta_0$,
\[
\widetilde\gamma(\beta_0)-\gamma
=
(n^{-1}W'W)^{-1}(n^{-1}W'\varepsilon).
\]
By the conditions of Lemma~\ref{lem:restricted},
\[
n^{-1}W'\varepsilon=o_p(1),
\qquad
\|(n^{-1}W'W)^{-1}\|=O_p(1).
\]
Hence
\[
\widetilde\gamma(\beta_0)-\gamma=o_p(1).
\]

The exact score difference is
\[
r_i
:=
\widetilde\psi_i(\beta_0)-\psi_i
=
-(\bar D-\mu_{D,n})\varepsilon_i
-
(D_i-\bar D)W_i'
\{\widetilde\gamma(\beta_0)-\gamma\}.
\]
Therefore
\[
\frac1n\sum_i r_i^2
\le
2(\bar D-\mu_{D,n})^2
\frac1n\sum_i\varepsilon_i^2
+
2\|\widetilde\gamma-\gamma\|^2 B_n,
\]
where
\[
B_n
=
\frac1n\sum_i(D_i-\bar D)^2\|W_i\|^2.
\]

It remains to verify $B_n=O_p(1)$ without imposing cross-sectional independence on the controls. Since
\[
(D_i-\bar D)^2
\le
2(D_i-\mu_{D,n})^2+2(\bar D-\mu_{D,n})^2,
\]
\[
B_n
\le
2B_{1n}
+
2(\bar D-\mu_{D,n})^2B_{2n},
\]
with
\[
B_{1n}
=
\frac1n\sum_i(D_i-\mu_{D,n})^2\|W_i\|^2,
\qquad
B_{2n}
=
\frac1n\sum_i\|W_i\|^2.
\]
By assumption,
\[
B_{2n}=\frac1n\sum_i\|W_i\|^2=O_p(1).
\]
Since the treatment vector is independent of the full control array,
\[
E(B_{1n}\mid W_1,\ldots,W_n)
=
\sigma_{D,n}^2 B_{2n}.
\]
Because $\sup_n\sigma_{D,n}^2<\infty$, conditional Markov's inequality implies $B_{1n}=O_p(1)$. Finally,
\[
\bar D-\mu_{D,n}=O_p(n^{-1/2})
\]
under iid assignment. Thus $B_n=O_p(1)$.

The assumed error second-moment bound gives
\[
\frac1n\sum_i\varepsilon_i^2=O_p(1).
\]
Combining the preceding results,
\[
\frac1n\sum_i r_i^2=o_p(1),
\]
which proves Lemma~\ref{lem:restricted}.

Lemma~\ref{lem:design} gives the absolute relation
\[
\widehat Q_n-\sigma_{D,n}^2=o_p(1).
\]
For the scale-free theorem, Assumption~\ref{ass:scalefree} supplies the relative design relation
\[
\frac{\widehat Q_n}{\sigma_{D,n}^2}\pto1
\]
and, for the score terms,
\[
\frac{n^{-1}\sum_i r_i^2}{s_{\#,n}^2}=o_p(1),
\qquad
\frac{n^{-1}\sum_i\psi_i^2}{s_{\#,n}^2}\pto1.
\]
The second display is an explicitly relative oracle LLN; Lemma~\ref{lem:diag} alone gives only absolute $o_p(1)$ concentration and is insufficient when $s_{\#,n}^2$ can shrink.
Now
\[
\widetilde\psi_i^2-\psi_i^2
=
2\psi_i r_i+r_i^2.
\]
By Cauchy--Schwarz,
\[
\left|
\frac{
n^{-1}\sum_i\psi_i r_i
}{
s_{\#,n}^2
}
\right|
\le
\left(
\frac{n^{-1}\sum_i\psi_i^2}{s_{\#,n}^2}
\right)^{1/2}
\left(
\frac{n^{-1}\sum_i r_i^2}{s_{\#,n}^2}
\right)^{1/2}
=o_p(1).
\]
Consequently,
\[
\frac{
n^{-1}\sum_i\widetilde\psi_i^2
}{
s_{\#,n}^2
}
\pto1.
\]
Since
\[
V_n^\#=\sigma_{D,n}^{-4}s_{\#,n}^2,
\qquad
\widetilde V_n^\#(\beta_0)
=
\widehat Q_n^{-2}
\frac1n\sum_i\widetilde\psi_i^2,
\]
we obtain
\[
\frac{\widetilde V_n^\#(\beta_0)}{V_n^\#}
=
\frac{\sigma_{D,n}^4}{\widehat Q_n^2}
\frac{n^{-1}\sum_i\widetilde\psi_i^2}{s_{\#,n}^2}
\pto1.
\]

Under Assumption~\ref{ass:stable}, both $\sigma_{D,n}^2$ and $s_{\#,n}^2$ are bounded away from zero. Hence Lemma~\ref{lem:design} implies the first clause of Assumption~\ref{ass:scalefree}, Lemma~\ref{lem:restricted} implies the second clause, and Lemma~\ref{lem:diag} implies the third clause. Moreover, the absolute $L^2$ replacement and Cauchy--Schwarz give
\[
\frac1n\sum_i\widetilde\psi_i^2
-
\frac1n\sum_i\psi_i^2
=o_p(1).
\]
Together with $\widehat Q_n\to_p\sigma_{D,n}^2$ and Lemma~\ref{lem:diag},
\[
\widetilde V_n^\#(\beta_0)-V_n^\#=o_p(1).
\]
The covariance-filter identity yields
\[
\widetilde V_n^\#(\beta_0)-V_n
=
-\sigma_{D,n}^{-4}\Delta_{A,n}+o_p(1).
\]
\qed

\paragraph{Primitive sufficient rates for the scale-free clauses.}
For the first clause, let
\[
Q_{0n}=\frac1n\sum_i(D_{i,n}-\mu_{D,n})^2.
\]
Then
\[
\frac{\widehat Q_n}{\sigma_{D,n}^2}-1
=
\frac{Q_{0n}-\sigma_{D,n}^2}{\sigma_{D,n}^2}
-
\frac{(\bar D_n-\mu_{D,n})^2}{\sigma_{D,n}^2}.
\]
Chebyshev gives
\[
P\!\left(
\left|\frac{Q_{0n}-\sigma_{D,n}^2}{\sigma_{D,n}^2}\right|>\eta
\right)
\le
\frac{E[(D_{1,n}-\mu_{D,n})^4]}{n\eta^2\sigma_{D,n}^4},
\]
and
\[
E\!\left[
\frac{(\bar D_n-\mu_{D,n})^2}{\sigma_{D,n}^2}
\right]=\frac1n.
\]
This proves the stated primitive treatment-design condition.

For the third clause, the conditional von Bahr--Esseen calculation in Lemma~\ref{lem:diag} yields
\[
\frac1n\sum_i
\left[
\psi_{i,n}^2-E(\psi_{i,n}^2\mid\mathcal F_{e,n})
\right]
=
O_p\!\left(n^{-\delta/(2+\delta)}\right).
\]
Thus one sufficient route to relative oracle concentration is
\[
n^{\delta/(2+\delta)}s_{\#,n}^2\to\infty
\]
together with the relative latent quadratic LLN
\[
\frac{
n^{-1}\sum_i E(\psi_{i,n}^2\mid\mathcal F_{e,n})-s_{\#,n}^2
}{
s_{\#,n}^2
}
=o_p(1).
\]
The relative latent LLN is an additional rate condition; the absolute latent LLN in Lemma~\ref{lem:diag} does not imply it when $s_{\#,n}^2$ shrinks.

For the second clause, the bound in Lemma~\ref{lem:restricted} shows that it is enough that
\[
\frac{(\bar D-\mu_{D,n})^2\,n^{-1}\sum_i\varepsilon_i^2}{s_{\#,n}^2}
=o_p(1)
\]
and
\[
\frac{\|\widetilde\gamma-\gamma\|^2 B_n}{s_{\#,n}^2}
=o_p(1).
\]
A simple sufficient pair is
\[
n s_{\#,n}^2\to\infty,
\qquad
\|\widetilde\gamma-\gamma\|=o_p(s_{\#,n}),
\]
together with $n^{-1}\sum_i\varepsilon_i^2=O_p(1)$.

\begin{remark}[Hierarchy of feasible-score conditions]
There are three relative requirements in a genuinely scale-free theorem: treatment-design concentration, feasible-score replacement, and oracle diagonal-score concentration. They are the three clauses of Assumption~\ref{ass:scalefree}. Under Assumption~\ref{ass:stable}, Lemma~\ref{lem:design} together with Lemmas~\ref{lem:restricted} and \ref{lem:diag} implies them automatically. If either the treatment-design scale or the diagonal score scale shrinks, the corresponding absolute $o_p(1)$ statement is not enough by itself.
\end{remark}

\subsection{Proof of Theorem~\ref{thm:phase}: bootstrap validity frontier}

Theorem~\ref{thm:feasible} and the definition of $R_{A,n}$ imply
\[
\frac{\widetilde V_n^\#(\beta_0)}{V_n}
=
\frac{\widetilde V_n^\#(\beta_0)}{V_n^\#}
\frac{V_n^\#}{V_n}
=
\{1+o_p(1)\}(1-R_{A,n})
\pto 1-r.
\]
The strict inequality $r<1$ is essential here. If $r=1$, then $V_n^\#/V_n\to0$ and the iid-standardized statistic generally has no finite nondegenerate Gaussian scale under this normalization. For $r<1$,
\[
\left\{
\frac{V_n}{\widetilde V_n^\#(\beta_0)}
\right\}^{1/2}
\pto
(1-r)^{-1/2}.
\]
Multiplying the standardized sampling statistic by this factor and applying Slutsky's theorem gives
\[
\frac{\sqrt n(\widehat\beta-\beta_0)}
{\{\widetilde V_n^\#(\beta_0)\}^{1/2}}
\dto
N\!\left(0,\frac1{1-r}\right).
\]
If $Z\sim N(0,1)$, the limiting rejection probability is
\[
P\{|Z|>z_{1-\alpha/2}\sqrt{1-r}\}
=
2\Phi\!\left(-z_{1-\alpha/2}\sqrt{1-r}\right).
\]
This function is strictly increasing in $r$ on $(-\infty,1)$, which proves the sign classification. \qed

\subsection{Proof of Corollary~\ref{cor:localpower}: equal-size local power}
Under $\beta_n=\beta_0+c/\sqrt n$,
\[
\frac{\sqrt n(\widehat\beta-\beta_0)}{V_n^{1/2}}
=
\frac{\sqrt n(\widehat\beta-\beta_n)}{V_n^{1/2}}
+
\frac{c}{V_n^{1/2}}
\dto N(\delta,1),
\qquad
\delta=\frac{c}{V^{1/2}}.
\]
The exact covariance-filter identity gives $V_n^\#/V_n=1-R_{A,n}\to1-r$, and the assumed local-sequence variance consistency gives
\[
\frac{\widetilde V_n^\#(\beta_0)}{V_n}
=
\frac{\widetilde V_n^\#(\beta_0)}{V_n^\#}
\frac{V_n^\#}{V_n}
\pto1-r.
\]
Hence the unadjusted iid test rejects asymptotically when the limiting $N(\delta,1)$ variable exceeds $z\sqrt{1-r}$ in absolute value, giving
\[
\Phi\!\left(-z\sqrt{1-r}-\delta\right)
+1-\Phi\!\left(z\sqrt{1-r}-\delta\right).
\]
For a projected procedure whose variance estimator is relatively consistent for $V_n$ along the local sequence, the threshold in sampling-standardized units is $z$, so its local-power curve is
\[
\Phi(-z-\delta)+1-\Phi(z-\delta).
\]
Finally, the oracle equal-size benchmark replaces the iid statistic's usual critical value $z$ by $z/\sqrt{1-r}$. Because the iid standard error is asymptotically $\sqrt{1-r}$ times the sampling standard deviation, this calibration again produces threshold $z$ in sampling-standardized units. The size-adjusted iid curve therefore coincides with the projected curve. \qed

\subsection{Proof of Corollary~\ref{cor:te}: binary-treatment identity}

Under Section~\ref{sec:binary},
\[
D_{i,n}\sim\operatorname{Bernoulli}(p_n),
\qquad
q_n=p_n(1-p_n),
\]
and
\[
e_{i,n}
=
\begin{pmatrix}
Y_{i,n}(0)-\alpha_n\\
\tau_{i,n}-\beta_n
\end{pmatrix},
\qquad
\lambda_n(D_{i,n})
=
\begin{pmatrix}
1\\D_{i,n}
\end{pmatrix}.
\]
First verify the regression target. Since assignment is independent of the potential-outcome array,
\[
\begin{aligned}
E[(D_{i,n}-p_n)Y_{i,n}]
&=
E[(D_{i,n}-p_n)Y_{i,n}(0)]\\
&\quad+
E[(D_{i,n}-p_n)D_{i,n}\tau_{i,n}]\\
&=
0+E[(D_{i,n}-p_n)D_{i,n}]E[\tau_{i,n}]\\
&=
q_n\beta_n.
\end{aligned}
\]
Thus the population slope is $\beta_n$, and the regression error is
\[
\varepsilon_{i,n}
=
Y_{i,n}(0)-\alpha_n+D_{i,n}(\tau_{i,n}-\beta_n)
=
\lambda_n(D_{i,n})'e_{i,n}.
\]

Next,
\[
A_{i,n}
=
(D_{i,n}-p_n)
\begin{pmatrix}
1\\D_{i,n}
\end{pmatrix},
\]
so
\[
\mu_{A,n}
=
E[A_{i,n}]
=
\begin{pmatrix}
0\\E[(D_{i,n}-p_n)D_{i,n}]
\end{pmatrix}
=
\begin{pmatrix}
0\\q_n
\end{pmatrix}.
\]
Therefore
\[
\mu_{A,n}'e_{i,n}
=
q_n(\tau_{i,n}-\beta_n).
\]
Since subtracting a deterministic common mean does not change covariance,
\[
\Cov(\mu_{A,n}'e_{i,n},\mu_{A,n}'e_{j,n})
=
q_n^2\Cov(\tau_{i,n},\tau_{j,n}).
\]
Hence, exactly for every $n$,
\[
\Delta_{A,n}
=
q_n^2
\frac1n\sum_{i\ne j}\Cov(\tau_{i,n},\tau_{j,n})
=
\sigma_{D,n}^4\Delta_{\tau,n}.
\]
Theorem~\ref{thm:filter} then gives
\[
V_n^\#-V_n
=
-\sigma_{D,n}^{-4}\Delta_{A,n}
=
-\Delta_{\tau,n}.
\]
If $V_n>0$, division by $V_n$ and the identity
\[
R_{A,n}=1-\frac{V_n^\#}{V_n}
\]
give
\[
R_{A,n}
=
\frac{\Delta_{\tau,n}}{V_n}.
\]
No asymptotic approximation is used in these reductions. \qed

\subsection{Proof of Corollary~\ref{cor:size}}
Corollary~\ref{cor:te} gives the exact finite-$n$ identity
\[
R_{A,n}
=
\frac{\Delta_{\tau,n}}{V_n}.
\]
Hence the maintained convergence $\Delta_{\tau,n}/V_n\to\rho$ is exactly the hypothesis
$R_{A,n}\to\rho$ in Theorem~\ref{thm:phase}. Applying that theorem gives the result. \qed

\subsection{General treatment probability in the benchmark data-generating process (DGP)}
For the stationary Monte Carlo benchmark, specialize to a fixed assignment probability $p_n\equiv p$. Let $Y_i(0)=u_i$, $Y_i(1)=u_i+\tau+b_i$, with $u_i$ independent of $b_i$ and $D_i\sim\mathrm{Bernoulli}(p)$. The score is
\[
\psi_i=(D_i-p)(u_i+D_i b_i).
\]
If $\Var(u_i)=\sigma_u^2$ and $\Var(b_i)=\gamma_b(0)$, then
\[
\E\psi_i^2=p(1-p)\{\sigma_u^2+(1-p)\gamma_b(0)\},
\]
so
\[
V^\#(p)=\frac{\sigma_u^2+\gamma_b(0)}p+\frac{\sigma_u^2}{1-p}.
\]
For $i\ne j$, conditioning on the latent process gives
\[
\E_D[\psi_i\mid u_i,b_i]=p(1-p)b_i,
\]
and therefore
\[
\E[\psi_i\psi_j]=[p(1-p)]^2\Cov(b_i,b_j).
\]
After ordinary least squares (OLS) normalization the factor cancels, yielding the $p$-invariant covariance correction.

\section{Ordinary pairs bootstrap: full theory}\label{app:pairs-full}

The main paper records only the covariance-filter implication. This section gives the full row-resampling construction, regularity conditions, distributional theorem, Hahn--Liao moment distinction, and binary second-moment result conditional on nondegenerate resamples.

\subsection{Row-resampling construction and first-order law}
The ordinary empirical pairs bootstrap resamples entire observations and recomputes OLS. Unlike the score bootstrap, row resampling perturbs both the score and the design matrix. Its validity therefore requires additional leverage regularity beyond the baseline CHLS assumptions.

Let $X_i=(D_i,W_i')'$ and $\widehat\theta=(\widehat\beta,\widehat\gamma')'$. Conditional on the observed sample, draw bootstrap counts
\[
(N_1^*,\ldots,N_n^*)\sim\operatorname{Multinomial}\left(n;\frac1n,\ldots,\frac1n\right),
\]
and let $\widehat\theta^*$ denote OLS computed from the resampled rows whenever the bootstrap design is nonsingular. On singular resamples, define $\widehat\theta^*=\widehat\theta$ purely as a bookkeeping convention. Under Assumption~\ref{ass:pairs}(i), the conditional probability of a singular bootstrap design is $o_p(1)$, so this convention does not affect the first-order law. Define
\[
\widehat Q_X=\frac1n\sum_i X_iX_i',\qquad
\widehat Q_X^*=\frac1n\sum_iN_i^*X_iX_i'.
\]
Because the original OLS residuals $\widehat\varepsilon_i=Y_i-X_i'\widehat\theta$ satisfy $\sum_iX_i\widehat\varepsilon_i=0$, the bootstrap normal equations give, on nonsingular resamples, the exact identity
\begin{equation}
\sqrt n(\widehat\theta^*-\widehat\theta)
=(\widehat Q_X^*)^{-1}\frac1{\sqrt n}\sum_i(N_i^*-1)X_i\widehat\varepsilon_i.
\label{eq:pairs-exact}
\end{equation}
No approximation has been used in \eqref{eq:pairs-exact}.

Let $\widehat v=M_WD$ denote the sample residual from regressing $D$ on $W$, and define
\[
\widehat Q_D=\frac1n\sum_i\widehat v_i^2,
\qquad
\widehat\phi_i=\widehat Q_D^{-1}\widehat v_i\widehat\varepsilon_i.
\]
The leading pairs-bootstrap influence statistic is
\[
Z_n^*=\frac1{\sqrt n}\sum_i(N_i^*-1)\widehat\phi_i.
\]
Since OLS residuals are orthogonal to both $D$ and $W$,
\[
\sum_i\widehat v_i\widehat\varepsilon_i=0
\]
exactly. Hence the conditional variance of $Z_n^*$ is
\begin{equation}
\Var^*(Z_n^*)
=\frac1n\sum_i\widehat\phi_i^2
=\widehat Q_D^{-2}\frac1n\sum_i\widehat v_i^2\widehat\varepsilon_i^2.
\label{eq:pairs-hc0}
\end{equation}
Thus the variance of the \emph{linearized} pairs-bootstrap statistic is exactly the heteroskedasticity-robust HC0 coefficient variance associated with the residualized randomized-regressor score. This identity does not, by itself, establish consistency of the conditional second moment of the nonlinear pairs-bootstrap OLS statistic.

\begin{assumption}[Pairs-bootstrap regularity]\label{ass:pairs}
Suppose:
\begin{enumerate}[label=(\roman*)]
\item with probability tending to one $X'X$ is nonsingular, the residualized treatment scale satisfies
\[
\frac{\widehat Q_D}{\sigma_{D,n}^2}\to_p1,
\]
and the maximum sample leverage satisfies
\[
\max_{1\le i\le n}h_{ii}=o_p(1),
\qquad
h_{ii}=X_i'(X'X)^{-1}X_i;
\]
\item the empirical influence scores satisfy
\[
\frac{\max_i\widehat\phi_i^2}{\sum_i\widehat\phi_i^2}=o_p(1);
\]
\item the unrestricted residualized score variance obeys
\[
\widehat Q_D^{-2}\frac1n\sum_i\widehat v_i^2\widehat\varepsilon_i^2-V_n^\#=o_p(1);
\]
\item the full OLS score has bounded conditional quadratic scale,
\[
\sum_{i=1}^n h_{ii}\widehat\varepsilon_i^2=O_p(1).
\]
\end{enumerate}
\end{assumption}
Condition (i) makes the original design and the coefficient-specific residualized treatment scale well behaved and controls the random bootstrap Hessian. Condition (ii) is the empirical Lindeberg condition for sampling from the empirical influence distribution. Condition (iii) is the feasible residualized-score variance requirement needed to identify the pairs-bootstrap target with $V_n^\#$. Condition (iv) controls the full bootstrap score in the design-normalized metric when the random Hessian is replaced by its sample counterpart. Under stronger conventional moment and uniform-integrability conditions, (iii)--(iv) can be derived from primitive assumptions; under the weaker CHLS control conditions we state them explicitly rather than conceal additional regularity.

\paragraph{Primitive sufficient conditions.}
Assumption~\ref{ass:pairs} is high level because the weakest CHLS conditions do not by themselves control unrestricted residual products or the random pairs-bootstrap Hessian. The Online Appendix gives a conventional primitive sufficient-condition route based on fixed dimension, fourth-plus-$\delta$ moments, a root-$n$ nuisance-score rate, and design nonsingularity. It also verifies the nuisance-score rate under $m$-dependence, martingale differences, strong mixing, and cluster structures. These results support Theorem~\ref{thm:pairs} but are not part of the paper's headline covariance-filter contribution.

\begin{theorem}[Ordinary pairs bootstrap]\label{thm:pairs}
Suppose the setup in Section~\ref{sec:setup}, Assumption~\ref{ass:stable}, and Assumption~\ref{ass:pairs} hold. Then, conditionally on the data,
\begin{equation}
\sqrt n(\widehat\beta^*-\widehat\beta)
=\frac1{\sqrt n}\sum_i(N_i^*-1)\widehat\phi_i+o_{P^*}(1)
\label{eq:pairs-lin}
\end{equation}
in probability, and
\[
\sup_{t\in\mathbb R}
\left|
P^*\!\left\{\frac{\sqrt n(\widehat\beta^*-\widehat\beta)}{\widehat V_{\mathrm{HC0},\beta}^{1/2}}\le t\right\}-\Phi(t)
\right|=o_p(1),
\]
where
\[
\widehat V_{\mathrm{HC0},\beta}
=\widehat Q_D^{-2}\frac1n\sum_i\widehat v_i^2\widehat\varepsilon_i^2.
\]
Moreover, by Assumption~\ref{ass:pairs}(iii),
\[
\widehat V_{\mathrm{HC0},\beta}-V_n^\#=o_p(1),
\]
and Proposition~\ref{prop:pairsprimitive} gives primitive sufficient conditions for this requirement together with Assumption~\ref{ass:pairs}(i)--(iv). Consequently, the ordinary pairs bootstrap has the same \emph{first-order distributional target} as the iid multiplier score bootstrap, and its linearized influence statistic has variance target $V_n^\#$.
\end{theorem}

\begin{corollary}[Ordinary-pairs size phase]\label{cor:pairsphase}
Suppose the conditions of Theorem~\ref{thm:pairs} hold and, in addition, the covariance-filter, positivity, standardized sampling-law, and relevance-index conditions of Theorem~\ref{thm:phase} hold with
\[
R_{A,n}\to r<1.
\]
For $a\in(0,1)$, let $q_{a,n}^*$ be any conditional $a$-quantile of $\sqrt n(\widehat\beta^*-\widehat\beta)$ under the ordinary pairs bootstrap, using the arbitrary bookkeeping convention on singular resamples employed in the proof of Theorem~\ref{thm:pairs}. Then the two-sided test that rejects when
\[
\sqrt n(\widehat\beta-\beta_0)<q_{\alpha/2,n}^*
\quad\text{or}\quad
\sqrt n(\widehat\beta-\beta_0)>q_{1-\alpha/2,n}^*
\]
has limiting rejection probability
\[
\pi_\alpha(r)
=
2\Phi\!\left(-z_{1-\alpha/2}\sqrt{1-r}\right).
\]
Hence ordinary pairs quantiles are anti-conservative for $r>0$, first-order correct for $r=0$, and conservative for $r<0$.
\end{corollary}

\paragraph{Proof of Corollary~\ref{cor:pairsphase}.}
Theorem~\ref{thm:pairs} and Assumption~\ref{ass:pairs}(iii) imply, conditionally on the data,
\[
\sup_t\left|
P^*\!\left\{
\frac{\sqrt n(\widehat\beta^*-\widehat\beta)}{(V_n^\#)^{1/2}}
\le t
\right\}-\Phi(t)
\right|=o_p(1),
\]
because $\widehat V_{\mathrm{HC0},\beta}/V_n^\#\to_p1$ under Assumption~\ref{ass:stable}. Continuity and strict monotonicity of $\Phi$ therefore give, for each fixed $a\in(0,1)$,
\[
\frac{q_{a,n}^*}{(V_n^\#)^{1/2}}\to_p z_a.
\]
The covariance-filter identity and $R_{A,n}\to r$ give
\[
\frac{V_n^\#}{V_n}\to1-r.
\]
Hence
\[
\frac{q_{\alpha/2,n}^*}{V_n^{1/2}}
\to_p-z_{1-\alpha/2}\sqrt{1-r},
\qquad
\frac{q_{1-\alpha/2,n}^*}{V_n^{1/2}}
\to_p z_{1-\alpha/2}\sqrt{1-r}.
\]
Combining these random-threshold limits with the standardized OLS sampling central limit theorem (CLT) in Theorem~\ref{thm:phase} yields the displayed rejection probability by Slutsky's theorem and continuity of the standard normal law. \qed

Combining Theorem~\ref{thm:pairs} with Theorem~\ref{thm:filter} yields the variance discrepancy for the \emph{linearized} pairs-bootstrap statistic,
\begin{equation}
\widehat V_{\mathrm{HC0},\beta}-V_n
=-\sigma_{D,n}^{-4}\Delta_{A,n}+o_p(1).
\label{eq:pairs-gap}
\end{equation}
For binary randomized treatment,
\begin{equation}
\widehat V_{\mathrm{HC0},\beta}-V_n
=-\Delta_{\tau,n}+o_p(1).
\label{eq:pairs-gap-te}
\end{equation}
Hence the linearized ordinary-pairs variance target is too small, correct, or too large according to the same aggregate covariance criterion. If, in addition, the sampling-law and relevance-index conditions of Theorem~\ref{thm:phase} hold with $R_{A,n}\to r<1$, a two-sided test based on ordinary-pairs quantiles of $\sqrt n(\widehat\beta^*-\widehat\beta)$ has limiting rejection probability $\pi_\alpha(r)$. It is therefore anti-conservative for $r>0$, first-order correct for $r=0$, and conservative for $r<0$. Without those sampling-law and relative-gap conditions, equations~\eqref{eq:pairs-gap}--\eqref{eq:pairs-gap-te} are variance-target statements only.

\begin{remark}[Pairs-bootstrap standard errors versus pairs-bootstrap laws]\label{rem:pairs-moments}
Let
\[
V_{\mathrm{pairs},n}^*
=\Var^*\!\left\{\sqrt n(\widehat\beta^*-\widehat\beta)\right\},
\]
whenever this conditional variance is well defined. Theorem~\ref{thm:pairs} does \emph{not} imply
\[
V_{\mathrm{pairs},n}^*-\widehat V_{\mathrm{HC0},\beta}=o_p(1).
\]
The theorem establishes conditional weak approximation after linearization. Convergence of conditional second moments additionally requires conditional uniform integrability or an equivalent moment bound on the nonlinear bootstrap estimator, which is the issue emphasized by \citet{HahnLiao2021}. Without such a condition, we do not claim that the actual pairs-bootstrap standard error is consistent, conservative, or anti-conservative.
\end{remark}

For regressions with controls, the Online Appendix also studies a regularized pairs construction that truncates rare ill-conditioned resamples. Under fourth-moment control it is $L^2$-close to the linearized pairs statistic and has conditional variance converging to $V_n^\#$. This auxiliary result closes the Hahn--Liao moment issue without changing the covariance-filter conclusion.

\paragraph{Binary treatment without additional controls.}
In the intercept-plus-binary-treatment regression, singular resamples arise only when a bootstrap sample contains one treatment status. Let
\[
n_1=\sum_iD_i,\qquad \widehat p=n_1/n,\qquad
\mathcal H_n=\{1\le n_1\le n-1\},
\]
and, on $\mathcal H_n$, let
\[
N_1^*\mid\mathcal D_n\sim\operatorname{Binomial}(n,\widehat p),
\qquad
\mathcal H_n^*=\{1\le N_1^*\le n-1\}.
\]
If $p_n\in[\underline p,1-\underline p]$, then the original-sample singular probability and
\[
P^*((\mathcal H_n^*)^c)
=
\widehat p^{\,n}+(1-\widehat p)^n
\]
both vanish exponentially. Define the nondegenerate-resample pairs procedure by conditioning on $\mathcal H_n^*$ and set
\[
V_{\mathrm{pairs},H,n}^*
=
\Var^*\!\left(
\sqrt n(\widehat\beta^*-\widehat\beta)
\,\middle|\,\mathcal H_n^*
\right).
\]

\begin{theorem}[Ordinary pairs-bootstrap standard error for binary treatment conditional on nondegenerate resamples]\label{thm:binarypairs}
Suppose the binary-treatment specialization in Section~\ref{sec:binary}, the conditions of Lemma~\ref{lem:diag}, and Assumption~\ref{ass:stable} hold. Assume also
\[
\sqrt n(\widehat\beta-\beta_n)=O_p(1),
\]
and the conditions of Lemma~\ref{lem:restricted}. Suppose
\[
p_n\in[\underline p,1-\underline p]
\]
for some $\underline p>0$, and
\[
\frac1n\sum_{i=1}^n |Y_{i,n}|^{4+\delta}=O_p(1),
\qquad
\frac1n\sum_{i=1}^n \varepsilon_{i,n}^2=O_p(1)
\]
for some $\delta>0$. Then, on $\mathcal H_n$,
\[
V_{\mathrm{pairs},H,n}^*
=
\widehat V_{\mathrm{HC0},\beta}+o_p(1)
=
V_n^\#+o_p(1),
\]
and because $P(\mathcal H_n)\to1$,
\[
\boxed{
V_{\mathrm{pairs},H,n}^*-V_n
=
-\Delta_{\tau,n}+o_p(1)
}
\]
in probability. Thus the ordinary pairs-bootstrap standard error conditional on $\mathcal H_n^*$ has the same first-order variance discrepancy as the iid multiplier score bootstrap.
\end{theorem}

The nondegenerate-resample convention affects only exponentially rare singular resamples under the maintained treatment-probability condition. The Online Appendix gives the exact finite-sample convention, inverse-binomial moment calculation, regularization argument, and leverage calculation for the general row-resampled Hessian. These pairs results concern iid row resampling and do not cover fixed-treatment-count complete randomization, matched-pair, covariate-adaptive, or cluster-level assignment.

\subsection{Proof of Theorem~\ref{thm:pairs}: ordinary pairs bootstrap}
Let $X_i=(D_i,W_i')'$ and $\widehat Q_X=n^{-1}\sum_iX_iX_i'$. Conditional on the observed sample, let $N_i^*$ denote multinomial bootstrap counts and set $\widehat Q_X^*=n^{-1}\sum_iN_i^*X_iX_i'$. On the event that $\widehat Q_X^*$ is nonsingular, $Y_i=X_i'\widehat\theta+\widehat\varepsilon_i$ and $\sum_iX_i\widehat\varepsilon_i=0$ imply exactly
\[
\widehat Q_X^*(\widehat\theta^*-\widehat\theta)
=\frac1n\sum_i(N_i^*-1)X_i\widehat\varepsilon_i.
\]
Multiplication by $\sqrt n$ yields equation~\eqref{eq:pairs-exact}.

We first control the bootstrap Hessian. Put
\[
B_i=\widehat Q_X^{-1/2}X_iX_i'\widehat Q_X^{-1/2}.
\]
Then $n^{-1}\sum_iB_i=I_k$ and
\[
\widehat Q_X^{-1/2}(\widehat Q_X^*-\widehat Q_X)\widehat Q_X^{-1/2}
=\frac1n\sum_i(N_i^*-1)B_i.
\]
For multinomial counts, direct covariance calculation gives
\[
E^*\left\|\frac1n\sum_i(N_i^*-1)B_i\right\|_F^2
\le \frac1{n^2}\sum_i\|B_i\|_F^2.
\]
Because $B_i$ is rank one,
\[
\|B_i\|_F^2=(X_i'\widehat Q_X^{-1}X_i)^2=n^2h_{ii}^2.
\]
Hence
\[
E^*\left\|\widehat Q_X^{-1/2}(\widehat Q_X^*-\widehat Q_X)\widehat Q_X^{-1/2}\right\|_F^2
\le\sum_i h_{ii}^2
\le k\max_i h_{ii}=o_p(1).
\]
Conditional Markov therefore gives
\[
\widehat Q_X^{-1/2}(\widehat Q_X^*-\widehat Q_X)\widehat Q_X^{-1/2}=o_{P^*}(1)
\]
in probability. Hence
\[
P^*(\widehat Q_X^*\text{ is singular})=o_p(1).
\]
The bookkeeping convention on singular draws is therefore first-order irrelevant. A Neumann-series argument yields, on the nonsingular high-probability event,
\[
(\widehat Q_X^*)^{-1}=\widehat Q_X^{-1}+o_{P^*}(1)
\]
in the corresponding $\widehat Q_X$-normalized operator norm. Because the singular event has conditional probability $o_p(1)$ and $\widehat\theta^*=\widehat\theta$ there by convention, the resulting first-order expansion extends to the full bootstrap probability space.

Next define
\[
G_n^*=\frac1{\sqrt n}\sum_i(N_i^*-1)X_i\widehat\varepsilon_i.
\]
A direct multinomial covariance calculation gives
\[
E^*\|\widehat Q_X^{-1/2}G_n^*\|^2
=\sum_i h_{ii}\widehat\varepsilon_i^2.
\]
Assumption~\ref{ass:pairs}(iv) therefore implies $S_n^*:=\widehat Q_X^{-1/2}G_n^*=O_{P^*}(1)$ in probability. For the coefficient on $D$, put
\[
a_n'=e_D'\widehat Q_X^{-1/2}.
\]
The partitioned-inverse/Frisch--Waugh--Lovell (FWL) identity gives
\[
\|a_n\|^2=e_D'\widehat Q_X^{-1}e_D=\widehat Q_D^{-1}=O_p(1)
\]
by Assumption~\ref{ass:pairs}(i). On the nonsingular high-probability event,
\[
\sqrt n(\widehat\beta^*-\widehat\beta)
=a_n'(I+A_n^*)^{-1}S_n^*,
\qquad
A_n^*=\widehat Q_X^{-1/2}(\widehat Q_X^*-\widehat Q_X)\widehat Q_X^{-1/2}.
\]
Since $A_n^*=o_{P^*}(1)$ in probability and $S_n^*=O_{P^*}(1)$, the Neumann bound and $\|a_n\|=O_p(1)$ yield
\[
\sqrt n(\widehat\beta^*-\widehat\beta)
=a_n'S_n^*+o_{P^*}(1).
\]
The same FWL identity gives
\[
a_n'S_n^*=\frac1{\sqrt n}\sum_i(N_i^*-1)\widehat\phi_i,
\]
and therefore
\[
\sqrt n(\widehat\beta^*-\widehat\beta)
=\frac1{\sqrt n}\sum_i(N_i^*-1)\widehat\phi_i+o_{P^*}(1),
\]
which proves \eqref{eq:pairs-lin}.

Because $\sum_i\widehat v_i\widehat\varepsilon_i=0$, we have $n^{-1}\sum_i\widehat\phi_i=0$. The multinomial bootstrap is equivalently the sum of $n$ iid draws from the empirical distribution placing mass $1/n$ on each $\widehat\phi_i$. Hence its conditional variance is exactly
\[
\frac1n\sum_i\widehat\phi_i^2
=\widehat V_{\mathrm{HC0},\beta}.
\]
Assumption~\ref{ass:pairs}(ii) is the Lindeberg condition for this empirical triangular array. Assumption~\ref{ass:pairs}(iii) and Assumption~\ref{ass:stable} imply that $\widehat V_{\mathrm{HC0},\beta}$ is bounded away from zero and infinity with probability tending to one. Hence the unstudentized $o_{P^*}(1)$ remainder in \eqref{eq:pairs-lin} is also negligible after studentization. The conditional Lindeberg--Feller theorem therefore yields
\[
\sup_t\left|P^*\left\{\frac{n^{-1/2}\sum_i(N_i^*-1)\widehat\phi_i}{\widehat V_{\mathrm{HC0},\beta}^{1/2}}\le t\right\}-\Phi(t)\right|=o_p(1).
\]
Combining this with the linearization proves the conditional normal approximation. The feasible-score condition used in the pairs-bootstrap linearization gives $\widehat V_{\mathrm{HC0},\beta}-V_n^\#=o_p(1)$, completing the proof. \qed

\paragraph{Second-moment warning.}
The preceding argument gives conditional weak approximation of the nonlinear pairs-bootstrap estimator by its linearized influence statistic. It does not establish convergence of
\[
E^*\!\left[n(\widehat\beta^*-\widehat\beta)^2\right]
\]
to the second moment of the linearized statistic. Such a conclusion would require, for example, conditional uniform integrability of
$n(\widehat\beta^*-\widehat\beta)^2$ or an $L^{2+\delta}$ bootstrap bound strong enough to control the random inverse Hessian. We therefore identify $\widehat V_{\mathrm{HC0},\beta}$ as the variance of the linearized pairs-bootstrap statistic, not as the actual conditional variance of the nonlinear pairs-bootstrap OLS estimator.

\subsection{Proof of Theorem~\ref{thm:binarypairs}: ordinary pairs bootstrap conditional on nondegenerate resamples}\label{app:binarypairs}

Work on the original-sample event
\[
\mathcal H_n=\{1\le n_1\le n-1\}.
\]
For the intercept-plus-binary-treatment regression,
\[
\widehat\beta=\bar Y_1-\bar Y_0.
\]
Represent the pairs bootstrap by iid indices
\[
I_1,\ldots,I_n
\]
drawn uniformly from $\{1,\ldots,n\}$ conditional on the observed sample, and put
\[
K=\sum_{r=1}^n D_{I_r}.
\]
Then
\[
K\mid\mathcal D_n\sim\operatorname{Binomial}(n,\widehat p).
\]

Conditional on a complete bootstrap treatment-status vector with exactly $k$ treated positions, the indices occupying treated positions are independent draws from the empirical treated distribution, while those occupying control positions are independent draws from the empirical control distribution. The corresponding conditional laws do not depend on the arrangement of the $k$ treated positions. Consequently, conditional on $K=k\in\{1,\ldots,n-1\}$,
\[
E^*(\widehat\beta^*-\widehat\beta\mid K=k)=0
\]
and
\[
\Var^*(\widehat\beta^*-\widehat\beta\mid K=k)
=
\frac{\widehat s_1^2}{k}
+
\frac{\widehat s_0^2}{n-k},
\]
where
\[
\widehat s_d^2
=
\frac1{n_d}\sum_{i:D_i=d}(Y_i-\bar Y_d)^2
\]
is the variance under the empirical distribution in arm $d$. Since the conditional mean is zero for every nondegenerate $k$, the law of total variance gives
\[
V_{\mathrm{pairs},H,n}^*
=
n\widehat s_1^2 E^*(K^{-1}\mid\mathcal H_n^*)
+
n\widehat s_0^2 E^*\{(n-K)^{-1}\mid\mathcal H_n^*\}.
\]

For the inverse-binomial expectations, define
\[
\mathcal E_{1,n}^*
=
\{|K-n\widehat p|\le n\widehat p/2\}.
\]
On the high-probability original-sample event
\[
\widehat p\in[\underline p/2,1-\underline p/2],
\]
a Chernoff bound gives
\[
P^*((\mathcal E_{1,n}^*)^c)\le2e^{-c n}.
\]
On $\mathcal E_{1,n}^*$, Taylor's theorem for $f(x)=x^{-1}$ around
$m=n\widehat p$ yields
\[
\frac1K
=
\frac1m-\frac{K-m}{m^2}
+\frac{(K-m)^2}{m^3}
+R_{n}^*,
\]
where, because $K\ge m/2$ on $\mathcal E_{1,n}^*$,
\[
|R_n^*|
\le
C\frac{|K-m|^3}{m^4}.
\]
For a binomial variable with $\widehat p$ bounded away from zero and one,
\[
E^*|K-m|^3=O(n^{3/2}),
\qquad
E^*(K-m)^2=n\widehat p(1-\widehat p).
\]
The truncation of the first centered moment to $\mathcal E_{1,n}^*$ changes it only by an exponentially small amount. Hence
\[
nE^*(K^{-1};\mathcal E_{1,n}^*)
=
\widehat p^{-1}
+
\frac{1-\widehat p}{n\widehat p^2}
+
O_p(n^{-3/2})
+
O_p(e^{-cn}).
\]
On $\{K\ge1\}\cap(\mathcal E_{1,n}^*)^c$,
\[
K^{-1}\le1,
\]
so
\[
nE^*(K^{-1};K\ge1,(\mathcal E_{1,n}^*)^c)
\le
nP^*((\mathcal E_{1,n}^*)^c)
=o_p(n^{-M})
\]
for every fixed $M$. Therefore
\[
nE^*(K^{-1};K\ge1)
=
\widehat p^{-1}+O_p(n^{-1}).
\]
Because
\[
P^*(K=0\text{ or }K=n)
=
(1-\widehat p)^n+\widehat p^n
=
O_p(e^{-cn}),
\]
conditioning on $\mathcal H_n^*=\{1\le K\le n-1\}$ changes this expectation only by $O_p(e^{-cn})$. Thus
\[
nE^*(K^{-1}\mid\mathcal H_n^*)
=
\widehat p^{-1}+O_p(n^{-1}).
\]
Applying the same argument to $n-K\sim\operatorname{Binomial}(n,1-\widehat p)$ gives
\[
nE^*\{(n-K)^{-1}\mid\mathcal H_n^*\}
=
(1-\widehat p)^{-1}+O_p(n^{-1}).
\]

The assumed empirical $(4+\delta)$ moment bound implies
\[
\widehat s_1^2+\widehat s_0^2=O_p(1)
\]
on $\mathcal H_n$ because $\widehat p$ is bounded away from zero and one with probability tending to one. Hence
\[
V_{\mathrm{pairs},H,n}^*
=
\frac{\widehat s_1^2}{\widehat p}
+
\frac{\widehat s_0^2}{1-\widehat p}
+
o_p(1).
\]

For a binary regressor with an intercept, exact algebra gives
\[
\widehat Q_D
=
\frac1n\sum_i(D_i-\widehat p)^2
=
\widehat p(1-\widehat p),
\]
and
\[
\begin{aligned}
\widehat V_{\mathrm{HC0},\beta}
&=
\widehat Q_D^{-2}
\frac1n\sum_i(D_i-\widehat p)^2\widehat\varepsilon_i^2\\
&=
\frac{\widehat s_1^2}{\widehat p}
+
\frac{\widehat s_0^2}{1-\widehat p}.
\end{aligned}
\]
Thus
\[
V_{\mathrm{pairs},H,n}^*
=
\widehat V_{\mathrm{HC0},\beta}+o_p(1).
\]

It remains to identify unrestricted HC0 with $V_n^\#$. In the intercept-plus-treatment regression,
\[
\widehat\varepsilon_i
=
\varepsilon_i-(\widehat\alpha-\alpha_n)-D_i(\widehat\beta-\beta_n).
\]
The intercept normal equation gives
\[
\widehat\alpha-\alpha_n
=
\frac1n\sum_i\varepsilon_i
-\bar D(\widehat\beta-\beta_n).
\]
The conditions of Lemma~\ref{lem:restricted}, applied to the intercept in $W_i$, imply
\[
n^{-1}\sum_i\varepsilon_i=o_p(1),
\]
while the explicit root-$n$ rate in the theorem gives
\[
\widehat\beta-\beta_n=O_p(n^{-1/2}).
\]
Therefore
\[
\widehat\alpha-\alpha_n=o_p(1)
\]
and, because $D_i$ is bounded,
\[
\frac1n\sum_i(\widehat\varepsilon_i-\varepsilon_i)^2=o_p(1).
\]
Together with $n^{-1}\sum_i\varepsilon_i^2=O_p(1)$, Cauchy--Schwarz gives
\[
\frac1n\sum_i(D_i-\widehat p)^2
(\widehat\varepsilon_i^2-\varepsilon_i^2)
=o_p(1).
\]
Moreover,
\[
(D_i-\widehat p)^2-(D_i-p_n)^2
=
(p_n-\widehat p)\{2D_i-p_n-\widehat p\},
\]
so boundedness of $D_i$,
\[
\widehat p-p_n=O_p(n^{-1/2}),
\]
and $n^{-1}\sum_i\varepsilon_i^2=O_p(1)$ imply
\[
\frac1n\sum_i
\left[(D_i-\widehat p)^2-(D_i-p_n)^2\right]\varepsilon_i^2
=o_p(1).
\]
Finally,
\[
\widehat Q_D^{-2}
=
q_n^{-2}+o_p(1)
=
\sigma_{D,n}^{-4}+o_p(1).
\]
Hence
\[
\widehat V_{\mathrm{HC0},\beta}
=
\sigma_{D,n}^{-4}
\frac1n\sum_i(D_i-p_n)^2\varepsilon_i^2
+o_p(1).
\]
Lemma~\ref{lem:diag} identifies the right-hand side with $V_n^\#+o_p(1)$. Corollary~\ref{cor:te} then gives
\[
V_n^\#-V_n=-\Delta_{\tau,n}.
\]
Combining the preceding displays proves the theorem. \qed

\paragraph{Moment condition.}
The explicit inverse-binomial calculation itself only requires
$\widehat s_1^2+\widehat s_0^2=O_p(1)$. We retain the displayed empirical $(4+\delta)$ moment condition in the theorem because it is a simple sufficient condition compatible with the stronger tail requirements used elsewhere in the pairs-bootstrap analysis. It is not claimed to be minimal.

\subsection{Regularized pairs-bootstrap standard errors}\label{app:regpairs}

\begin{assumption}[Bootstrap fourth-moment control]\label{ass:pairsmoment}
Let
\[
A_n^*=\widehat Q_X^{-1/2}(\widehat Q_X^*-\widehat Q_X)\widehat Q_X^{-1/2},
\qquad
S_n^*=\widehat Q_X^{-1/2}\frac1{\sqrt n}\sum_i(N_i^*-1)X_i\widehat\varepsilon_i.
\]
Suppose
\[
E^*\|A_n^*\|_{\mathrm{op}}^4=o_p(1),
\qquad
E^*\|S_n^*\|^4=O_p(1).
\]
\end{assumption}

Define
\[
\mathcal G_n^*
=
\left\{
\left\|
\widehat Q_X^{-1/2}
(\widehat Q_X^*-\widehat Q_X)
\widehat Q_X^{-1/2}
\right\|_{\mathrm{op}}
\le\frac12
\right\}.
\]
The regularized pairs estimator equals ordinary pairs OLS on $\mathcal G_n^*$ and equals the original estimator on its complement:
\[
\widetilde\theta^*
=
\widehat\theta
+
1\{\mathcal G_n^*\}
(\widehat Q_X^*)^{-1}
\frac1n\sum_i(N_i^*-1)X_i\widehat\varepsilon_i.
\]
Let $\widetilde\beta^*$ denote its coefficient on $D$. This definition makes the estimator measurable and finite on every bootstrap draw.

\begin{proposition}[Regularized pairs $L^2$ and second-moment approximation]\label{prop:regpairs}
Suppose the conditions of Theorem~\ref{thm:pairs} and Assumption~\ref{ass:pairsmoment} hold. Let
\[
Z_n^*=\frac1{\sqrt n}\sum_i(N_i^*-1)\widehat\phi_i
\]
be the linearized pairs statistic. Then
\[
E^*\!\left[
\left\{
\sqrt n(\widetilde\beta^*-\widehat\beta)-Z_n^*
\right\}^2
\right]=o_p(1),
\]
\[
E^*\!\left[\sqrt n(\widetilde\beta^*-\widehat\beta)\right]=o_p(1),
\]
and
\[
\Var^*\!\left\{\sqrt n(\widetilde\beta^*-\widehat\beta)\right\}
=\widehat V_{\mathrm{HC0},\beta}+o_p(1)
=V_n^\#+o_p(1).
\]
\end{proposition}

\paragraph{Proof of Proposition~\ref{prop:regpairs}.}
Write
\[
S_n^*
=
\widehat Q_X^{-1/2}G_n^*,
\qquad
A_n^*
=
\widehat Q_X^{-1/2}
(\widehat Q_X^*-\widehat Q_X)
\widehat Q_X^{-1/2}.
\]
On $\mathcal G_n^*$,
\[
(\widehat Q_X^*)^{-1}
=
\widehat Q_X^{-1/2}(I+A_n^*)^{-1}\widehat Q_X^{-1/2}.
\]
Let
\[
a_n'
=
e_D'\widehat Q_X^{-1/2}.
\]
The regularized coefficient displacement can therefore be written as
\[
\sqrt n(\widetilde\beta^*-\widehat\beta)
=
1\{\mathcal G_n^*\}
a_n'(I+A_n^*)^{-1}S_n^*.
\]
By the FWL identity, the linearized statistic in Proposition~\ref{prop:regpairs} also satisfies
\[
Z_n^*=a_n'S_n^*.
\]
Hence
\[
R_n^*
:=
\sqrt n(\widetilde\beta^*-\widehat\beta)-Z_n^*
\]
equals
\[
1\{\mathcal G_n^*\}
a_n'\{(I+A_n^*)^{-1}-I\}S_n^*
-
1\{(\mathcal G_n^*)^c\}a_n'S_n^*.
\]

By Assumption~\ref{ass:pairs}(i),
$\|a_n\|^2=\widehat Q_D^{-1}=O_p(1)$.  On $\mathcal G_n^*$,
\[
\|(I+A_n^*)^{-1}-I\|_{\mathrm{op}}
\le
2\|A_n^*\|_{\mathrm{op}}.
\]
Therefore, by conditional Cauchy--Schwarz,
\[
E^*\!\left[
1\{\mathcal G_n^*\}|R_n^*|^2
\right]
\le
C\|a_n\|^2
\{E^*\|A_n^*\|_{\mathrm{op}}^4\}^{1/2}
\{E^*\|S_n^*\|^4\}^{1/2}
=o_p(1).
\]
For the bad-design event, conditional Cauchy--Schwarz and Markov give
\[
\begin{aligned}
E^*\!\left[
1\{(\mathcal G_n^*)^c\}|a_n'S_n^*|^2
\right]
&\le
\|a_n\|^2
\{E^*\|S_n^*\|^4\}^{1/2}
\{P^*((\mathcal G_n^*)^c)\}^{1/2}\\
&\le
C\|a_n\|^2
\{E^*\|S_n^*\|^4\}^{1/2}
\{E^*\|A_n^*\|_{\mathrm{op}}^4\}^{1/2}
=o_p(1).
\end{aligned}
\]
Thus
\[
E^*[(R_n^*)^2]=o_p(1).
\]

Since $E^*Z_n^*=0$ and
\[
E^*[(Z_n^*)^2]=\widehat V_{\mathrm{HC0},\beta},
\]
Cauchy--Schwarz yields
\[
\left|
E^*\!\left[n(\widetilde\beta^*-\widehat\beta)^2\right]
-
\widehat V_{\mathrm{HC0},\beta}
\right|
=o_p(1).
\]
Moreover,
\[
|E^*\sqrt n(\widetilde\beta^*-\widehat\beta)|
=
|E^*R_n^*|
\le
\{E^*(R_n^*)^2\}^{1/2}
=o_p(1),
\]
so centering by the conditional mean changes the second moment by $o_p(1)$. Hence
\[
\widetilde V_{\mathrm{pairs},n}^*
=
\widehat V_{\mathrm{HC0},\beta}+o_p(1).
\]
Assumption~\ref{ass:pairs}(iii) gives
$\widehat V_{\mathrm{HC0},\beta}=V_n^\#+o_p(1)$, and Theorem~\ref{thm:filter} gives the stated covariance-filter discrepancy. Proposition~\ref{prop:pairsprimitive} supplies a primitive route to Assumption~\ref{ass:pairs}. \qed

\paragraph{A convenient sufficient route for Assumption~\ref{ass:pairsmoment}.}
The required conditional fourth moments can be verified directly. Conditional on the data, represent multinomial resampling by iid indices $I_1,\ldots,I_n$ uniform on $\{1,\ldots,n\}$. Define
\[
B_i=\widehat Q_X^{-1/2}X_iX_i'\widehat Q_X^{-1/2},
\qquad
U_i=\widehat Q_X^{-1/2}X_i\widehat\varepsilon_i.
\]
Then
\[
A_n^*=\frac1n\sum_{r=1}^n(B_{I_r}-I_k),
\qquad
S_n^*=\frac1{\sqrt n}\sum_{r=1}^n U_{I_r},
\]
because $n^{-1}\sum_iB_i=I_k$ and $n^{-1}\sum_iU_i=0$ by the OLS normal equations.

For fixed $k$, applying the scalar fourth-moment formula coordinatewise and using norm equivalence gives a constant $C_k$ such that
\[
E^*\|A_n^*\|_F^4
\le
C_k\left[
\sum_i h_{ii}^4+
\left(\sum_i h_{ii}^2\right)^2+
n^{-2}
\right].
\]
Indeed, $\|B_i\|_F=n h_{ii}$ because $B_i$ is rank one. Hence
\[
\sum_i h_{ii}^2=o_p(1)
\]
implies
\[
E^*\|A_n^*\|_{\mathrm{op}}^4
\le E^*\|A_n^*\|_F^4=o_p(1).
\]

Similarly,
\[
E^*\|S_n^*\|^4
\le
C_k\left[
\sum_i h_{ii}^2\widehat\varepsilon_i^4+
\left(\sum_i h_{ii}\widehat\varepsilon_i^2\right)^2
\right],
\]
because
\[
\|U_i\|^2=n h_{ii}\widehat\varepsilon_i^2.
\]
Therefore Assumption~\ref{ass:pairsmoment} follows from
\[
\sum_i h_{ii}^2=o_p(1),
\qquad
\sum_i h_{ii}\widehat\varepsilon_i^2=O_p(1),
\qquad
\sum_i h_{ii}^2\widehat\varepsilon_i^4=o_p(1).
\] Together with Proposition~\ref{prop:ui-moments}(iii), these same conditions also imply conditional squared uniform integrability of the regularized pairs coefficient. Thus the tail assumptions used to justify its bootstrap standard error are sufficient in practice for the Hahn--Liao moment requirement; no separate UI assumption is needed.
The first condition follows from vanishing maximum leverage since $\sum_i h_{ii}=k$. The final condition is a genuine tail requirement and is intentionally not inferred from weak bootstrap validity alone.

\subsection{Why the pairs-bootstrap variance target equals the diagonal assignment target}
Theorem~\ref{thm:pairs} isolates the only remaining feasible-score issue in Assumption~\ref{ass:pairs}(iii). Under the stable-scale regime, a sufficient primitive route is:
\begin{enumerate}[label=(\roman*)]
\item $n^{-1}\sum_i\{\widehat v_i-(D_i-\mu_{D,n})\}^2=o_p(1)$, which follows from independence of $D_i$ and $W_i$, nonsingularity of $n^{-1}W'W$, and Lemma~\ref{lem:design};
\item $n^{-1}\sum_i(\widehat\varepsilon_i-\varepsilon_i)^2=o_p(1)$ together with a product-negligibility condition sufficient to replace $\widehat v_i^2\widehat\varepsilon_i^2$ by $(D_i-\mu_{D,n})^2\varepsilon_i^2$ in average;
\item Lemma~\ref{lem:diag}.
\end{enumerate}
Under stronger conventional moment and uniform-integrability assumptions these replacement steps can be verified directly. Under the weaker CHLS control assumptions, however, the required products need not be controlled by second moments alone. We therefore retain Assumption~\ref{ass:pairs}(iii), together with the score-scale condition in Assumption~\ref{ass:pairs}(iv), rather than conceal additional regularity.

\section{Primitive sufficient conditions for pairs resampling}\label{app:primitive-pairs}

\begin{assumption}[Primitive sufficient conditions for pairs resampling]\label{ass:pairsprimitive}
Let $k=\dim(X_i)$ be fixed. In addition to the setup and covariance-filter conditions of Theorem~\ref{thm:filter} and Assumption~\ref{ass:stable}, suppose that, for some $\delta>0$,
\begin{enumerate}[label=(P\arabic*)]
\item the sample design is uniformly well conditioned: for some $c>0$,
\[
P\{\lambda_{\min}(n^{-1}X'X)\ge c\}\to1,
\]
and
\[
\sup_{i,n}\E\|X_i\|^{4+\delta}<\infty;
\]
\item the regression errors satisfy
\[
\sup_{i,n}\E|\varepsilon_i|^{4+\delta}<\infty;
\]
no conditional moment restriction on $\varepsilon_i$ given the full design is imposed;
\item the control-score vector obeys the root-$n$ rate
\[
\frac1{\sqrt n}W'\varepsilon=O_p(1);
\]
this strengthens the CHLS requirement $n^{-1}W'\varepsilon=o_p(1)$ only by imposing the conventional stochastic rate needed for full-vector root-$n$ OLS estimation;
\item for some $\delta_e\in(0,2]$, the assignment loading satisfies
\[
\sup_n\sup_{d\in\operatorname{supp}(D_{1,n})}\|\lambda_n(d)\|<\infty,
\]
and the latent array satisfies
\[
\frac1n\sum_{i=1}^n\|e_{i,n}\|^{2+\delta_e}=O_p(1),
\qquad
\left\|
\frac1n\sum_{i=1}^n
\{e_{i,n}e_{i,n}'-E[e_{i,n}e_{i,n}']\}
\right\|=o_p(1).
\]
Together with (P1), these are the latent-array conditions used by Lemma~\ref{lem:diag} to deliver oracle diagonal-score concentration.
\end{enumerate}
\end{assumption}

Assumption~\ref{ass:pairsprimitive} is a conventional set of sufficient conditions and is stronger than the baseline CHLS conditions. It is included to show that Assumption~\ref{ass:pairs} can be supported by primitive restrictions rather than left entirely high level. The fourth-plus-$\delta$ moments provide uniform integrability for the score products, while (P4) supplies the latent-array law of large numbers needed for oracle diagonal-score concentration. The oracle diagonal-score nondegeneracy needed below is already part of Assumption~\ref{ass:stable} and is therefore not repeated here. No conditional second-moment bound on the errors given the full design is required: the design-normalized score is controlled by a weighted Markov argument using only unconditional moments. Nor do we impose root-$n$ consistency of the full OLS vector directly. Proposition~\ref{prop:pairsprimitive} derives it from the control-score rate in (P3), the exact randomized-score variance decomposition, and design nonsingularity. The root-$n$ rate for the sample projection of $D$ on $W$ is likewise derived from random assignment and the design conditions.

\begin{proposition}[Primitive conditions imply pairs-bootstrap regularity]\label{prop:pairsprimitive}
Suppose Assumption~\ref{ass:pairsprimitive} holds. Then Assumption~\ref{ass:pairs}(i)--(iv) holds. Consequently, the conclusions of Theorem~\ref{thm:pairs} follow from Assumption~\ref{ass:pairsprimitive}.
\end{proposition}

\subsection{Concrete sufficient conditions for the control-score rate}

Condition (P3) is a rate restriction on the nuisance score rather than a dependence assumption per se. It follows from the following covariance-growth condition.

\begin{proposition}[Control-score covariance criterion]\label{prop:controlscore}
Let $Z_{i,n}=W_{i,n}\varepsilon_{i,n}$ and $\bar Z_{i,n}=Z_{i,n}-E Z_{i,n}$. If
\[
\left\|\frac1{\sqrt n}\sum_{i=1}^n E Z_{i,n}\right\|\le C
\qquad\text{and}\qquad
\frac1n E\left\|\sum_{i=1}^n \bar Z_{i,n}\right\|^2\le C
\]
for all sufficiently large $n$, then
\[
\frac1{\sqrt n}W'\varepsilon=O_p(1).
\]
\end{proposition}

The covariance criterion is implied by several familiar dependence structures. The following corollary records four useful examples; none is needed for the abstract results above.

\begin{corollary}[Primitive dependence classes implying (P3)]\label{cor:controlscoreclasses}
Suppose $k_W=\dim(W_i)$ is fixed and
\[
\left\|n^{-1/2}\sum_i E Z_{i,n}\right\|=O(1).
\]
Write $\bar Z_{i,n}=Z_{i,n}-E Z_{i,n}$. Condition (P3) follows under any one of the following sets of conditions.
\begin{enumerate}[label=(\alph*)]
\item \emph{$m$-dependence:} $\{\bar Z_{i,n}\}$ is $m$-dependent for a fixed $m$ and $\sup_{i,n}E\|\bar Z_{i,n}\|^2<\infty$.
\item \emph{Martingale differences:} for some filtration $\{\mathcal F_{i,n}\}$, $\bar Z_{i,n}$ is $\mathcal F_{i,n}$-measurable, $E(\bar Z_{i,n}\mid\mathcal F_{i-1,n})=0$, and $\sup_{i,n}E\|\bar Z_{i,n}\|^2<\infty$.
\item \emph{Strong mixing:} each coordinate of $\bar Z_{i,n}$ has uniformly bounded $(2+\eta)$ moments for some $\eta>0$, the centered array is uniformly strongly mixing with coefficients $\alpha_n(h)$ satisfying
\[
\sup_n\sum_{h=1}^{\infty}\alpha_n(h)^{\eta/(2+\eta)}<\infty.
\]
\item \emph{Independent clusters:} the centered observations can be partitioned into independent clusters $\mathcal C_{g,n}$, $\sup_{i,n}E\|\bar Z_{i,n}\|^2<\infty$, and
\[
\sum_g |\mathcal C_{g,n}|^2=O(n).
\]
Within-cluster dependence may otherwise be arbitrary.
\end{enumerate}
\end{corollary}

The conditions in Corollary~\ref{cor:controlscoreclasses} are sufficient rather than necessary. In particular, the pairs-bootstrap theorem itself only needs the rate in (P3); it does not require a mixing, cluster, martingale, or finite-dependence model for the regression errors.

The arbitrary-within-cluster condition in Corollary~\ref{cor:controlscoreclasses}(d) is intentionally conservative: it forces $\sum_g m_{g,n}^2=O(n)$, where $m_{g,n}=|\mathcal C_{g,n}|$. Growing clusters are allowed under weaker within-cluster dependence. The following condition is convenient.

\begin{corollary}[Growing clusters with summable within-cluster covariance]\label{cor:growingclusters}
Suppose the centered control scores can be partitioned into clusters $\mathcal C_{g,n}$ that are independent across $g$. Let $\bar Z_{i,n}=Z_{i,n}-E Z_{i,n}$ and suppose
\[
\sup_{n}\sup_i \sum_{j:\,j\in \mathcal C_{g(i),n}}
\left\|\Cov(\bar Z_{i,n},\bar Z_{j,n})\right\|_{\mathrm{op}}\le C,
\]
where $g(i)$ denotes the cluster containing observation $i$. If
\[
\left\|n^{-1/2}\sum_i E Z_{i,n}\right\|=O(1),
\]
then
\[
n^{-1/2}W'\varepsilon=O_p(1).
\]
No bound on $\max_g m_{g,n}$ is required beyond what is compatible with the uniform covariance-row-sum condition.
\end{corollary}

Corollary~\ref{cor:growingclusters} allows $m_{g,n}\to\infty$. Its content is not that arbitrary dependence inside a growing cluster is harmless. Rather, aggregate covariance per observation must remain uniformly bounded. Under unrestricted within-cluster dependence the stronger size condition in Corollary~\ref{cor:controlscoreclasses}(d) remains necessary for the simple second-moment argument used here.

\subsection{Proof of Proposition~\ref{prop:pairsprimitive}: primitive sufficient conditions}
We verify Assumption~\ref{ass:pairs}(i)--(iv) in turn. Throughout, $k=\dim(X_i)$ is fixed. Assumption~\ref{ass:stable} supplies the oracle diagonal-score nondegeneracy required below.

\paragraph{Preliminary step: root-$n$ rate of the full OLS vector.}
We first show that (P3) implies the root-$n$ rate needed for the feasible residual and leverage arguments. Since the first component of $W_i$ is a nonzero constant, (P3) implies
\[
\frac1{\sqrt n}\sum_i\varepsilon_i=O_p(1).
\tag{A.P0a}
\]
Moreover, write $(D_i-\mu_{D,n})\varepsilon_i=A_i'e_i=(A_i-\mu_A)'e_i+\mu_A'e_i$. Independence of the iid assignment sequence from the latent array makes all cross-observation covariance terms involving $A_i-\mu_A$ vanish, and the within-observation cross term with $\mu_A'e_i$ is zero by $E(A_i-\mu_A\mid e_i)=0$. Hence the exact score-variance decomposition is
\[
\Var\!\left\{\frac1{\sqrt n}\sum_i(D_i-\mu_{D,n})\varepsilon_i\right\}
=\sigma_{e,1,n}^2+\sigma_{e,2,n}^2
=\sigma_{D\varepsilon,n}^2.
\]
Assumption~\ref{ass:stable} bounds this variance uniformly, while the score is centered under the CHLS random-assignment conditions. Chebyshev's inequality therefore yields
\[
\frac1{\sqrt n}\sum_i(D_i-\mu_{D,n})\varepsilon_i=O_p(1).
\tag{A.P0b}
\]
Therefore
\[
\frac1{\sqrt n}D'\varepsilon
=
\frac1{\sqrt n}\sum_i(D_i-\mu_{D,n})\varepsilon_i
+\mu_{D,n}\frac1{\sqrt n}\sum_i\varepsilon_i
=O_p(1).
\tag{A.P0c}
\]
Combining (P3) with (A.P0c) yields
\[
\frac1{\sqrt n}X'\varepsilon=O_p(1).
\tag{A.P0d}
\]
Since
\[
\sqrt n(\widehat\theta-\theta)
=(n^{-1}X'X)^{-1}\frac1{\sqrt n}X'\varepsilon
\]
and (P1) makes $(n^{-1}X'X)^{-1}=O_p(1)$, we conclude
\[
\sqrt n(\widehat\theta-\theta)=O_p(1).
\tag{A.P0e}
\]
Thus the full-vector root-$n$ rate is a consequence of the primitive control-score rate rather than a separate assumption. Notice that bare CHLS orthogonality $n^{-1}W'\varepsilon=o_p(1)$ is not enough for this conclusion; the $n^{-1/2}$ rate in (P3) is a genuine strengthening.

\paragraph{Step 1: maximum leverage.}
Since $X'X=n\widehat Q_X$,
\[
h_{ii}=X_i'(X'X)^{-1}X_i
=\frac1nX_i'\widehat Q_X^{-1}X_i
\le \frac{\|X_i\|^2}{n\lambda_{\min}(\widehat Q_X)}.
\]
By (P1), $\lambda_{\min}(\widehat Q_X)$ is bounded away from zero with probability tending to one. Moreover, for every $\eta>0$,
\[
\Pr\!\left(\max_i\|X_i\|^2>\eta n\right)
\le \sum_i\frac{\E\|X_i\|^{4+\delta}}{(\eta n)^{2+\delta/2}}
\le C_\eta n^{-1-\delta/2}\to0.
\]
Hence $\max_i\|X_i\|^2/n=o_p(1)$ and therefore
\[
\max_i h_{ii}=o_p(1).
\]
Together with (P1), this proves the nonsingularity and leverage parts of Assumption~\ref{ass:pairs}(i). Cross-sectional independence of the controls is not used.

\paragraph{Step 2: the sample FWL residual and empirical influence Lindeberg condition.}
Let $\pi_0$ denote the population projection coefficient of $D$ on $W$. Random assignment and the intercept in $W$ imply $W_i'\pi_0=\mu_{D,n}$, so the population residual is
\[
v_i=D_i-\mu_{D,n}.
\]
The sample projection coefficient satisfies
\[
\widehat\pi-\pi_0
=(W'W)^{-1}W'(D-\mu_{D,n}\mathbf 1).
\]
Conditional on $W=(W_1,\ldots,W_n)$,
\[
\Var\!\left(\frac1{\sqrt n}\sum_i W_i(D_i-\mu_{D,n})\middle|W\right)
=\sigma_{D,n}^2\frac1nW'W.
\]
By (P1), $n^{-1}W'W=O_p(1)$ and its minimum eigenvalue is bounded away from zero with probability tending to one (as a principal submatrix of $n^{-1}X'X$). Conditional Chebyshev therefore gives
\[
\sqrt n(\widehat\pi-\pi_0)=O_p(1).
\tag{A.P1}
\]
Writing $a_i=\widehat v_i-v_i=-W_i'(\widehat\pi-\pi_0)$, (A.P1) and (P1) imply
\[
\frac1n\sum_i a_i^4
\le \|\widehat\pi-\pi_0\|^4\frac1n\sum_i\|W_i\|^4
=o_p(1),
\tag{A.P2}
\]
and, since $\max_i\|W_i\|/\sqrt n=o_p(1)$,
\[
\max_i|a_i|=o_p(1).
\tag{A.P3}
\]
Thus the FWL replacement conditions used in the high-level theorem are consequences of random assignment and the primitive design assumptions rather than separate restrictions. They also deliver the residualized-treatment scale required in Assumption~\ref{ass:pairs}(i). Indeed,
\[
\widehat Q_D=\frac1n\sum_i(v_i+a_i)^2.
\]
Lemma~\ref{lem:design} gives $n^{-1}\sum_i v_i^2/\sigma_{D,n}^2\to_p1$. Moreover, (A.P1) and $n^{-1}W'W=O_p(1)$ imply $n^{-1}\sum_i a_i^2=O_p(n^{-1})$, and Cauchy--Schwarz makes the cross term negligible relative to the stable treatment scale. Hence
\[
\frac{\widehat Q_D}{\sigma_{D,n}^2}\to_p1.
\tag{A.P3b}
\]
This completes Assumption~\ref{ass:pairs}(i).

Let $\phi_i=\sigma_D^{-2}v_i\varepsilon_i$ denote the oracle influence score. For $r=2+\delta/2$, H\"older's inequality and (P1)--(P2) give
\[
\sup_{i,n}\E|v_i\varepsilon_i|^r<\infty,
\]
because $2r=4+\delta$. Hence
\[
\frac{\max_i\phi_i^2}{n}=o_p(1).
\tag{A.P4}
\]
Put $b_i=\widehat\varepsilon_i-\varepsilon_i=-X_i'(\widehat\theta-\theta)$. By (A.P0e) and Step 1,
\[
\max_i|b_i|
\le \|\widehat\theta-\theta\|\max_i\|X_i\|
=o_p(1),
\tag{A.P5}
\]
and
\[
\frac1n\sum_i b_i^4
\le \|\widehat\theta-\theta\|^4\frac1n\sum_i\|X_i\|^4
=o_p(1).
\tag{A.P6}
\]
Using $\widehat v_i\widehat\varepsilon_i-v_i\varepsilon_i=a_i\varepsilon_i+v_ib_i+a_ib_i$, (A.P2)--(A.P6), H\"older's inequality, and the $4+\delta$ moments imply
\[
\frac1n\sum_i(\widehat v_i\widehat\varepsilon_i-v_i\varepsilon_i)^2=o_p(1)
\tag{A.P7}
\]
and
\[
\frac{\max_i|\widehat v_i\widehat\varepsilon_i-v_i\varepsilon_i|}{\sqrt n}=o_p(1).
\tag{A.P8}
\]
Since $\widehat Q_D\to_p\sigma_{D,n}^2$, the same statements hold for the feasible and oracle influence scores. By Assumption~\ref{ass:stable}, Lemma~\ref{lem:diag}, and (A.P7),
\[
\frac1n\sum_i\widehat\phi_i^2
\]
is bounded away from zero and infinity in probability, while (A.P4) and (A.P8) give $\max_i\widehat\phi_i^2/n=o_p(1)$. Therefore
\[
\frac{\max_i\widehat\phi_i^2}{\sum_i\widehat\phi_i^2}=o_p(1),
\]
which proves Assumption~\ref{ass:pairs}(ii).

\paragraph{Step 3: feasible HC0 score variance.}
Using $a_i=\widehat v_i-v_i$ and $b_i=\widehat\varepsilon_i-\varepsilon_i$,
\[
\widehat v_i^2\widehat\varepsilon_i^2-v_i^2\varepsilon_i^2
=
2v_i a_i\varepsilon_i^2+a_i^2\varepsilon_i^2
+
2v_i^2\varepsilon_i b_i+v_i^2b_i^2
+
4v_i a_i\varepsilon_i b_i
+
2v_i a_i b_i^2
+
2a_i^2\varepsilon_i b_i
+
a_i^2b_i^2.
\]
We verify that every sample average on the right is $o_p(1)$.

By H\"older,
\[
\left|\frac1n\sum_i v_i a_i\varepsilon_i^2\right|
\le
\left(\frac1n\sum_i a_i^4\right)^{1/4}
\left(
\frac1n\sum_i |v_i|^{4/3}|\varepsilon_i|^{8/3}
\right)^{3/4}
=o_p(1),
\]
and
\[
\frac1n\sum_i a_i^2\varepsilon_i^2
\le
\left(\frac1n\sum_i a_i^4\right)^{1/2}
\left(\frac1n\sum_i\varepsilon_i^4\right)^{1/2}
=o_p(1).
\]
Similarly,
\[
\left|\frac1n\sum_i v_i^2\varepsilon_i b_i\right|
\le
\left(\frac1n\sum_i b_i^4\right)^{1/4}
\left(
\frac1n\sum_i |v_i|^{8/3}|\varepsilon_i|^{4/3}
\right)^{3/4}
=o_p(1),
\]
and
\[
\frac1n\sum_i v_i^2b_i^2
\le
\left(\frac1n\sum_i b_i^4\right)^{1/2}
\left(\frac1n\sum_i v_i^4\right)^{1/2}
=o_p(1).
\]

For the four mixed remainder terms,
\[
\frac1n\sum_i |v_i a_i\varepsilon_i b_i|
\le
\left(\frac1n\sum_i a_i^4\right)^{1/4}
\left(\frac1n\sum_i b_i^4\right)^{1/4}
\left(\frac1n\sum_i v_i^2\varepsilon_i^2\right)^{1/2}
=o_p(1),
\]
\[
\frac1n\sum_i |v_i a_i|b_i^2
\le
\left(\frac1n\sum_i a_i^4\right)^{1/4}
\left(\frac1n\sum_i b_i^4\right)^{1/2}
\left(\frac1n\sum_i v_i^4\right)^{1/4}
=o_p(1),
\]
\[
\frac1n\sum_i a_i^2|\varepsilon_i b_i|
\le
\left(\frac1n\sum_i a_i^4\right)^{1/2}
\left(\frac1n\sum_i\varepsilon_i^2b_i^2\right)^{1/2}
=o_p(1),
\]
where
\[
\frac1n\sum_i\varepsilon_i^2b_i^2
\le
\left(\frac1n\sum_i\varepsilon_i^4\right)^{1/2}
\left(\frac1n\sum_i b_i^4\right)^{1/2}
=o_p(1),
\]
and finally
\[
\frac1n\sum_i a_i^2b_i^2
\le
\left(\frac1n\sum_i a_i^4\right)^{1/2}
\left(\frac1n\sum_i b_i^4\right)^{1/2}
=o_p(1).
\]
The empirical moments multiplying the vanishing $a_i$ and $b_i$ factors are $O_p(1)$ by the uniform $(4+\delta)$ assumptions and Markov's inequality. Hence
\[
\frac1n\sum_i\widehat v_i^2\widehat\varepsilon_i^2
-
\frac1n\sum_i v_i^2\varepsilon_i^2=o_p(1).
\]
Since $\widehat Q_D\to_p\sigma_{D,n}^2$ and Assumption~\ref{ass:stable} bounds $\sigma_{D,n}^2$ away from zero,
\[
\widehat Q_D^{-2}\frac1n\sum_i\widehat v_i^2\widehat\varepsilon_i^2
-
\sigma_{D,n}^{-4}\frac1n\sum_i v_i^2\varepsilon_i^2
=o_p(1).
\]
Lemma~\ref{lem:diag} and the lower bound on $\sigma_{D,n}^2$ identify the second term with $V_n^\#+o_p(1)$. This proves Assumption~\ref{ass:pairs}(iii).

\paragraph{Step 4: design-normalized full score without a conditional error-moment bound.}
On the event $\lambda_{\min}(\widehat Q_X)\ge c/2$,
\[
h_{ii}
\le \frac{C}{n}\|X_i\|^2.
\]
Therefore
\[
\sum_i h_{ii}\varepsilon_i^2
\le
C\frac1n\sum_i\|X_i\|^2\varepsilon_i^2.
\tag{A.P9}
\]
No independence between $X_i$ and $\varepsilon_i$ is needed. By Cauchy--Schwarz and (P1)--(P2),
\[
\sup_{i,n}\E(\|X_i\|^2\varepsilon_i^2)
\le
\sup_{i,n}\{\E\|X_i\|^4\}^{1/2}
\sup_{i,n}\{\E|\varepsilon_i|^4\}^{1/2}
<\infty.
\]
Hence Markov's inequality applied to the nonnegative average on the right of (A.P9) gives
\[
\sum_i h_{ii}\varepsilon_i^2=O_p(1).
\tag{A.P10}
\]
This weighted argument is valid under arbitrary cross-sectional dependence.

It remains to replace $\varepsilon_i$ by $\widehat\varepsilon_i$. From $b_i=-X_i'(\widehat\theta-\theta)$,
\[
\sum_i h_{ii}b_i^2
\le
C\|\widehat\theta-\theta\|^2\frac1n\sum_i\|X_i\|^4
=o_p(1)
\]
by (P1) and (A.P0e). Cauchy--Schwarz together with (A.P10) then controls the cross term:
\[
\left|\sum_i h_{ii}\varepsilon_i b_i\right|
\le
\left(\sum_i h_{ii}\varepsilon_i^2\right)^{1/2}
\left(\sum_i h_{ii}b_i^2\right)^{1/2}
=o_p(1).
\]
Consequently,
\[
\sum_i h_{ii}\widehat\varepsilon_i^2=O_p(1),
\]
which proves Assumption~\ref{ass:pairs}(iv) and completes the proof. \qed

\paragraph{Scope of the primitive sufficient conditions.}
The primitive theorem uses fourth-plus-$\delta$ moments and the root-$n$ control-score condition $n^{-1/2}W'\varepsilon=O_p(1)$, so its sufficient conditions are stronger than the baseline CHLS conditions. The full-vector rate $\sqrt n(\widehat\theta-\theta)=O_p(1)$ is derived from the stated conditions rather than imposed separately. No conditional bound of the form $E(\varepsilon_i^2\mid X_1,\ldots,X_n)\le C$ is needed: the weighted leverage term is controlled by unconditional moments through (A.P9)--(A.P10), without any cross-sectional independence assumption. Likewise, the root-$n$ rate and the fourth-power/maximal replacement for the sample FWL residual follow from independent random assignment and the displayed design conditions.

\subsection{Proof of Proposition~\ref{prop:controlscore} and Corollary~\ref{cor:controlscoreclasses}}

Let $Z_{i,n}=W_{i,n}\varepsilon_{i,n}$, $\bar Z_{i,n}=Z_{i,n}-E Z_{i,n}$, and write
\[
S_n=\sum_{i=1}^n Z_{i,n}
=\sum_{i=1}^n E Z_{i,n}+\sum_{i=1}^n\bar Z_{i,n}.
\]
The deterministic first term is $O(\sqrt n)$ by assumption. For the centered term,
\[
E\left\|\frac1{\sqrt n}\sum_i\bar Z_{i,n}\right\|^2\le C.
\]
Hence Markov's inequality gives
\[
\frac1{\sqrt n}\sum_i\bar Z_{i,n}=O_p(1).
\]
Adding the deterministic mean term proves $n^{-1/2}W'\varepsilon=O_p(1)$. \qed

We verify the four primitive dependence classes for the centered array $\bar Z_{i,n}$ in turn.

\paragraph{$m$-dependence.}
If $\bar Z_{i,n}$ is $m$-dependent, with fixed $m$ and $\sup_{i,n}E\|\bar Z_{i,n}\|^2\le C$, then
\[
E\left\|\sum_i\bar Z_{i,n}\right\|^2
=\sum_i E\|\bar Z_{i,n}\|^2
+2\sum_{i<j}E(\bar Z_{i,n}'\bar Z_{j,n}).
\]
All terms with $|i-j|>m$ vanish. For the remaining terms, Cauchy--Schwarz gives
\[
|E(\bar Z_{i,n}'\bar Z_{j,n})|
\le \{E\|\bar Z_{i,n}\|^2 E\|\bar Z_{j,n}\|^2\}^{1/2}\le C.
\]
There are at most $mn$ nonzero off-diagonal pairs, so the second moment is $O(n)$.

\paragraph{Martingale differences.}
Suppose $\bar Z_{i,n}$ is adapted to $\mathcal F_{i,n}$ and $E(\bar Z_{i,n}\mid\mathcal F_{i-1,n})=0$. If $j<i$, then $\bar Z_{j,n}$ is $\mathcal F_{i-1,n}$-measurable, and therefore
\[
E(\bar Z_{i,n}'\bar Z_{j,n})=0.
\]
Thus
\[
E\left\|\sum_i\bar Z_{i,n}\right\|^2
=\sum_iE\|\bar Z_{i,n}\|^2=O(n).
\]

\paragraph{Strong mixing.}
Let $\bar Z_{i,n}^{(r)}$ denote coordinate $r$ of $\bar Z_{i,n}$. Under the stated $(2+\eta)$ moment bound and uniform strong-mixing condition, the standard mixing covariance inequality yields
\[
|\operatorname{Cov}(\bar Z_{i,n}^{(r)},\bar Z_{i+h,n}^{(s)})|
\le C\alpha_n(h)^{\eta/(2+\eta)}
\]
uniformly in $n,i,h,r,s$. Since $k_W$ is fixed and
\[
\sup_n\sum_{h\ge1}\alpha_n(h)^{\eta/(2+\eta)}<\infty,
\]
we obtain
\[
\frac1nE\left\|\sum_i\bar Z_{i,n}\right\|^2=O(1).
\]
No stationarity is required beyond the uniform moment and mixing bounds used in this covariance estimate.

\paragraph{Independent clusters.}
Let $\bar S_{g,n}=\sum_{i\in\mathcal C_{g,n}}\bar Z_{i,n}$. Independence across clusters and centering imply
\[
E\left\|\sum_i\bar Z_{i,n}\right\|^2=\sum_gE\|\bar S_{g,n}\|^2.
\]
If $m_{g,n}=|\mathcal C_{g,n}|$, then
\[
\|\bar S_{g,n}\|^2
\le m_{g,n}\sum_{i\in\mathcal C_{g,n}}\|\bar Z_{i,n}\|^2.
\]
Hence
\[
E\left\|\sum_i\bar Z_{i,n}\right\|^2
\le C\sum_gm_{g,n}^2=O(n).
\]
This permits arbitrary dependence within each cluster. \qed

\paragraph{Growing clusters with summable within-cluster covariance.}
For cluster $g$, write $\bar S_{g,n}=\sum_{i\in\mathcal C_{g,n}}\bar Z_{i,n}$. Independence across clusters gives
\[
E\left\|\sum_i\bar Z_{i,n}\right\|^2
=\sum_g E\|\bar S_{g,n}\|^2.
\]
For each cluster,
\[
E\|\bar S_{g,n}\|^2
=\sum_{i,j\in\mathcal C_{g,n}}\operatorname{tr}\{\Cov(\bar Z_{i,n},\bar Z_{j,n})\}.
\]
Since $k_W$ is fixed,
\[
\left|\operatorname{tr}\{\Cov(\bar Z_{i,n},\bar Z_{j,n})\}\right|
\le k_W\left\|\Cov(\bar Z_{i,n},\bar Z_{j,n})\right\|_{\mathrm{op}}.
\]
The uniform covariance-row-sum condition therefore yields
\[
E\|\bar S_{g,n}\|^2
\le C\sum_{i\in\mathcal C_{g,n}}1
=Cm_{g,n}.
\]
Summing over clusters,
\[
E\left\|\sum_i\bar Z_{i,n}\right\|^2
\le C\sum_gm_{g,n}=Cn.
\]
The covariance-growth criterion in Proposition~\ref{prop:controlscore} gives $n^{-1/2}W'\varepsilon=O_p(1)$. This proof permits $\max_gm_{g,n}\to\infty$; what is ruled out is unrestricted accumulation of within-cluster covariance. \qed

\paragraph{Scope of the dependence examples.}
The four cases above are only sufficient devices for verifying (P3); they are not imposed by the pairs-bootstrap theorem itself. The deterministic mean requirement $\|n^{-1/2}\sum_i E(W_i\varepsilon_i)\|=O(1)$ prevents a first-order deterministic drift in the nuisance score, while exact observation-by-observation orthogonality is not required. The cluster condition $\sum_gm_{g,n}^2=O(n)$ rules out a small number of very large clusters under arbitrary within-cluster dependence; allowing growing clusters requires additional within-cluster covariance control. Likewise, the strong-mixing result depends on the stated uniform $(2+\eta)$ moment and mixing-rate summability conditions and does not follow from mixing alone.

\section{Proof of the covariance-quotient minimality theorem}\label{app:minimality}

\subsection{Proof of Theorem~\ref{thm:minimality}}

Theorem~\ref{thm:filter} gives
\[
V_n^\#-V_n
=
-\sigma_{D,n}^{-4}\mu_{A,n}'H_n\mu_{A,n},
\]
hence
\[
c_n:=V_n-V_n^\#
=
\sigma_{D,n}^{-4}\mathcal L_n(H_n).
\]

For part (i), the comparison is only over model-compatible off-diagonal covariance operators with the assignment law, diagonal target $V_n^\#$, $\mu_{A,n}$, and $\sigma_{D,n}^2$ held fixed. If $H_n-\widetilde H_n\in\mathcal N_n$, then
\[
\mathcal L_n(H_n)=\mathcal L_n(\widetilde H_n),
\]
so the two off-diagonal operators make the same contribution $c_n$ to the first-order variance. No claim is made that every matrix in $\mathbb S^L$ is itself generated by a stochastic process satisfying the maintained model conditions.

For part (ii), if $\mu_{A,n}\neq0$, then
\[
\mathcal L_n(\mu_{A,n}\mu_{A,n}')=\|\mu_{A,n}\|^4>0.
\]
Thus $\mathcal L_n$ has rank one, and rank--nullity implies that $\mathcal N_n$ has codimension one and the ambient quotient $\mathbb S^L/\mathcal N_n$ is one-dimensional. The model-compatible covariance class inherits the induced equivalence relation by restriction to the corresponding cosets. If $\mu_{A,n}=0$, then $\mathcal L_n$ is identically zero and $\mathcal N_n=\mathbb S^L$.

For part (iii),
\[
\frac{\widehat V_n^{\,*}-V_n}{V_n}
=
\frac{\widetilde V_n^\#-V_n^\#}{V_n}
+
\frac{\widehat c_n-c_n}{V_n}.
\]
The first term is $o_p(1)$ by assumption. Hence
\[
\boxed{
\frac{\widehat V_n^{\,*}}{V_n}\pto1
\quad\Longleftrightarrow\quad
\frac{\widehat c_n-c_n}{V_n}\pto0.
}
\]

For sufficiency, on the event $\{\widehat V_n^{\,*}>0\}$, conditional centered Gaussianity gives
\[
P^*\{T_n^*\le x\}
=
\Phi\!\left(\frac{x}{(\widehat V_n^{\,*})^{1/2}}\right).
\]
The complementary event has probability $o(1)$ by assumption and therefore does not affect convergence in probability of the conditional Kolmogorov distance.
Writing $z=x/V_n^{1/2}$,
\[
\sup_x
\left|
\Phi\!\left(\frac{x}{(\widehat V_n^{\,*})^{1/2}}\right)
-
\Phi\!\left(\frac{x}{V_n^{1/2}}\right)
\right|
=
\sup_z
\left|
\Phi\!\left(z\sqrt{\frac{V_n}{\widehat V_n^{\,*}}}\right)-\Phi(z)
\right|.
\]
If $\widehat V_n^{\,*}/V_n\to_p1$, Lemma~\ref{lem:gaussianscale} makes the right-hand side $o_p(1)$. Let
\[
F_n(z)
=
P\!\left\{\frac{\sqrt n(\widehat\beta-\beta_0)}{V_n^{1/2}}\le z\right\}.
\]
The standardized sampling CLT gives $F_n(z)\to\Phi(z)$ at every $z$. Because the limiting cdf $\Phi$ is continuous, P\'olya's theorem upgrades this pointwise cdf convergence to
\[
\sup_z|F_n(z)-\Phi(z)|\to0.
\]
Changing variables $z=x/V_n^{1/2}$ therefore yields
\[
\sup_x\left|
P\{\sqrt n(\widehat\beta-\beta_0)\le x\}
-
\Phi(x/V_n^{1/2})
\right|\to0.
\]
The triangle inequality proves bootstrap validity.

For necessity, suppose the conditional Kolmogorov distance between bootstrap and sampling laws converges to zero. The standardized sampling CLT gives
\[
P\{\sqrt n(\widehat\beta-\beta_0)\le V_n^{1/2}\}\to\Phi(1).
\]
On $\{\widehat V_n^{\,*}>0\}$, evaluating bootstrap validity at the deterministic point $x_n=V_n^{1/2}$ yields
\[
\Phi\!\left(\sqrt{\frac{V_n}{\widehat V_n^{\,*}}}\right)\pto\Phi(1).
\]
The exceptional event has probability $o(1)$. Because $\Phi$ is continuous and strictly increasing and $P(\widehat V_n^{\,*}>0)\to1$,
\[
\frac{\widehat V_n^{\,*}}{V_n}\pto1.
\]
The boxed equivalence then gives
\[
\frac{\widehat c_n-c_n}{V_n}\pto0.
\]

Finally, if
\[
\widehat c_n
=
\sigma_{D,n}^{-4}\mathcal L_n(\widehat H_n),
\]
then
\[
\frac{\widehat c_n-c_n}{V_n}
=
\frac{\sigma_{D,n}^{-4}\mathcal L_n(\widehat H_n-H_n)}{V_n}.
\]
No restriction is needed on errors lying in $\mathcal N_n$. \qed

\begin{remark}[Scope and local-to-zero sequences]
The necessity direction uses conditional centered Gaussianity so that the bootstrap law is indexed by its conditional variance alone. No assumption that $V_n$ converges to a positive constant is needed: the theorem is invariant to deterministic rescaling and remains valid when $V_n\to0$ or $V_n\to\infty$, provided the standardized sampling CLT holds and the bootstrap variance is positive with probability approaching one.

If $\mu_{A,n}\to0$, the algebraic quotient remains one-dimensional whenever $\mu_{A,n}\neq0$, but the correction can be locally negligible:
\[
\frac{\sigma_{D,n}^{-4}\mu_{A,n}'H_n\mu_{A,n}}{V_n}\to0.
\]
A sufficient condition is
\[
\frac{\sigma_{D,n}^{-4}\|\mu_{A,n}\|^2\|H_n\|_{\mathrm{op}}}{V_n}\to0.
\]
Thus $\|\mu_{A,n}\|\to0$ by itself is not enough if $\|H_n\|_{\mathrm{op}}/V_n$ diverges. For non-Gaussian bootstrap laws, projected variance matching remains an important necessary scale condition under suitable moment control, but variance matching alone need not characterize the entire conditional law.
\end{remark}

\subsection{Proof of Proposition~\ref{prop:rbound}}
The covariance-filter identity gives
\[
1-R_{A,n}=\frac{V_n^{\#}}{V_n}=\frac{s_{\#,n}^2}{\sigma_{D\varepsilon,n}^2}.
\]
Because $s_{\#,n}^2=n^{-1}\sum_iE[\psi_{i,n}^2]$, the equality $s_{\#,n}^2=0$ would imply $\psi_{i,n}=0$ almost surely for every $i$ and hence $\sigma_{D\varepsilon,n}^2=0$, contradicting $V_n>0$. Thus $s_{\#,n}^2>0$ and $R_{A,n}<1$ for every $n$ satisfying the stated conditions. Under Assumption~\ref{ass:stable}, $s_{\#,n}^2\ge c$ and $\sigma_{D\varepsilon,n}^2\le C$, so the stronger uniform bound $1-R_{A,n}\ge c/C$ follows. \qed

\subsection{Proof of Lemma~\ref{lem:sharpquotient}}
Corollary~\ref{cor:te} gives $\mathcal L_n(H_n)=q^2\Delta_{\tau,n}$. For $\tau_i=\eta_i+\theta\eta_{i-1}$, $\gamma_\tau(1)=\theta$ and $\gamma_\tau(h)=0$ for $h\ge2$. Hence
\[
\Delta_{\tau,n}=2(1-n^{-1})\theta,
\]
which proves the display. Every fixed $\theta$ yields a stationary finite-second-moment process independent of assignment, and variation in $\theta$ produces a continuum through zero and both signs. \qed

\subsection{Proof of Theorem~\ref{thm:marginalimpossibility}}
For either sign, $\tau_i^{\pm}$ is Gaussian with mean zero and variance
\[
\Var(\tau_i^{\pm})=1+\theta^2.
\]
Hence the one-observation treatment-effect law is the same under the two designs. Because $u_i$ and $D_i$ have the same laws and are independent of the treatment-effect process under both designs,
\[
\mathcal L(D_i,Y_i^+)=\mathcal L(D_i,Y_i^-).
\]
This proves part (i).

The population regression has intercept zero and treatment coefficient zero. Its error is
\[
\varepsilon_i^{\pm}=u_i+D_i\tau_i^{\pm},
\]
and the leading score is
\[
\psi_i^{\pm}=(D_i-p)(u_i+D_i\tau_i^{\pm}).
\]
Put $q=p(1-p)$. Since $D_i$ is independent of $(u_i,\tau_i^{\pm})$,
\[
\begin{aligned}
E[(\psi_i^{\pm})^2]
&=(1-p)p^2\sigma_u^2
+p(1-p)^2\{\sigma_u^2+(1+\theta^2)\}\\
&=q\sigma_u^2+q(1-p)(1+\theta^2).
\end{aligned}
\]
Because $\sigma_D^2=q$, the diagonal target is therefore
\[
V^{\#}
=q^{-2}E[(\psi_i^{\pm})^2]
=
\frac{\sigma_u^2}{p(1-p)}+
\frac{1+\theta^2}{p},
\]
which is the same under both designs. This proves part (ii).

The only nonzero off-diagonal autocovariance of the treatment-effect process is at lag one:
\[
\Cov(\tau_i^{\pm},\tau_{i-1}^{\pm})=\pm\theta,
\qquad
\Cov(\tau_i^{\pm},\tau_{i-h}^{\pm})=0\quad (|h|\ge2).
\]
Hence the stationary formula in Section~\ref{sec:binary} gives
\[
\Delta_{\tau,n}^{\pm}
=
2\left(1-\frac1n\right)(\pm\theta).
\]
Corollary~\ref{cor:te} yields
\[
V_n^{\pm}=V^{\#}+\Delta_{\tau,n}^{\pm},
\]
which proves part (iii). Notice that $V_n^-$ remains strictly positive because
\[
V^{\#}-2\theta
\ge
\frac{1+\theta^2}{p}-2\theta
>
(1-\theta)^2\ge0,
\]
where strictness follows from $p<1$.

The score process $\{\psi_i^{\pm}\}$ is stationary and one-dependent: observations separated by more than one index depend on disjoint sets of the iid sequences $(D_i,u_i,\eta_i)$. It has moments of every order. The standard CLT for stationary $m$-dependent sequences therefore applies to the leading score. The transfer to OLS is direct in this intercept-only construction. Writing
\[
 \widehat Q_{D,n}=\frac1n\sum_{i=1}^n(D_i-\bar D)^2,
\]
the Bernoulli law of large numbers gives $\widehat Q_{D,n}\to_p q$. Moreover,
\[
 \frac1{\sqrt n}\sum_{i=1}^n(D_i-\bar D)Y_i^{\pm}
 =
 \frac1{\sqrt n}\sum_{i=1}^n(D_i-p)Y_i^{\pm}
 -\sqrt n(\bar D-p)\,\bar Y_n^{\pm}.
\]
Here $\sqrt n(\bar D-p)=O_p(1)$ and $\bar Y_n^{\pm}=O_p(n^{-1/2})$ because $\{Y_i^{\pm}\}$ is stationary one-dependent with finite second moment and mean zero. Hence the last product is $o_p(1)$, and
\[
 \sqrt n\,\widehat\beta_n^{\pm}
 =q^{-1}\frac1{\sqrt n}\sum_{i=1}^n\psi_i^{\pm}+o_p(1).
\]
Thus the same $m$-dependent CLT yields the stated sampling CLT for the OLS coefficient without any additional high-level linearization assumption.

Now suppose $\widehat V_n^m$ is asymptotically marginal. By part (i), the one-observation laws under the $+$ and $-$ designs coincide. Hence the deterministic sequence
\[
a_n:=\nu_n\!\left(\mathcal L(D_i,Y_i)\right)
\]
is the same under both designs, and $\widehat V_n^m-a_n\to_p0$ under each. Because
\[
V_n^{\pm}\to V^{\#}\pm2\theta
\]
and $V^{\#}-2\theta>(1-\theta)^2>0$, both variance sequences are bounded away from zero. Thus $\widehat V_n^m-a_n\to_p0$ also implies $(\widehat V_n^m-a_n)/V_n^{\pm}\to_p0$ under the corresponding design. If relative consistency held under both, the deterministic sequence would therefore satisfy
\[
\frac{a_n}{V_n^+}\to1
\qquad\text{and}\qquad
\frac{a_n}{V_n^-}\to1.
\]
These two relations would imply $V_n^+/V_n^-\to1$. But part (iii) gives
\[
\frac{V_n^+}{V_n^-}
\to
\frac{V^{\#}+2\theta}{V^{\#}-2\theta}
\neq1,
\]
a contradiction. Thus no asymptotically marginal variance estimator is relatively consistent under both designs. This argument allows the marginal functional itself to vary with $n$.

For a conditionally centered-Gaussian bootstrap with conditional variance $\widehat V_n^*$, the OLS sampling CLT established above and Lemma~\ref{lem:gaussianscale} imply directly that first-order validity forces $\widehat V_n^*/V_n^{\pm}\to_p1$ under the corresponding design. Therefore such a bootstrap cannot be valid under both designs when its conditional variance is asymptotically marginal. For a more general bootstrap satisfying the conditional squared-uniform-integrability condition of Theorem~\ref{thm:ui}, first-order validity forces its conditional second moment to converge to the corresponding $V_n^{\pm}$. The same common-marginal contradiction applies if that second moment is asymptotically marginal. \qed

\section{Weak validity, uniform integrability, and filtered second moments}\label{app:ui}

\subsection{Proof of Theorem~\ref{thm:ui}}
Let $F_n^*(x)=P^*(Z_n^*\le x)$. By assumption,
\[
\|F_n^*-\Phi\|_\infty\pto0.
\]
Fix $M>0$ and define the bounded continuous truncation
\[
g_M(z)=z^2\wedge M^2.
\]
Conditional weak convergence in probability implies
\[
E^*[g_M(Z_n^*)]\pto E[g_M(Z)],\qquad Z\sim N(0,1).
\]
Indeed, from any subsequence one can extract a further subsequence along which the conditional Kolmogorov distance converges almost surely to zero; ordinary weak convergence of the resulting conditional laws then yields convergence of bounded continuous expectations almost surely, which proves convergence in probability along the original sequence.

Now
\[
\begin{aligned}
\left|E^*[(Z_n^*)^2]-1\right|
&\le \left|E^*[g_M(Z_n^*)]-E[g_M(Z)]\right|\\
&\quad +E^*[(Z_n^*)^2 1\{|Z_n^*|>M\}] +E[Z^2 1\{|Z|>M\}].
\end{aligned}
\]
For fixed $M$ the first term is $o_p(1)$; the third vanishes as $M\to\infty$; and conditional uniform integrability makes the second negligible in probability uniformly for large $n$. Hence
\[
E^*[(Z_n^*)^2]\pto1.
\]
Squared uniform integrability implies first-moment uniform integrability by Cauchy--Schwarz, so the same truncation argument applied to $z$ gives
\[
E^*[Z_n^*]\pto0.
\]
Therefore
\[
\frac{\Var^*(T_n^*)}{V_n}=E^*[(Z_n^*)^2]-\{E^*[Z_n^*]\}^2\pto1.
\]

Under the second-moment decomposition,
\[
\frac{E^*[(T_n^*)^2]-V_n}{V_n}
=\frac{\widetilde V_n^\#-V_n^\#}{V_n}+\frac{\widehat c_n-c_n}{V_n}.
\]
The left side tends to zero because $E^*[(T_n^*)^2]/V_n\to_p1$, and the first term on the right is $o_p(1)$. Thus
\[
\frac{\widehat c_n-c_n}{V_n}\pto0.
\]
Theorem~\ref{thm:minimality} gives $c_n=\sigma_{D,n}^{-4}\mathcal L_n(H_n)$. \qed

\paragraph{Contrapositive (the Hahn--Liao channel).}
Under the maintained conditions other than uniform integrability, if weak validity holds but $(\widehat c_n-c_n)/V_n\not\to_p0$, then conditional squared uniform integrability cannot hold. Otherwise Theorem~\ref{thm:ui} would imply $(\widehat c_n-c_n)/V_n\to_p0$, a contradiction.

\begin{remark}[Why no converse is claimed]
For a non-Gaussian bootstrap, $E^*[(T_n^*)^2]/V_n\to_p1$ does not determine the conditional distribution. A bootstrap may match the filtered second moment and still have incorrect skewness, tail behavior, or other first-order distributional features. The Gaussian theorem is an equivalence because a centered Gaussian law is indexed by its variance; Theorem~\ref{thm:ui} is only a necessity result because general laws are not.
\end{remark}

\begin{proposition}[Primitive moment routes to conditional uniform integrability]\label{prop:ui-moments}
Let $Z_n^*=T_n^*/V_n^{1/2}$ and suppose $V_n>0$.
\begin{enumerate}[label=(\roman*)]
\item If $E^*|Z_n^*|^{2+\eta}=O_p(1)$ for some $\eta>0$, then $\{(Z_n^*)^2\}$ is conditionally uniformly integrable in probability.
\item If $T_n^*$ is conditionally centered Gaussian with variance $\widehat V_n^*$ and $\widehat V_n^*/V_n=O_p(1)$, then (i) holds for every fixed $\eta>0$.
\item Under the conditions of Theorem~\ref{thm:pairs} and Assumption~\ref{ass:pairsmoment}, the regularized pairs estimator satisfies
\[
E^*\!\left[\left|\sqrt n(\widetilde\beta^*-\widehat\beta)\right|^4\right]=O_p(1).
\]
\item Under the conditions of Theorem~\ref{thm:binarypairs}, conditional on the nondegenerate-resample event $\mathcal H_n^*$, the treatment-probability and empirical $(4+\delta)$ moment conditions imply
\[
E_H^*\!\left[\left|\sqrt n(\widehat\beta^*-\widehat\beta)\right|^4\right]=O_p(1).
\]
\end{enumerate}
\end{proposition}

\subsection{Proof of Proposition~\ref{prop:ui-moments}}

\paragraph{Part (i).}
For every $M>0$,
\[
(Z_n^*)^2 1\{|Z_n^*|>M\}
\le
M^{-\eta}|Z_n^*|^{2+\eta}.
\]
Hence
\[
E^*[(Z_n^*)^2 1\{|Z_n^*|>M\}]
\le
M^{-\eta}E^*|Z_n^*|^{2+\eta}.
\]
Let $R_n=E^*|Z_n^*|^{2+\eta}=O_p(1)$. For arbitrary $\varepsilon,\rho>0$, choose $C<\infty$ so that
\[
\limsup_n P(R_n>C)<\rho.
\]
Then choose $M$ so large that $C M^{-\eta}<\varepsilon$. It follows that
\[
\limsup_n
P\!\left(
E^*[(Z_n^*)^2 1\{|Z_n^*|>M\}]>\varepsilon
\right)
\le \rho.
\]
Since $\rho$ is arbitrary, the conditional squared laws are uniformly integrable in probability.

\paragraph{Part (ii).}
If $T_n^*\mid\mathcal D_n\sim N(0,\widehat V_n^*)$, then for any fixed $q=2+\eta$,
\[
E^*|Z_n^*|^q
=
E|N(0,1)|^q
\left(\frac{\widehat V_n^*}{V_n}\right)^{q/2}.
\]
Tightness of $\widehat V_n^*/V_n$ therefore implies
\[
E^*|Z_n^*|^{2+\eta}=O_p(1),
\]
and part (i) applies. The Gaussian projected-dependence bootstrap is conditionally centered Gaussian by construction.

\paragraph{Part (iii).}
For the regularized pairs estimator,
\[
\sqrt n(\widetilde\beta^*-\widehat\beta)
=
1\{\mathcal G_n^*\}
a_n'(I+A_n^*)^{-1}S_n^*.
\]
On $\mathcal G_n^*$,
\[
\|(I+A_n^*)^{-1}\|_{\mathrm{op}}\le2.
\]
Thus
\[
\left|\sqrt n(\widetilde\beta^*-\widehat\beta)\right|^4
\le
16\|a_n\|^4\|S_n^*\|^4.
\]
Assumption~\ref{ass:pairs}(i) gives $\|a_n\|^2=\widehat Q_D^{-1}=O_p(1)$ and Assumption~\ref{ass:pairsmoment} gives
\[
E^*\|S_n^*\|^4=O_p(1).
\]
Therefore
\[
E^*\!\left[\left|\sqrt n(\widetilde\beta^*-\widehat\beta)\right|^4\right]
=O_p(1).
\]
Since the stable-scale regime bounds $V_n$ away from zero, the standardized fourth moment is $O_p(1)$ and part (i) applies with $\eta=2$.

\paragraph{Part (iv).}
Work on the original-sample event $\mathcal H_n$ and write $K=N_1^*$. Let
\[
X_{1,j}^*=Y_{1,j}^*-\bar Y_1,
\qquad
X_{0,j}^*=Y_{0,j}^*-\bar Y_0
\]
for iid draws from the empirical treated and control distributions. Conditional on $K=k\in\{1,\ldots,n-1\}$,
\[
\sqrt n(\widehat\beta^*-\widehat\beta)
=
\sqrt n\left\{
 k^{-1}\sum_{j=1}^k X_{1,j}^*
-
(n-k)^{-1}\sum_{j=1}^{n-k}X_{0,j}^*
\right\}.
\]
For centered iid variables, the exact fourth-moment formula gives
\[
E^*\left|k^{-1}\sum_{j=1}^k X_{1,j}^*\right|^4
\le
C\left(
\frac{\widehat m_{4,1}}{k^3}
+
\frac{\widehat s_1^4}{k^2}
\right),
\]
where $\widehat m_{4,1}=n_1^{-1}\sum_{D_i=1}|Y_i-\bar Y_1|^4$, and analogously for the control arm.

Let
\[
\mathcal C_n^*=\{K\in[cn,(1-c)n]\}
\]
for a fixed $c>0$ smaller than $\underline p/4$. Since $\widehat p$ is bounded away from zero and one with probability tending to one, a Chernoff bound gives
\[
P^*((\mathcal C_n^*)^c\mid\mathcal H_n^*)\le Ce^{-c_0n}.
\]
On $\mathcal C_n^*$, the preceding fourth-moment formula and the empirical $(4+\delta)$ moment condition imply
\[
E^*\!\left[
\left|\sqrt n(\widehat\beta^*-\widehat\beta)\right|^4
1\{\mathcal C_n^*\}
\,\middle|\,\mathcal H_n^*
\right]
=O_p(1).
\]
On $(\mathcal C_n^*)^c\cap\mathcal H_n^*$,
\[
|\widehat\beta^*-\widehat\beta|
\le
4\max_{i\le n}|Y_i|,
\]
so
\[
\left|\sqrt n(\widehat\beta^*-\widehat\beta)\right|^4
\le
C n^2\max_i|Y_i|^4.
\]
The empirical $(4+\delta)$ moment condition implies
\[
\max_i|Y_i|^4
\le
\left(\sum_i|Y_i|^{4+\delta}\right)^{4/(4+\delta)}
=O_p\!\left(n^{4/(4+\delta)}\right).
\]
Multiplication by the exponentially small tail probability therefore yields an $o_p(1)$ contribution. Consequently,
\[
E_H^*\!\left[\left|\sqrt n(\widehat\beta^*-\widehat\beta)\right|^4\right]
=O_p(1).
\]
Again, the stable-scale condition bounds $V_n$ away from zero, so part (i) applies with $\eta=2$. \qed

\begin{remark}[Literal pairs OLS with controls]
Part (iii) is not extended to unrestricted ordinary pairs OLS with general controls. Assumption~\ref{ass:pairs} controls the bootstrap Hessian only in probability and therefore does not rule out rare near-singular draws with large inverse-Hessian moments. A conditional $(2+\eta)$ bound for the nonlinear ordinary-pairs statistic would require additional inverse-Hessian tail assumptions. The regularized construction removes exactly this obstruction.
\end{remark}

\section{Projected-dependence bootstrap: proof and implementations}\label{app:repair}

\subsection{Proof of Theorem~\ref{thm:pd}}

Let
\[
a_n=\{\widetilde V_{\mathrm{PD},n}(\beta_0)\}^{1/2},
\qquad b_n=V_n^{1/2}.
\]
By assumption, $a_n/b_n\to_p1$. Conditional asymptotic normality gives
\[
\sup_z\left|
P^*\{T_{\mathrm{PD},n}^*/a_n\le z\}-\Phi(z)
\right|\to_p0,
\]
while the sampling CLT gives
\[
\sup_z\left|
P\{\sqrt n(\widehat\beta-\beta_0)/b_n\le z\}-\Phi(z)
\right|\to0.
\]
Lemma~\ref{lem:gaussianscale} converts $a_n/b_n\to_p1$ into uniform convergence of the corresponding centered Gaussian cdfs; the triangle inequality then gives the asserted Kolmogorov-distance convergence. \qed

\subsection{Proof of Theorem~\ref{thm:fixedb}}
Write
\[
Z_n=\frac{n^{-1/2}\sum_i\psi_{i,n}}{\Lambda_n^{1/2}},
\qquad
J_n^{(b)}=\frac{n^{-1}\psi_n'K_{n,b}\psi_n}{\Lambda_n}.
\]
The maintained joint fixed-bandwidth limit gives
$(Z_n,J_n^{(b)})\dto\{W(1),Q_{k,b}(W)\}$. The treatment-scale consistency and the first negligible-remainder condition imply
\[
\frac{\sqrt n(\widehat\beta-\beta_0)}{\sigma_{D,n}^{-2}\Lambda_n^{1/2}}
=Z_n+o_p(1).
\]
Likewise, by the definition of the feasible projected quadratic form, the feasible-score replacement condition and $\widehat Q_n/\sigma_{D,n}^2\pto1$ give
\[
\frac{\widetilde V_{\mathrm{PD},n}^{(b)}(\beta_0)}{\sigma_{D,n}^{-4}\Lambda_n}
=J_n^{(b)}+o_p(1).
\]
Because $Q_{k,b}(W)>0$ almost surely, the continuous-mapping theorem yields
\[
\frac{\sqrt n(\widehat\beta-\beta_0)}
{\{\widetilde V_{\mathrm{PD},n}^{(b)}(\beta_0)\}^{1/2}}
\dto
\frac{W(1)}{\{Q_{k,b}(W)\}^{1/2}}.
\]
The stated rejection probability follows at any continuity-point critical value $c_{1-\alpha}(b,k)$ satisfying the tail-probability definition in Theorem~\ref{thm:fixedb}. \qed

\begin{remark}[A scale-correct sufficient route to fixed-bandwidth feasible-score replacement]\label{rem:fixedbreplace}
The fixed-bandwidth replacement condition is intentionally stated separately from the small-bandwidth heteroskedasticity-and-autocorrelation-consistent (HAC) conditions. A convenient sufficient route for the Bartlett kernel is process-level rather than operator-norm based. Let
$d_{i,n}=\widetilde\psi_i-\psi_i$. If
\[
\sup_{0\le r\le1}
\left|
\frac{1}{(n\Lambda_n)^{1/2}}
\sum_{i\le nr}d_{i,n}
\right|=o_p(1)
\]
and the standardized oracle partial-sum process
\[
\left\{
\frac{1}{(n\Lambda_n)^{1/2}}
\sum_{i\le nr}\psi_{i,n}:0\le r\le1
\right\}
\]
is tight, then the zero-padded rolling-sum representation of the Bartlett quadratic form implies
\[
\frac{1}{n\Lambda_n}
\left|
\widetilde\psi'K_{n,b}\widetilde\psi-
\psi'K_{n,b}\psi
\right|=o_p(1).
\]
Indeed, if $m_n$ is the Bartlett kernel denominator, so the weights are $1-h/m_n$ for $h=1,\ldots,m_n-1$, the Bartlett quadratic is $m_n^{-1}$ times a sum of squared zero-padded rolling-window sums of length $m_n$; every feasible-minus-oracle window sum is the difference of two partial-sum replacement errors. In the simulation notation $m_n=\ell+1$, where $\ell$ is the largest included positive lag. The standardized conditions above control the two cross terms and the squared replacement term uniformly at the $n\Lambda_n$ scale. Under a joint functional CLT for the standardized fixed-dimensional augmented score/nuisance process and root-$n$ nuisance estimation, the exact low-rank score decomposition used in the cumulant proof is one route to the displayed uniform partial-sum replacement. Thus the fixed-bandwidth assumption is verifiable under familiar process-level strengthening, but it is not a consequence of the small-bandwidth rate argument alone. If $\Lambda_n$ is bounded away from zero and infinity, these conditions reduce to the unscaled $n^{-1/2}$ formulation.
\end{remark}

\begin{remark}[Recentering the null-imposed covariance score]\label{rem:recenter}
Let $\widetilde\psi=(\widetilde\psi_1,\ldots,\widetilde\psi_n)'$, $\bar{\widetilde\psi}=n^{-1}\mathbf 1'\widetilde\psi$, and $\widetilde\psi^c=\widetilde\psi-\bar{\widetilde\psi}\mathbf 1$. For any symmetric kernel matrix $K_n$,
\[
\frac1n\{(\widetilde\psi^c)'K_n\widetilde\psi^c-\widetilde\psi'K_n\widetilde\psi\}
=
-\frac{2\bar{\widetilde\psi}}{n}\mathbf 1'K_n\widetilde\psi
+
\frac{\bar{\widetilde\psi}^{\,2}}{n}\mathbf 1'K_n\mathbf 1.
\]
Consequently, if
\[
\bar{\widetilde\psi}=O_p(n^{-1/2}),\qquad
\mathbf 1'K_n\widetilde\psi=O_p(m_n\sqrt n),\qquad
\mathbf 1'K_n\mathbf 1=O(nm_n),
\]
then the centered and uncentered quadratic forms differ by $O_p(m_n/n)$. Thus, when the relevant variance scale is bounded away from zero and $m_n=o(n)$, recentering is first-order negligible. These displayed orders are sufficient conditions, not consequences of Theorem~\ref{thm:pdmixing} by themselves; a particular implementation must verify them.

At fixed bandwidth, $m_n/n\to b>0$, the same algebra gives only an $O_p(1)$ difference. Under the usual functional central limit theorem, replacing the score by its sample-demeaned version changes the partial-sum limit from $W(r)$ to the Brownian bridge $B(r)=W(r)-rW(1)$ and therefore changes the kernel quadratic reference functional. For this reason, the fixed-bandwidth theorem in the main paper uses the nonrecentered null-imposed score. A recentered fixed-bandwidth procedure is feasible, but its critical values must be derived from the corresponding bridge-based joint limit rather than imported from Theorem~\ref{thm:fixedb}.
\end{remark}

\begin{proposition}[Uniform size transfer for the projected normal test]\label{prop:pduniform}
For each $n$, let $\mathcal P_n$ be a class of data-generating processes. Suppose that
\[
\sup_{P\in\mathcal P_n}\sup_{x\in\mathbb R}
\left|
P\!\left\{\frac{\sqrt n(\widehat\beta-\beta_0)}{V_n(P)^{1/2}}\le x\right\}
-\Phi(x)
\right|\longrightarrow0,
\]
that for every $\eta>0$,
\[
\sup_{P\in\mathcal P_n}
P\!\left(
\left|\frac{\widetilde V_{\mathrm{PD},n}(\beta_0)}{V_n(P)}-1\right|>\eta
\right)\longrightarrow0,
\]
and that
\[
\inf_{P\in\mathcal P_n}
P\{\widetilde V_{\mathrm{PD},n}(\beta_0)>0\}\longrightarrow1.
\]
Then, for every fixed $\alpha\in(0,1)$,
\[
\sup_{P\in\mathcal P_n}
\left|
P\!\left\{
\frac{|\sqrt n(\widehat\beta-\beta_0)|}
{\{\widetilde V_{\mathrm{PD},n}(\beta_0)\}^{1/2}}
>z_{1-\alpha/2}
\right\}-\alpha
\right|\longrightarrow0.
\]
This is a transfer result: it does not by itself supply primitive uniform mixing, cumulant, or cluster conditions for the displayed hypotheses.
\end{proposition}

\paragraph{Proof.}
If the conclusion failed, there would exist $\epsilon>0$, a subsequence, and $P_n\in\mathcal P_n$ along that subsequence for which the absolute size error is at least $\epsilon$. Along this selected sequence, the three uniform assumptions imply the corresponding pointwise conclusions: the standardized sampling cdf converges uniformly to $\Phi$, the projected variance ratio converges to one in probability, and the feasible projected variance is positive with probability tending to one. Slutsky's theorem therefore gives a standard-normal limit for the feasible studentized statistic along the selected sequence, so its two-sided rejection probability converges to $\alpha$, a contradiction. \qed

\begin{remark}[Scope of the uniform statement]\label{rem:pduniformscope}
The proposition does not make the dependence-ignorant iid procedure uniformly valid over classes on which the relevance index can drift. In particular, if a class permits $R_{A,n}(P)$ to approach one, then $V_n^\#(P)/V_n(P)=1-R_{A,n}(P)$ can approach zero and the iid test can approach certain rejection under the phase formula. Uniform validity therefore requires uniform control of the variance target actually used by the procedure.
\end{remark}

\subsection{Proof of Corollary~\ref{cor:pdbinary}}
The identity $V_n^\#-V_n=-\Delta_{\tau,n}$ is Corollary~\ref{cor:te}. If the projected-dependence bootstrap satisfies the conditions of Theorem~\ref{thm:pd}, its scalar projected variance is consistent for $V_n$ and its standardized conditional law is asymptotically standard normal. Theorem~\ref{thm:pd} therefore gives first-order validity. No condition requiring reproduction of the full potential-outcome dependence vector enters that argument. \qed

\subsection{A primitive stationary strong-mixing implementation}

\begin{theorem}[Projected-HAC validity under stationary strong mixing]\label{thm:pdmixing}
Under $H_0$, suppose the setup and iid assignment structure of Section~\ref{sec:setup} hold. Let $\psi_{i,n}$ denote the oracle scalar score for the randomized-regressor coefficient and suppose, for each $n$, that $\{\psi_{i,n}:i\in\mathbb Z\}$ is a strictly stationary, mean-zero process. Assume uniformly in $n$:
\begin{enumerate}[label=(\roman*)]
\item for some $\delta>0$, $E|\psi_{0,n}|^{4+\delta}\le C$;
\item its strong-mixing coefficients satisfy
\[
\sum_{h=1}^\infty h\,\alpha_n(h)^{\delta/(4+\delta)}\le C;
\]
\item the long-run variance
\[
\Lambda_n=\sum_{h=-\infty}^{\infty}\Cov(\psi_{0,n},\psi_{h,n})
\]
is bounded away from zero;
\item $k$ is an even, bounded, continuous lag kernel with $k(0)=1$ and compact support such that $K_n=(k((i-j)/m_n))_{i,j\le n}$ is positive semidefinite for every $n$, and the kernel denominator satisfies
\[
m_n\to\infty,\qquad m_n=o(n^{1/3});
\]
\item the fixed-dimensional null-imposed nuisance regression satisfies
\[
\|(n^{-1}W'W)^{-1}\|=O_p(1),\qquad n^{-1/2}W'\varepsilon=O_p(1),
\]
\[
n^{-1}\sum_i\varepsilon_i^2=O_p(1),\qquad
n^{-1}\sum_i(D_i-\bar D)^2\|W_i\|^2=O_p(1),
\]
and
\[
\widehat Q_n/\sigma_{D,n}^2\pto1,
\qquad
0<c\le \sigma_{D,n}^2\le C.
\]
\end{enumerate}
The mixing exponent $\delta/(4+\delta)$ is intentionally stronger than the pairwise exponent needed only for covariance summability. The HAC consistency proof also controls covariances of product terms such as $\psi_i\psi_{i+h}$; applying Davydov's inequality to that product process leads to the displayed exponent. Thus one mixing condition supports both the long-run-variance edge bound and quadratic-form concentration.

Set
\[
\kappa_{ij,n}=k\!\left(\frac{i-j}{m_n}\right)
\]
and use Gaussian projected multipliers with covariance matrix $K_n=(\kappa_{ij,n})$. Define the finite-sample score variance
\[
\Lambda_n^{\mathrm{fs}}
:=
\Var\!\left(n^{-1/2}\sum_{i=1}^n\psi_{i,n}\right)
=
\sum_{|h|<n}\left(1-\frac{|h|}{n}\right)\varrho_{\psi,n}(h),
\qquad
\varrho_{\psi,n}(h)=\Cov(\psi_{0,n},\psi_{h,n}).
\]
By the definition in Section~\ref{sec:setup},
\[
V_n=\sigma_{D,n}^{-4}\Lambda_n^{\mathrm{fs}}
\]
exactly for the leading-score variance. Conditions (i)--(ii) imply uniform first-moment absolute summability of $\varrho_{\psi,n}(h)$, and therefore
\[
|\Lambda_n^{\mathrm{fs}}-\Lambda_n|
\le
\frac1n\sum_{|h|<n}|h|\,|\varrho_{\psi,n}(h)|
+
\sum_{|h|\ge n}|\varrho_{\psi,n}(h)|
=O(n^{-1}).
\]
Conditions (i)--(ii) give $|\Lambda_n^{\mathrm{fs}}-\Lambda_n|=O(n^{-1})$, while condition (iii) gives $\Lambda_n\ge c>0$; hence
\[
\frac{\Lambda_n^{\mathrm{fs}}}{\Lambda_n}\to1.
\]
Together with condition (v) and the implied uniform upper bound on $\Lambda_n$, these conditions keep $V_n$ bounded away from zero and infinity. Suppose in addition that the CHLS linearization
\[
\sqrt n(\widehat\beta-\beta_0)
=
\sigma_{D,n}^{-2}n^{-1/2}\sum_{i=1}^n\psi_{i,n}+o_p(1)
\]
holds under $H_0$ and that the scalar sampling CLT
\[
\frac{n^{-1/2}\sum_{i=1}^n\psi_{i,n}}{\Lambda_n^{1/2}}\dto N(0,1)
\]
holds. Then the $o_p(1)$ remainder in the CHLS linearization is also $o_p(V_n^{1/2})$, and
\[
\widetilde V_{\mathrm{PD},n}(\beta_0)-V_n\pto0,
\qquad
\frac{\widetilde V_{\mathrm{PD},n}(\beta_0)}{V_n}\pto1.
\]
Because $K_n$ is positive semidefinite, $\widetilde V_{\mathrm{PD},n}(\beta_0)\ge0$ identically. The preceding ratio convergence and the lower bound on $V_n$ imply
\[
P\{\widetilde V_{\mathrm{PD},n}(\beta_0)>0\}\to1.
\]
Consequently, all positivity and variance-ratio conditions in Theorem~\ref{thm:pd} hold, and the projected-dependence bootstrap is first-order valid:
\[
\sup_x\left|
P^*\{T_{\mathrm{PD},n}^*(\beta_0)\le x\}
-
P\{\sqrt n(\widehat\beta-\beta_0)\le x\}
\right|\pto0.
\]
\end{theorem}

The theorem leaves the scalar triangular-array CLT as an explicit assumption. The next proposition shows that in a fixed-process stationary specialization the displayed mixing and moment conditions already imply that CLT. The kernel-denominator rate $m_n=o(n^{1/3})$ is the conservative rate delivered by the self-contained overlap-aware proof below; Section~\ref{app:cumulant-sharp} gives the sharper $m_n=o(n)$ route under fourth-order cumulant summability.

\subsection{Fixed-process closure of the scalar sampling CLT}

\begin{proposition}[Fixed-process strong-mixing CLT]\label{prop:mixingclt}
Suppose $\psi_{i,n}=\psi_i$ for all $n$, where $\{\psi_i:i\in\mathbb Z\}$ is strictly stationary and mean zero. Suppose the fixed-process analogues of conditions (i)--(iii) of Theorem~\ref{thm:pdmixing} hold. Then
\[
\frac{n^{-1/2}\sum_{i=1}^n\psi_i}{\Lambda^{1/2}}\dto N(0,1),
\qquad
\Lambda=\sum_{h=-\infty}^{\infty}\Cov(\psi_0,\psi_h).
\]
Thus the separate scalar sampling CLT in Theorem~\ref{thm:pdmixing} is redundant in this fixed-process specialization. This proposition is not asserted for a genuinely row-varying triangular array.
\end{proposition}

\paragraph{Proof.}
Put $r=2+\delta$. Condition (i) of Theorem~\ref{thm:pdmixing} gives
\[
E|\psi_0|^{2+r}=E|\psi_0|^{4+\delta}<\infty.
\]
A classical stationary strong-mixing CLT with this moment exponent uses
\[
\sum_{h\ge1}\alpha(h)^{r/(2+r)}
=
\sum_{h\ge1}\alpha(h)^{(2+\delta)/(4+\delta)}
<\infty.
\]
Condition (ii) is stronger. Since $0\le\alpha(h)\le1$,
\[
\alpha(h)^{(2+\delta)/(4+\delta)}
\le
\alpha(h)^{\delta/(4+\delta)},
\]
so
\[
\sum_{h\ge1}\alpha(h)^{(2+\delta)/(4+\delta)}
\le
\sum_{h\ge1}h\,\alpha(h)^{\delta/(4+\delta)}
<\infty.
\]
Davydov's inequality gives absolute covariance summability, hence finiteness of
\[
\Lambda=\sum_{h=-\infty}^{\infty}\Cov(\psi_0,\psi_h).
\]
Condition (iii) supplies the remaining nondegeneracy, $\Lambda>0$. The classical stationary strong-mixing CLT therefore gives
\[
\frac{\sum_{i=1}^n\psi_i}{\sqrt{n\Lambda}}\dto N(0,1).
\]

If the law of the stationary process changes with $n$, however, a fixed-sequence CLT cannot simply
be applied row by row to infer a triangular-array limit. One must either maintain the scalar
triangular-array CLT, as Theorem~\ref{thm:pdmixing} does, or impose the additional uniform/Lindeberg
conditions of a triangular-array strong-mixing CLT. \qed

\subsection{Proof of Theorem~\ref{thm:pdmixing}}

Write
\[
\varrho_{\psi,n}(h)=\Cov(\psi_{0,n},\psi_{h,n}),
\qquad
\widehat\varrho_{\psi,n}(h)
=
\frac1n\sum_{i=1}^{n-|h|}
\psi_{i,n}\psi_{i+|h|,n}.
\]
By Davydov's covariance inequality, assumptions (i)--(ii) imply, uniformly in $n$,
\[
|\varrho_{\psi,n}(h)|
\le
C\alpha_n(h)^{\delta/(4+\delta)},
\]
and therefore
\[
\sup_n\sum_{h=-\infty}^{\infty}(1+|h|)|\varrho_{\psi,n}(h)|<\infty.
\]
In particular, $\Lambda_n$ is well defined. To control the kernel bias uniformly along the triangular array, fix $H<\infty$. For $|h|\le H$,
\[
\sup_{|h|\le H}
\left|
k(h/m_n)\left(1-\frac{|h|}{n}\right)-1
\right|\to0.
\]
For the tail, the weighted summability gives
\[
\sup_n\sum_{|h|>H}|\varrho_{\psi,n}(h)|
\le \frac{C}{H+1}.
\]
Therefore, letting first $n\to\infty$ and then $H\to\infty$,
\[
\sum_{|h|<n}
k(h/m_n)\left(1-\frac{|h|}{n}\right)\varrho_{\psi,n}(h)
-\Lambda_n
\longrightarrow0.
\]
This finite-lag/uniform-tail argument is the triangular-array justification.

Define the oracle kernel quadratic form
\[
\widehat\Lambda_n^{\,o}
=
\frac1n\sum_{i,j=1}^n
k\!\left(\frac{i-j}{m_n}\right)
\psi_{i,n}\psi_{j,n}.
\]
Its expectation is the preceding kernel-weighted covariance sum, so it remains to control its centered part. For $h\ge0$, put
\[
X_{i,h,n}=\psi_{i,n}\psi_{i+h,n}-E(\psi_{i,n}\psi_{i+h,n}).
\]
Assumption (i) gives a uniform $2+\delta/2$ moment bound. The overlap region must be treated explicitly: if $|i-j|\le h$, then
\[
|\Cov(X_{i,h,n},X_{j,h,n})|\le C.
\]
If $|i-j|>h$, the two product blocks are separated by at least $|i-j|-h$, and Davydov's inequality gives
\[
|\Cov(X_{i,h,n},X_{j,h,n})|
\le C\alpha_n(|i-j|-h)^{\delta/(4+\delta)}.
\]
Therefore
\[
\sum_j|\Cov(X_{i,h,n},X_{j,h,n})|\le C(1+h)
\]
uniformly in $n$ and $h$, so
\[
\Var\!\left(n^{-1}\sum_{i=1}^{n-h}X_{i,h,n}\right)
\le C\frac{1+h}{n}.
\]
Consequently,
\[
\left\|\widehat\Lambda_n^{\,o}-E\widehat\Lambda_n^{\,o}\right\|_2
\le \frac{C}{\sqrt n}\sum_{h=0}^{Cm_n}(1+h)^{1/2}
\le C\frac{m_n^{3/2}}{\sqrt n}
=o(1)
\]
under $m_n=o(n^{1/3})$. Hence $\widehat\Lambda_n^{\,o}-\Lambda_n\pto0$.

To pass from oracle to feasible scores, use the exact identity
\[
r_{i,n}:=\widetilde\psi_i-\psi_{i,n}
=-(\bar D-\mu_{D,n})\varepsilon_i-(D_i-\bar D)W_i'(\widetilde\gamma-\gamma).
\]
Condition (v) gives
\[
\widetilde\gamma-\gamma=(n^{-1}W'W)^{-1}(n^{-1}W'\varepsilon)=O_p(n^{-1/2}),
\]
and iid assignment gives $\bar D-\mu_{D,n}=O_p(n^{-1/2})$. Hence
\[
n^{-1}\sum_i r_{i,n}^2
\le2(\bar D-\mu_{D,n})^2n^{-1}\sum_i\varepsilon_i^2+
2\|\widetilde\gamma-\gamma\|^2
n^{-1}\sum_i(D_i-\bar D)^2\|W_i\|^2
=O_p(n^{-1}).
\]
Thus the empirical $L^2$ score error is $O_p(n^{-1/2})$. Because $k$ is bounded and compactly supported, every row of $K_n$ contains at most $Cm_n$ nonzero entries of uniformly bounded magnitude. Hence
\[
\|K_n\|_{\rm op}
\le
\max_i\sum_j|K_{ij,n}|
=
O(m_n).
\]
Stationarity and the uniform second-moment bound implied by condition (i) give
\[
n^{-1}\|\psi\|^2=O_p(1),
\qquad
n^{-1}\|r\|^2=O_p(n^{-1}).
\]
Now
\[
\frac1n\widetilde\psi'K_n\widetilde\psi
-
\frac1n\psi'K_n\psi
=
\frac2n\psi'K_nr+\frac1n r'K_nr.
\]
Therefore
\[
\left|\frac1n\psi'K_nr\right|
\le
\|K_n\|_{\rm op}
\left(\frac{\|\psi\|^2}{n}\right)^{1/2}
\left(\frac{\|r\|^2}{n}\right)^{1/2}
=
O_p(m_n/\sqrt n),
\]
and
\[
\left|\frac1n r'K_nr\right|
\le
\|K_n\|_{\rm op}\frac{\|r\|^2}{n}
=
O_p(m_n/n).
\]
Thus
\[
|n^{-1}\widetilde\psi'K_n\widetilde\psi-n^{-1}\psi'K_n\psi|
=
O_p(m_n/\sqrt n)+O_p(m_n/n)
=o_p(1),
\]
because $m_n=o(n^{1/3})$ implies $m_n=o(n^{1/2})$. Since $\widehat Q_n/\sigma_{D,n}^2\to_p1$ and $\sigma_{D,n}^2$ is bounded away from zero and infinity,
\[
\widehat Q_n^{-2}
\left(n^{-1}\widetilde\psi'K_n\widetilde\psi\right)
=
\sigma_{D,n}^{-4}\Lambda_n+o_p(1).
\]
The generic first-order variance $V_n$ in Section~\ref{sec:setup}, however, is based on the finite-sample score variance
\[
\Lambda_n^{\mathrm{fs}}
:=
\Var\!\left(n^{-1/2}\sum_{i=1}^n\psi_{i,n}\right)
=
\sum_{|h|<n}\left(1-\frac{|h|}{n}\right)\varrho_{\psi,n}(h),
\]
so that
\[
V_n=\sigma_{D,n}^{-4}\Lambda_n^{\mathrm{fs}}
\]
exactly. The weighted covariance summability established above gives
\[
\begin{aligned}
|\Lambda_n^{\mathrm{fs}}-\Lambda_n|
&\le
\frac1n\sum_{|h|<n}|h|\,|\varrho_{\psi,n}(h)|
+
\sum_{|h|\ge n}|\varrho_{\psi,n}(h)|\\
&\le
\frac2n\sum_{h\in\mathbb Z}(1+|h|)|\varrho_{\psi,n}(h)|
=O(n^{-1})
\end{aligned}
\]
uniformly in $n$. Therefore
\[
\widetilde V_{\rm PD,n}(\beta_0)-V_n=o_p(1).
\]
Because $\sigma_{D,n}^2$ and $\Lambda_n$ are bounded away from zero and infinity and $\Lambda_n^{\mathrm{fs}}-\Lambda_n=O(n^{-1})$, $V_n$ is bounded away from zero and infinity for all sufficiently large $n$. Hence the $o_p(1)$ remainder in the CHLS linearization is also
\[
o_p(V_n^{1/2}).
\]
Moreover $\Lambda_n^{\mathrm{fs}}/\Lambda_n\to1$, so the assumed scalar CLT and Slutsky's theorem imply
\[
\frac{\sqrt n(\widehat\beta-\beta_0)}{V_n^{1/2}}
\dto N(0,1).
\]
Gaussian projected multipliers are conditionally Gaussian with conditional variance
\[
\widetilde V_{\rm PD,n}(\beta_0)\ge0,
\]
because $K_n$ is positive semidefinite. Since
\[
\widetilde V_{\rm PD,n}(\beta_0)/V_n\pto1
\]
and $V_n$ is bounded away from zero, the projected variance is strictly positive with probability tending to one. On that event, standardization by its own conditional standard deviation gives an exact $N(0,1)$ law. Thus the standardized conditional-law condition of Theorem~\ref{thm:pd} holds (with zero approximation error on the positive-variance event), while the preceding arguments verify its sampling-CLT and variance-ratio conditions. Theorem~\ref{thm:pd} therefore yields the asserted conditional Kolmogorov validity. \qed

\paragraph{Primitive routes to the nuisance-score rate.}
For $Z_{i,n}=W_{i,n}\varepsilon_{i,n}$, it is enough that
\[
\|n^{-1/2}\sum_iEZ_{i,n}\|=O(1),\qquad
n^{-1}E\|\sum_i(Z_{i,n}-EZ_{i,n})\|^2=O(1).
\]
With fixed-dimensional controls this follows, for example, under fixed-order $m$-dependence,
martingale differences, standard strong-mixing covariance summability with a $2+\eta$ moment,
or independent clusters satisfying $\sum_g|G_{g,n}|^2=O(n)$. These are sufficient routes to the
nuisance-score rate, not additional restrictions on the projected score.

\section{Sharper projected-HAC bandwidth under cumulant summability}\label{app:cumulant-sharp}
The next result provides a complementary route to the preceding strong-mixing theorem.

\begin{theorem}[Sharper bandwidth under fourth-order cumulant summability]\label{thm:pdcumulant}
Under $H_0$, suppose the setup and iid assignment structure of Section~\ref{sec:setup} hold. Let $\{\psi_{i,n}:i\in\mathbb Z\}$ be a mean-zero strictly stationary scalar triangular array. Write
\[
\varrho_{\psi,n}(h)=\Cov(\psi_{0,n},\psi_{h,n}),
\]
and let
\[
\kappa_n(a,b,c)
=
\operatorname{cum}(\psi_{0,n},\psi_{a,n},\psi_{b,n},\psi_{c,n})
\]
denote its fourth-order cumulant function. Suppose uniformly in $n$:
\[
\sum_{h\in\mathbb Z}(1+|h|)|\varrho_{\psi,n}(h)|\le C,
\qquad
\sum_{a,b,c\in\mathbb Z}|\kappa_n(a,b,c)|\le C,
\]
and
\[
0<c\le \Lambda_n:=\sum_{h\in\mathbb Z}\varrho_{\psi,n}(h)\le C.
\]
Let $k$ be even, bounded, continuous, and compactly supported with $k(0)=1$, and suppose
\[
K_n=\left(k\!\left(\frac{i-j}{m_n}\right)\right)_{i,j\le n}
\]
is positive semidefinite for every $n$. Let
\[
m_n\to\infty,\qquad m_n=o(n).
\]
The positive-semidefinite requirement is needed only for the Gaussian multiplier implementation below; the oracle and feasible quadratic-form consistency arguments use only evenness, boundedness, continuity, and compact support of $k$.

Then the oracle projected-HAC quadratic form
\[
\widehat\Lambda_n^{\,o}
=
\frac1n\sum_{i,j=1}^n
k\!\left(\frac{i-j}{m_n}\right)\psi_{i,n}\psi_{j,n}
\]
satisfies
\[
E\widehat\Lambda_n^{\,o}-\Lambda_n\to0,
\qquad
\Var(\widehat\Lambda_n^{\,o})=O(m_n/n),
\]
and hence
\[
\widehat\Lambda_n^{\,o}-\Lambda_n\pto0.
\]
If
\[
\Lambda_n^{\mathrm{fs}}
:=
\Var\!\left(n^{-1/2}\sum_{i=1}^n\psi_{i,n}\right)
=
\sum_{|h|<n}\left(1-\frac{|h|}{n}\right)\varrho_{\psi,n}(h),
\]
then the weighted covariance summability assumption also gives
\[
|\Lambda_n^{\mathrm{fs}}-\Lambda_n|=O(n^{-1}).
\]
For feasible inference, put
\[
U_{i,n}=(D_{i,n}-\mu_{D,n})W_{i,n},
\qquad
S_{i,n}
=
\begin{pmatrix}
\psi_{i,n}\\
\varepsilon_{i,n}\\
U_{i,n}
\end{pmatrix}.
\]
Under the CHLS centering and random-assignment conditions, every component of $S_{i,n}$ is mean zero. Suppose, in addition to the restricted-OLS conditions in Theorem~\ref{thm:pdmixing}(v), that $\{S_{i,n}\}$ is strictly stationary within row and, for every component pair $(j,k)$ and component quadruple $(j_1,j_2,j_3,j_4)$,
\[
\sup_n\sum_{h\in\mathbb Z}
\left|
\Cov(S_{0,n}^{(j)},S_{h,n}^{(k)})
\right|<\infty,
\]
and
\[
\sup_n\sum_{a,b,c\in\mathbb Z}
\left|
\operatorname{cum}
(S_{0,n}^{(j_1)},S_{a,n}^{(j_2)},S_{b,n}^{(j_3)},S_{c,n}^{(j_4)})
\right|<\infty.
\]
Also suppose
\[
\frac1n\sum_{i=1}^n\|W_{i,n}\|^2=O_p(1).
\]
Then, under the same kernel-denominator sequence
\[
m_n\to\infty,\qquad m_n=o(n),
\]
the feasible projected-HAC variance satisfies
\[
\widetilde V_{\mathrm{PD},n}(\beta_0)-\sigma_{D,n}^{-4}\Lambda_n\pto0.
\]
Since the generic leading-score variance in Section~\ref{sec:setup} is
\[
V_n=\sigma_{D,n}^{-4}\Lambda_n^{\mathrm{fs}},
\]
the $O(n^{-1})$ finite-sample/long-run-variance bridge and the treatment-scale bounds in Theorem~\ref{thm:pdmixing}(v) also give
\[
\widetilde V_{\mathrm{PD},n}(\beta_0)-V_n\pto0,
\qquad
\frac{\widetilde V_{\mathrm{PD},n}(\beta_0)}{V_n}\pto1.
\]
If, in addition, the CHLS linearization
\[
\sqrt n(\widehat\beta-\beta_0)
=
\sigma_{D,n}^{-2}n^{-1/2}\sum_{i=1}^n\psi_{i,n}+o_p(1)
\]
holds under $H_0$ and the scalar sampling CLT normalized by $\Lambda_n$ holds, then Gaussian projected multipliers yield a first-order valid projected-dependence bootstrap.
\end{theorem}

\begin{proposition}[A primitive route to augmented cumulant summability]\label{prop:Scumulant}
Let $\{S_{i,n}:i\in\mathbb Z\}$ be a fixed-dimensional, mean-zero, strictly stationary triangular array with strong-mixing coefficients $\alpha_{S,n}(h)$. Suppose for some $\delta>0$,
\[
\sup_n E\|S_{0,n}\|^{4+\delta}<\infty,
\qquad
\sup_n\sum_{h=1}^\infty
h^2\alpha_{S,n}(h)^{\delta/(4+\delta)}
<\infty.
\]
Then, uniformly in $n$, every pairwise cross-covariance sequence among the components of $S_{i,n}$ is first-moment absolutely summable,
\[
\sum_{h\in\mathbb Z}(1+|h|)
\left|\Cov(S_{0,n}^{(j)},S_{h,n}^{(k)})\right|\le C,
\]
and every fourth-order joint cumulant sequence is absolutely summable. Hence these primitive strong-mixing conditions imply the dependence assumptions used in the feasible part of Theorem~\ref{thm:pdcumulant}.
\end{proposition}

\begin{remark}[Primitive CHLS route]
A sufficient, but stronger, route is to assume that the fixed-dimensional latent/control process $(e_{i,n}',W_{i,n}')'$ is stationary strong mixing within row with the moment and weighted-mixing bounds in Proposition~\ref{prop:Scumulant}, while assignment is iid and independent of that process and the assignment transformations are uniformly bounded. Then $S_{i,n}$ is a measurable function of the augmented iid-assignment/latent process and inherits the required stationarity, moments, and mixing rate. This route is convenient but not necessary: Theorem~\ref{thm:pdcumulant} itself only needs the smaller score/nuisance process $S_{i,n}$.
\end{remark}

\begin{remark}[Two bandwidth routes]\label{rem:twobandwidth}
Theorems~\ref{thm:pdmixing} and \ref{thm:pdcumulant} are complementary. The strong-mixing theorem gives primitive HAC-consistency conditions in terms of moments and mixing coefficients but uses the conservative kernel-denominator rate $m_n=o(n^{1/3})$. The cumulant theorem analyzes the full HAC quadratic form directly and gives the standard oracle order $O(m_n/n)$. For feasible inference, joint covariance and cumulant summability are imposed only on the mean-zero augmented score vector $S_{i,n}=(\psi_{i,n},\varepsilon_{i,n},U_{i,n}')'$. The noncentered control vector $W_i$ enters only through higher-order nuisance terms, which can be handled by the crude kernel-operator bound. The resulting feasible replacement is
\[
O_p(n^{-1/2})+O_p(m_n/n)=o_p(1)
\]
under the same $m_n=o(n)$ kernel-denominator rate.
\end{remark}

\subsection{Proof of Theorem~\ref{thm:pdcumulant}: direct quadratic-form variance}

Let
\[
K_{ij,n}
=
k\!\left(\frac{i-j}{m_n}\right),
\qquad
\widehat\Lambda_n^{\,o}
=
\frac1n\sum_{i,j=1}^n K_{ij,n}\psi_{i,n}\psi_{j,n}.
\]
Because $k$ is bounded and compactly supported,
\[
|K_{ij,n}|\le C\,\mathbf 1\{|i-j|\le Cm_n\}.
\]

\paragraph{Bias.}
Stationarity gives
\[
E\widehat\Lambda_n^{\,o}
=
\sum_{|h|<n}
k(h/m_n)\left(1-\frac{|h|}{n}\right)\varrho_{\psi,n}(h).
\]
A triangular-array bias argument needs uniform tail control because covariance mass could otherwise drift to lags that increase with $n$. Fix $H$. For $|h|\le H$,
\[
\sup_{|h|\le H}\left|
k(h/m_n)\left(1-\frac{|h|}{n}\right)-1
\right|\to0.
\]
For the tail,
\[
\sup_n\sum_{|h|>H}|\varrho_{\psi,n}(h)|
\le
\frac{1}{H+1}
\sup_n\sum_h(1+|h|)|\varrho_{\psi,n}(h)|
\le \frac{C}{H+1}.
\]
Letting first $n\to\infty$ and then $H\to\infty$ yields
\[
E\widehat\Lambda_n^{\,o}-\Lambda_n\to0.
\]

\paragraph{Variance.}
For mean-zero variables,
\[
\Cov(\psi_i\psi_j,\psi_r\psi_s)
=
\operatorname{cum}(\psi_i,\psi_j,\psi_r,\psi_s)
+
\varrho_{\psi,n}(i-r)\varrho_{\psi,n}(j-s)
+
\varrho_{\psi,n}(i-s)\varrho_{\psi,n}(j-r).
\]
Therefore
\[
\Var(\widehat\Lambda_n^{\,o})
=
n^{-2}(C_{4,n}+P_{1,n}+P_{2,n}),
\]
where the three terms correspond to the fourth cumulant and the two Gaussian pairings.

For the cumulant term, boundedness of the kernel and stationarity imply
\[
\begin{aligned}
|C_{4,n}|
&\le
C\sum_{i,j,r,s}
\left|
\operatorname{cum}(\psi_i,\psi_j,\psi_r,\psi_s)
\right|\\
&\le
Cn\sum_{a,b,c\in\mathbb Z}
|\kappa_n(a,b,c)|
\le Cn.
\end{aligned}
\]
Hence its contribution is $O(n^{-1})$.

For the first pairing term,
\[
|P_{1,n}|
\le
C\sum_{i,j,r,s}
\mathbf 1\{|i-j|\le Cm_n\}
\mathbf 1\{|r-s|\le Cm_n\}
|\varrho_{\psi,n}(i-r)|\,|\varrho_{\psi,n}(j-s)|.
\]
Set
\[
u=i-r,\qquad v=j-s.
\]
For fixed $(i,u,v)$, the number of indices $j$ for which both kernel indicators are nonzero is at most $Cm_n$: the two restrictions require $j-i$ to lie in the intersection of two intervals of length $O(m_n)$. Thus
\[
|P_{1,n}|
\le
Cnm_n
\sum_{u,v\in\mathbb Z}
|\varrho_{\psi,n}(u)|\,|\varrho_{\psi,n}(v)|
\le
Cnm_n.
\]
The same argument gives $|P_{2,n}|\le Cnm_n$. Consequently,
\[
\boxed{
\Var(\widehat\Lambda_n^{\,o})
\le
C\left(\frac1n+\frac{m_n}{n}\right)
=
O(m_n/n).
}
\]
Since $m_n=o(n)$,
\[
\widehat\Lambda_n^{\,o}-\Lambda_n\pto0.
\]

\paragraph{Centered cross-bilinear quadratic-form bound.}
Let $X_i$ and $Y_i$ be any two scalar components of the mean-zero vector
\[
S_i=(\psi_i,\varepsilon_i,U_i')',
\qquad
U_i=(D_i-\mu_{D,n})W_i.
\]
Define
\[
\widehat C_{XY,n}
=
\frac1n\sum_{i,j=1}^nK_{ij,n}X_iY_j.
\]
Because $X_i$ and $Y_i$ are centered,
\[
E\widehat C_{XY,n}
=
\sum_{|h|<n}
k(h/m_n)(1-|h|/n)\gamma_{XY,n}(h)
=
O(1)
\]
under uniform absolute summability of the cross-covariance sequence. Moreover,
\[
\Cov(X_iY_j,X_rY_s)
=
\operatorname{cum}(X_i,Y_j,X_r,Y_s)
+
\Cov(X_i,X_r)\Cov(Y_j,Y_s)
+
\Cov(X_i,Y_s)\Cov(Y_j,X_r).
\]
Uniform absolute summability of all componentwise cross-covariances and fourth-order joint cumulants gives the same counting bounds as in the scalar proof:
\[
\Var(\widehat C_{XY,n})
=
O(n^{-1})+O(m_n/n)
=
O(m_n/n).
\]
Since the dimension of $S_i$ is fixed and $m_n=o(n)$,
\[
\boxed{\widehat C_{XY,n}=O_p(1)}
\]
simultaneously for all component pairs needed below.

\paragraph{Feasible score: exploiting low-rank nuisance structure.}
Let
\[
a_n=\bar D-\mu_{D,n},
\qquad
b_n=\widetilde\gamma(\beta_0)-\gamma,
\qquad
U_i=(D_i-\mu_{D,n})W_i.
\]
Then
\[
r_i:=\widetilde\psi_i-\psi_i
=
-a_n\varepsilon_i-U_i'b_n+a_nW_i'b_n.
\]
The restricted-OLS conditions imply
\[
a_n=O_p(n^{-1/2}),
\qquad
\|b_n\|=O_p(n^{-1/2}).
\]

The leading two terms involve only centered components of $S_i$:
\[
-a_n\,\frac1n\psi'K_n\varepsilon
-
b_n'\frac1n U'K_n'\psi
=
O_p(n^{-1/2}).
\]
The raw control vector $W_i$ need not be centered (and typically contains an intercept), so we do \emph{not} apply the centered cumulant decomposition to it. Instead, bounded kernel support implies
\[
\|K_n\|_{\mathrm{op}}=O(m_n).
\]
Together with
\[
n^{-1}\|\psi\|^2=O_p(1),
\qquad
n^{-1}\|W\|_F^2=O_p(1),
\]
we obtain
\[
\left|
\frac1n\psi'K_nW
\right|
=
O_p(m_n).
\]
Since $a_nb_n=O_p(n^{-1})$,
\[
a_nb_n'\frac1nW'K_n'\psi
=
O_p(m_n/n)
=
o_p(1).
\]
Therefore
\[
\boxed{
\frac1n\psi'K_nr
=
O_p(n^{-1/2})+O_p(m_n/n)
=
o_p(1).
}
\]

For the quadratic remainder, the terms containing only $\varepsilon$ and $U$ are $O_p(n^{-1})$ by the centered cross-bilinear bound. Every term involving $W$ has coefficient of order at most $n^{-3/2}$, except the $W$--$W$ term, whose coefficient is $O_p(n^{-2})$. To justify the crude bounds involving the noncentered controls, first note that the restricted-OLS conditions give
\[
\frac1n\|\varepsilon\|^2=O_p(1),
\qquad
\frac1n\|W\|_F^2=O_p(1).
\]
Moreover,
\[
U_i=(D_i-\mu_{D,n})W_i=(D_i-\bar D)W_i+a_nW_i,
\]
so
\[
\frac1n\|U\|_F^2
\le
2\frac1n\sum_i(D_i-\bar D)^2\|W_i\|^2
+2a_n^2\frac1n\sum_i\|W_i\|^2
=O_p(1).
\]
Because $\|K_n\|_{\rm op}=O(m_n)$,
\[
\left\|\frac1n\varepsilon'K_nW\right\|
\le
\|K_n\|_{\rm op}
\left(\frac{\|\varepsilon\|^2}{n}\right)^{1/2}
\left(\frac{\|W\|_F^2}{n}\right)^{1/2}
=O_p(m_n),
\]
\[
\left\|\frac1nU'K_nW\right\|
\le
\|K_n\|_{\rm op}
\left(\frac{\|U\|_F^2}{n}\right)^{1/2}
\left(\frac{\|W\|_F^2}{n}\right)^{1/2}
=O_p(m_n),
\]
and
\[
\left\|\frac1nW'K_nW\right\|_{\rm op}
\le
\|K_n\|_{\rm op}\frac{\|W\|_F^2}{n}
=O_p(m_n).
\]
Hence
\[
\boxed{
\frac1n r'K_nr
=
O_p(n^{-1})
+
O_p(m_n/n^{3/2})
+
O_p(m_n/n^2)
=
o_p(1)
}
\]
under $m_n=o(n)$.

Consequently,
\[
\frac1n\widetilde\psi'K_n\widetilde\psi
-
\frac1n\psi'K_n\psi
=
2\frac1n\psi'K_nr
+
\frac1n r'K_nr
=
O_p(n^{-1/2})+O_p(m_n/n)
=
o_p(1).
\]
Thus the feasible projected-HAC estimator inherits the oracle bandwidth
\[
m_n\to\infty,\qquad m_n=o(n),
\]
without incorrectly treating noncentered controls as centered score components. Combining this replacement with oracle consistency gives
\[
n^{-1}\widetilde\psi'K_n\widetilde\psi
=
\Lambda_n+o_p(1).
\]
Since $\widehat Q_n/\sigma_{D,n}^2\to_p1$ and $\sigma_{D,n}^2$ is bounded away from zero,
\[
\widetilde V_{\mathrm{PD},n}(\beta_0)
-
\sigma_{D,n}^{-4}\Lambda_n=o_p(1).
\]
By weighted covariance summability,
\[
|\Lambda_n^{\mathrm{fs}}-\Lambda_n|=O(n^{-1}),
\qquad
V_n=\sigma_{D,n}^{-4}\Lambda_n^{\mathrm{fs}},
\]
so
\[
\widetilde V_{\mathrm{PD},n}(\beta_0)-V_n=o_p(1),
\qquad
\widetilde V_{\mathrm{PD},n}(\beta_0)/V_n\to_p1.
\]
Under the stated CHLS linearization and scalar CLT, $\Lambda_n^{\mathrm{fs}}/\Lambda_n\to1$ and Slutsky's theorem give the standardized OLS limit at the generic variance scale $V_n$. Gaussian projected multipliers are conditionally Gaussian with the feasible variance above, so Lemma~\ref{lem:gaussianscale} proves first-order bootstrap validity. \qed

\paragraph{Why the improvement is possible.}
The generic operator-norm argument treats the entire perturbation $r_n$ as arbitrary and pays a factor $O(m_n)$. The exact restricted-OLS representation isolates two leading mean-zero nuisance-score processes, $\varepsilon_i$ and $U_i$, whose kernel cross-products with $\psi_i$ are $O_p(1)$. The noncentered raw control $W_i$ appears only multiplied by the second-order coefficient $a_nb_n=O_p(n^{-1})$, so its cruder $O_p(m_n)$ bound is still negligible under $m_n=o(n)$.

\subsection{Proof of Proposition~\ref{prop:Scumulant}: strong mixing implies joint cumulant summability}

Fix $n$ and suppress the row subscript. Let $X_t^{(j)}$ denote any scalar component of $S_t$. Davydov's covariance inequality and the uniform $4+\delta$ moment imply, for any two components,
\[
|\Cov(X_0^{(j)},X_h^{(k)})|
\le
C\alpha_S(h)^{(2+\delta)/(4+\delta)}
\le
C\alpha_S(h)^{\delta/(4+\delta)}.
\]
Because $\alpha_S(h)\in[0,1]$ and the weighted mixing assumption uses $h^2$,
\[
\sum_{h\ge1}(1+h)|\Cov(X_0^{(j)},X_h^{(k)})|
\le
C\sum_{h\ge1}(1+h)\alpha_S(h)^{\delta/(4+\delta)}
<\infty.
\]
This proves the required first-moment absolute summability of every pairwise cross-covariance sequence.

For fourth-order cumulants, choose arbitrary component indices and order the times as
\[
t_1\le t_2\le t_3\le t_4.
\]
Let
\[
g=\max\{t_2-t_1,t_3-t_2,t_4-t_3\}.
\]
We show that
\[
\left|
\operatorname{cum}(X_{t_1}^{(j_1)},X_{t_2}^{(j_2)},X_{t_3}^{(j_3)},X_{t_4}^{(j_4)})
\right|
\le
C\alpha_S(g)^{\delta/(4+\delta)}.
\]

Suppose first that the largest gap is $t_2-t_1$; the case $t_4-t_3$ is symmetric. Since all components are mean zero,
\[
\begin{aligned}
\operatorname{cum}(X_1,X_2,X_3,X_4)
={}&\Cov(X_1,X_2X_3X_4)\\
&-\Cov(X_1,X_2)\Cov(X_3,X_4)\\
&-\Cov(X_1,X_3)\Cov(X_2,X_4)\\
&-\Cov(X_1,X_4)\Cov(X_2,X_3),
\end{aligned}
\]
where $X_r$ abbreviates the corresponding ordered component. Davydov applied across the gap $g$ to $X_1$ and $X_2X_3X_4$, with exponents $4+\delta$ and $(4+\delta)/3$, gives
\[
|\Cov(X_1,X_2X_3X_4)|
\le C\alpha_S(g)^{\delta/(4+\delta)}.
\]
Each product of pairwise covariances contains a covariance linking $X_1$ to one of $X_2,X_3,X_4$. The pairwise Davydov bound has exponent $(2+\delta)/(4+\delta)>\delta/(4+\delta)$, so each product is also bounded by $C\alpha_S(g)^{\delta/(4+\delta)}$.

If the largest gap is the middle gap $t_3-t_2$, write instead
\[
\begin{aligned}
\operatorname{cum}(X_1,X_2,X_3,X_4)
={}&\Cov(X_1X_2,X_3X_4)\\
&-\Cov(X_1,X_3)\Cov(X_2,X_4)\\
&-\Cov(X_1,X_4)\Cov(X_2,X_3).
\end{aligned}
\]
Davydov applied to the two pair-products with exponent $(4+\delta)/2$ on each side again yields
\[
|\Cov(X_1X_2,X_3X_4)|
\le C\alpha_S(g)^{\delta/(4+\delta)}.
\]
Every covariance in the remaining two products crosses the middle gap, so the same (indeed stronger) pairwise Davydov bound applies. Thus the displayed cumulant bound holds in every case.

By stationarity, set the first time index equal to zero. Parameterize the remaining ordered locations by their three nonnegative successive gaps. The number of gap triples whose maximum equals $m$ is
\[
(m+1)^3-m^3=3m^2+3m+1=O((m+1)^2).
\]
Accounting for the finitely many orderings and component choices,
\[
\sum_{a,b,c\in\mathbb Z}
\left|
\operatorname{cum}
(X_0^{(j_1)},X_a^{(j_2)},X_b^{(j_3)},X_c^{(j_4)})
\right|
\le
C+C\sum_{m=1}^{\infty}(m+1)^2
\alpha_S(m)^{\delta/(4+\delta)}
<\infty.
\]
The bound is uniform in $n$, proving the proposition. \qed

\subsection{Compatibility of projected-score assumptions with the CHLS array}
\begin{proposition}[Compatibility with the CHLS random-assignment array]\label{prop:pdcompat}
Suppose within each row that $\{e_{i,n}\}$ is strictly stationary and strongly mixing, and $\{D_{i,n}\}$ is iid and independent of the full latent process. If $\psi_{i,n}=A_{i,n}'e_{i,n}$, then $\{\psi_{i,n}\}$ is strictly stationary and
\[
\alpha_{\psi,n}(h)\le\alpha_{e,n}(h).
\]
If $\|A_{i,n}\|\le C$ almost surely uniformly in $n$ and $\sup_nE\|e_{0,n}\|^{4+\delta}<\infty$, then $\sup_nE|\psi_{0,n}|^{4+\delta}<\infty$.
\end{proposition}
Stationarity is within row; the row law may vary with $n$.

\subsection{Proof of Proposition~\ref{prop:pdcompat}}
Fix a row $n$. Because the assignment and latent sequences are independent, the strong-mixing coefficients of the joint process satisfy
\[
\alpha_{(D,e),n}(h)\le\alpha_{D,n}(h)+\alpha_{e,n}(h)
\]
\citep[see][Theorem~5.2]{Bradley2005}. Iid assignment gives $\alpha_{D,n}(h)=0$ for $h\ge1$. Since $\psi_{i,n}=A_{i,n}'e_{i,n}$ is a measurable function of $(D_{i,n},e_{i,n})$, its past and future sigma-fields are sub-sigma-fields of those of the joint process. Hence
\[
\alpha_{\psi,n}(h)\le\alpha_{(D,e),n}(h)\le\alpha_{e,n}(h).
\]
If $\|A_{i,n}\|\le C$, then
\[
|\psi_{i,n}|^{4+\delta}\le C\|e_{i,n}\|^{4+\delta},
\]
which proves the moment claim. \qed

\paragraph{Relation to standard HAC theory.}
The kernel-denominator rate $m_n=o(n^{1/3})$ in the strong-mixing theorem is a conservative consequence of its self-contained lag-by-lag HAC-consistency bound. Proposition~\ref{prop:mixingclt} closes the scalar sampling CLT in the fixed-process stationary case; for a genuinely row-varying triangular array, the scalar CLT remains a separate maintained condition. Theorem~\ref{thm:pdcumulant} shows directly why the full quadratic form has the sharper $O(m_n/n)$ oracle variance under fourth-order cumulant summability. This is consistent with classical HAC theory; see \citet{NeweyWest1987}, \citet{Andrews1991HAC}, and \citet{deJong2000HAC}.

\paragraph{Primitive CHLS route to Proposition~\ref{prop:Scumulant}.}
Suppose the fixed-dimensional process $G_{i,n}=(e_{i,n}',W_{i,n}')'$ is strictly stationary and strong mixing within row, assignment is iid and independent of $G_{i,n}$, and the assignment transformations entering $\psi_i$, $\varepsilon_i$, and $U_i$ are uniformly bounded. Then
\[
S_{i,n}=(\psi_{i,n},\varepsilon_{i,n},U_{i,n}')'
\]
is a measurable function of $(D_{i,n},G_{i,n})$ and hence is stationary with
\[
\alpha_{S,n}(h)\le \alpha_{G,n}(h).
\]
A uniform $4+\delta$ moment for $G_{0,n}$ gives the corresponding moment for $S_{0,n}$. Thus
\[
\sup_n\sum_{h\ge1}
h^2\alpha_{G,n}(h)^{\delta/(4+\delta)}
<\infty
\]
is one primitive sufficient route to all pairwise covariance and fourth-order cumulant summability conditions used in the feasible cumulant theorem. This route is stronger than necessary because the theorem itself only requires these conditions for $S_i$.

\subsection{Gaussian implementation}

If $K_n$ is positive semidefinite, write $K_n=L_nL_n'$ and generate $g^*\sim N(0,I_n)$ conditionally on the data. Set $\xi_n^*=L_ng^*$. Then
\[
T_{\mathrm{PD},n}^*
=
\widehat Q_n^{-1}n^{-1/2}g^{*'}L_n'\widetilde\psi
\]
is conditionally Gaussian with variance
\[
\widehat Q_n^{-2}n^{-1}\widetilde\psi'K_n\widetilde\psi.
\]
Thus standardized conditional Gaussianity is exact whenever this variance is positive; the substantive requirement is consistency of the scalar quadratic form.

\subsection{Primitive projected implementations}

For an ordered sample, use a lag kernel for which the finite matrix $K_n=(\kappa_{ij,n})$ is symmetric positive semidefinite, with
\[
\kappa_{ij,n}=k\left(\frac{i-j}{m_n}\right).
\]
Here $m_n$ denotes the kernel denominator. For the Bartlett convention used in the simulations, if $\ell$ is the largest included positive lag, then $m_n=\ell+1$. The resulting quadratic form is a scalar kernel long-run-variance estimator of the restricted score. Under standard kernel-HAC consistency conditions on that projected score, Theorem~\ref{thm:pd} applies.

For clusters $G_1,\ldots,G_{C_n}$, setting
\[
\kappa_{ij,n}=\mathbf 1\{i,j\text{ are in the same cluster}\}
\]
gives
\[
\widetilde V_{\mathrm{PD},n}
=
\widehat Q_n^{-2}\frac1n\sum_{c=1}^{C_n}
\left(\sum_{i\in G_c}\widetilde\psi_i\right)^2.
\]
One Gaussian multiplier per cluster implements the corresponding projected cluster bootstrap.

This appendix records standard implementations of the high-level repair principle in Section~\ref{sec:repair}. For time-series dependence one may use a positive-semidefinite kernel or dependent multiplier covariance matrix applied to the randomized-regressor score. For known dependence clusters under individual-level random assignment, one may use cluster multipliers on sums of the randomized score. These constructions are standard; their role here is to recover the covariance component identified by the random-assignment filter rather than to constitute a new bootstrap technology.

\begin{proposition}[Projected-cluster validity under independent score clusters]\label{prop:pdcluster}
Under $H_0$, suppose the setup and iid assignment structure of Section~\ref{sec:setup} hold. Partition the observations into known clusters $G_{1,n},\ldots,G_{C_n,n}$ and define the oracle cluster-score sums
\[
S_{c,n}=\sum_{i\in G_{c,n}}\psi_{i,n}.
\]
Suppose the cluster-score sums are independent across $c$, $ES_{c,n}=0$, and
\[
\Lambda_{C,n}:=\frac1n\sum_{c=1}^{C_n}E S_{c,n}^2
\]
satisfies $0<c\le\Lambda_{C,n}\le C$. Assume the cluster fourth moments are diffuse in the sense that
\[
\frac1{n^2}\sum_{c=1}^{C_n}E S_{c,n}^4\longrightarrow0.
\]
Let $\widetilde S_{c,n}=\sum_{i\in G_{c,n}}\widetilde\psi_i(\beta_0)$ and suppose
\[
\frac1n\sum_{c=1}^{C_n}(\widetilde S_{c,n}-S_{c,n})^2=o_p(1),
\qquad
\frac{\widehat Q_n}{\sigma_{D,n}^2}\to_p1,
\qquad
0<c_D\le\sigma_{D,n}^2\le C_D.
\]
If the CHLS linearization in \eqref{eq:chlsrep} holds under $H_0$, then
\[
V_n=\sigma_{D,n}^{-4}\Lambda_{C,n}
\]
for the leading-score variance and the projected cluster estimator
\[
\widetilde V_{C,n}
=\widehat Q_n^{-2}\frac1n\sum_{c=1}^{C_n}\widetilde S_{c,n}^2
\]
satisfies
\[
\frac{\widetilde V_{C,n}}{V_n}\to_p1.
\]
Moreover,
\[
\frac{n^{-1/2}\sum_{c=1}^{C_n}S_{c,n}}{\Lambda_{C,n}^{1/2}}\dto N(0,1),
\]
so one independent standard Gaussian multiplier per cluster yields a first-order valid projected-dependence bootstrap.
\end{proposition}

\paragraph{Proof.}
Independence and centering imply
\[
\Var\left(n^{-1/2}\sum_i\psi_{i,n}\right)
=\frac1n\sum_c E S_{c,n}^2=\Lambda_{C,n}.
\]
The oracle cluster quadratic form $\widehat\Lambda_{C,n}^o=n^{-1}\sum_cS_{c,n}^2$ is unbiased and
\[
\Var(\widehat\Lambda_{C,n}^o)
\le\frac1{n^2}\sum_c E S_{c,n}^4=o(1),
\]
so $\widehat\Lambda_{C,n}^o-\Lambda_{C,n}=o_p(1)$. The same fourth-moment condition gives the Lindeberg condition for the independent triangular array of cluster sums: for every $\epsilon>0$,
\[
\frac1{n\Lambda_{C,n}}\sum_c
E\left[S_{c,n}^2\mathbf 1\{|S_{c,n}|>\epsilon\sqrt{n\Lambda_{C,n}}\}\right]
\le
\frac1{\epsilon^2 n^2\Lambda_{C,n}^2}\sum_c E S_{c,n}^4\to0.
\]
Hence Lindeberg--Feller yields the displayed cluster-sum CLT. For feasible replacement, Cauchy--Schwarz gives
\[
\begin{aligned}
&\left|\frac1n\sum_c\widetilde S_{c,n}^2-\frac1n\sum_cS_{c,n}^2\right|\\
&\quad\le
2\left(\frac1n\sum_cS_{c,n}^2\right)^{1/2}
\left(\frac1n\sum_c(\widetilde S_{c,n}-S_{c,n})^2\right)^{1/2}\\
&\qquad\quad
+\frac1n\sum_c(\widetilde S_{c,n}-S_{c,n})^2
=o_p(1).
\end{aligned}
\]
The treatment-scale conditions therefore imply $\widetilde V_{C,n}/V_n\to_p1$. Because $V_n$ is bounded away from zero and infinity, the $o_p(1)$ remainder in \eqref{eq:chlsrep} is negligible after standardization. Finally, the within-cluster-one multiplier covariance matrix is positive semidefinite, so the projected cluster bootstrap is conditionally centered Gaussian with variance $\widetilde V_{C,n}$; Theorem~\ref{thm:pd} completes the argument. \qed

For example, if cluster sizes are uniformly bounded, cluster-score fourth moments are uniformly bounded, and the individual feasible-score replacement obeys $n^{-1}\sum_i(\widetilde\psi_i-\psi_i)^2=o_p(1)$, then the two cluster-level diffuseness/replacement conditions above follow immediately. Growing clusters are also allowed whenever the displayed cluster-sum conditions themselves hold.

\subsection{Feasible relevance-index consistency}
\begin{proposition}[Feasible relevance index]\label{prop:rhat}
Suppose the covariance-filter conditions of Theorem~\ref{thm:filter} and Assumption~\ref{ass:stable} hold,
\[
\frac{\widetilde V_n^{\#}(\beta_0)}{V_n^{\#}}\pto1,
\qquad
\frac{\widetilde V_{\mathrm{PD},n}(\beta_0)}{V_n}\pto1.
\]
Define $\widehat r_n=1-\widetilde V_n^{\#}(\beta_0)/\widetilde V_{\mathrm{PD},n}(\beta_0)$ when the denominator is positive and assign any fixed bookkeeping value otherwise. Then
\[
P\{\widetilde V_{\mathrm{PD},n}(\beta_0)>0\}\to1,
\qquad
\widehat r_n-R_{A,n}\pto0.
\]
Consequently, if $R_{A,n}\to r<1$, then $P(\widehat r_n<1)\to1$. On that event define
\[
\widehat\pi_\alpha=2\Phi\!\left(-z_{1-\alpha/2}\sqrt{1-\widehat r_n}\right),
\]
with an arbitrary value on the complement. Then
$\widehat\pi_\alpha\pto\pi_\alpha(r)$, where
$\pi_\alpha(r)=2\Phi(-z_{1-\alpha/2}\sqrt{1-r})$. Under the additional standardized sampling-law conditions of Theorem~\ref{thm:phase}, $\pi_\alpha(r)$ is the limiting rejection probability of the iid multiplier score-bootstrap test.
\end{proposition}

\paragraph{Proof.}
Since $V_n$ is bounded away from zero under Assumption~\ref{ass:stable}, the assumed ratio $\widetilde V_{\mathrm{PD},n}/V_n\pto1$ implies
\[
P\{\widetilde V_{\mathrm{PD},n}>0\}\to1.
\]
On that event, the covariance-filter identity gives
\[
1-\widehat r_n
=\frac{\widetilde V_n^{\#}}{\widetilde V_{\mathrm{PD},n}}
=\frac{\widetilde V_n^{\#}}{V_n^{\#}}
\frac{V_n^{\#}}{V_n}
\frac{V_n}{\widetilde V_{\mathrm{PD},n}}.
\]
Under Assumption~\ref{ass:stable}, $V_n^{\#}/V_n=s_{\#,n}^2/\sigma_{D\varepsilon,n}^2$ is bounded away from zero and infinity. The two assumed variance ratios converge to one, so $1-\widehat r_n-(1-R_{A,n})=o_p(1)$ on an event whose probability tends to one; the arbitrary bookkeeping value on the complement is asymptotically irrelevant. If $R_{A,n}\to r<1$, consistency gives $P(\widehat r_n<1)\to1$, and the plug-in statement follows by continuity. \qed

\section{Additional Monte Carlo implementation details}\label{app:mcprojected}
The projected-dependence Monte Carlo in Section~\ref{sec:mcprojected} uses the same treatment-effect dependence designs as the baseline grid but an independent, fixed Monte Carlo stream. Under $H_0:\beta=0$, the feasible score is
\[
\widetilde\psi_i=(D_i-\bar D)\widetilde\varepsilon_i.
\]
The projected variance uses lag truncation $\ell=\lfloor n^{1/4}\rceil$ and Bartlett/Newey--West weights $1-h/(\ell+1)$, equivalently $k(h/(\ell+1))$ for $k(x)=(1-|x|)_+$. The oracle benchmark uses the same weights on the latent randomized-regressor score. The script \texttt{simulate\_projected\_bootstrap.py} reproduces Main Table~\ref{tab:mcprojected}.

\subsection{Sample-size stress test and diagnostic reliability}
The first stress exercise varies $n\in\{100,250,500,1000\}$ at $p=0.5$ and $\ell=\lfloor n^{1/4}\rceil$, using 20,000 Monte Carlo replications per row. Table~\ref{tab:stressn} reports the true relevance index, iid and projected rejection rates, the mean projected variance, and the sampling behavior of the feasible relevance diagnostic. The projected test moves rejection toward the nominal $5\%$ level as $n$ grows, but the diagnostic is substantially noisier. In the positive MA(1) design its sign is wrong in about $31\%$ of replications at $n=100$ and $8\%$ at $n=1000$; in the negative MA(1) design the corresponding wrong-sign frequencies are about $25\%$ and $5\%$.

\begin{table}[!htbp]\centering
\caption{Projected-HAC sensitivity to sample size ($p=0.5$, $L=\lfloor n^{1/4}\rceil$, 20,000 replications per row).}\label{tab:stressn}
\scriptsize\begin{tabular}{lrrrrrrr}\toprule
Design & $n$ & $L$ & IID reject. & Projected reject. & $V_n$ & $E[\widetilde V_{\mathrm{PD},n}]$ & Mean $\widehat r_n$\\\midrule
Positive MA(1) & 100 & 3 & 6.77\% & 5.00\% & 7.490 & 7.218 & 0.068 \\
Positive MA(1) & 250 & 4 & 6.69\% & 5.03\% & 7.496 & 7.287 & 0.091 \\
Positive MA(1) & 500 & 5 & 6.91\% & 5.16\% & 7.498 & 7.315 & 0.102 \\
Positive MA(1) & 1000 & 6 & 7.11\% & 5.32\% & 7.499 & 7.359 & 0.110 \\
Cancellation MA(2) & 100 & 3 & 5.15\% & 4.37\% & 8.510 & 8.753 & -0.005 \\
Cancellation MA(2) & 250 & 4 & 5.22\% & 4.59\% & 8.504 & 8.698 & 0.005 \\
Cancellation MA(2) & 500 & 5 & 4.85\% & 4.45\% & 8.502 & 8.681 & 0.009 \\
Cancellation MA(2) & 1000 & 6 & 4.87\% & 4.66\% & 8.501 & 8.642 & 0.009 \\
Negative MA(1) & 100 & 3 & 3.31\% & 3.88\% & 5.510 & 5.781 & -0.164 \\
Negative MA(1) & 250 & 4 & 3.42\% & 4.61\% & 5.504 & 5.710 & -0.160 \\
Negative MA(1) & 500 & 5 & 3.33\% & 4.47\% & 5.502 & 5.667 & -0.160 \\
Negative MA(1) & 1000 & 6 & 3.23\% & 4.33\% & 5.501 & 5.646 & -0.160 \\
\bottomrule\end{tabular}\end{table}

The finite-sample behavior of $\widehat r_n$ should not be reduced to a one-term Jensen story. Write $X_n=\widetilde V_n^\#$, $Y_n=\widetilde V_{\mathrm{PD},n}$, and let $\mu_X=E X_n$, $\mu_Y=E Y_n$. Under the usual second-order delta-method conditions and with the denominator bounded away from zero, a ratio expansion gives
\[
E\!\left[\frac{X_n}{Y_n}\right]
\approx
\frac{\mu_X}{\mu_Y}
-\frac{\Cov(X_n,Y_n)}{\mu_Y^2}
+\frac{\mu_X\Var(Y_n)}{\mu_Y^3},
\]
up to higher-order terms, while $\mu_Y$ itself carries lag-window bias. Thus $E\widehat r_n$ contains both a numerator--denominator covariance term and a denominator-variance term, not merely the convexity term from $E(1/Y_n)$. Their relative signs and magnitudes depend on bandwidth and dependence geometry. This is why the diagnostic can be attenuated or overshoot even when the projected long-run-variance estimator moves closer to its target. A bias adjustment based only on $\Var(Y_n)$ is therefore not generally second-order complete without also estimating the joint covariance term. The paper treats $\widehat r_n$ as a descriptive ratio rather than a sign pretest.

\subsection{Bandwidth sensitivity}
Table~\ref{tab:stressL} holds $n=1000$ and $p=0.5$ fixed and varies the Bartlett bandwidth from $\ell=2$ to $\ell=25$, using the same 20,000 Monte Carlo samples within each design across bandwidth choices. Increasing $\ell$ generally moves the projected variance toward the population long-run variance, but the mean diagnostic is not monotone and its sign uncertainty can increase because the denominator becomes noisier. The projected test itself remains much more stable than the diagnostic.

\begin{table}[!htbp]\centering
\caption{Bandwidth sensitivity at $n=1000$, $p=0.5$ (20,000 common replications within each design across bandwidths). For the cancellation design the sign-error column is omitted conceptually because $R_{A,n}$ is approximately zero.}\label{tab:stressL}
\scriptsize\begin{tabular}{lrrrrr}\toprule
Design & $L$ & Projected reject. & $E[\widetilde V_{\mathrm{PD},n}]$ & Mean $\widehat r_n$ & Sign error\\\midrule
Positive MA(1) & 2 & 5.45\% & 7.158 & 0.090 & 2.0\% \\
Positive MA(1) & 4 & 5.20\% & 7.291 & 0.104 & 5.1\% \\
Positive MA(1) & 6 & 5.07\% & 7.348 & 0.109 & 8.2\% \\
Positive MA(1) & 10 & 4.91\% & 7.400 & 0.111 & 13.4\% \\
Positive MA(1) & 16 & 4.71\% & 7.433 & 0.108 & 19.2\% \\
Positive MA(1) & 25 & 4.49\% & 7.451 & 0.099 & 24.7\% \\
Cancellation MA(2) & 2 & 4.64\% & 8.831 & 0.036 & -- \\
Cancellation MA(2) & 4 & 4.72\% & 8.698 & 0.018 & -- \\
Cancellation MA(2) & 6 & 4.72\% & 8.638 & 0.009 & -- \\
Cancellation MA(2) & 10 & 4.67\% & 8.586 & -0.002 & -- \\
Cancellation MA(2) & 16 & 4.64\% & 8.553 & -0.014 & -- \\
Cancellation MA(2) & 25 & 4.53\% & 8.531 & -0.029 & -- \\
Negative MA(1) & 2 & 4.35\% & 5.835 & -0.116 & 1.2\% \\
Negative MA(1) & 4 & 4.58\% & 5.702 & -0.145 & 2.9\% \\
Negative MA(1) & 6 & 4.69\% & 5.643 & -0.160 & 4.8\% \\
Negative MA(1) & 10 & 4.79\% & 5.591 & -0.177 & 8.3\% \\
Negative MA(1) & 16 & 4.75\% & 5.559 & -0.193 & 12.7\% \\
Negative MA(1) & 25 & 4.62\% & 5.535 & -0.213 & 16.8\% \\
\bottomrule\end{tabular}\end{table}

\subsection{Imbalanced assignment}
Table~\ref{tab:stressp} varies the assignment probability at $n=1000$ and $\ell=6$, with 20,000 Monte Carlo replications per row. The positive, cancellation, and negative covariance phases remain visible at $p=0.25$ and $p=0.75$, and the projected test again moves rejection toward nominal size. This exercise is a sensitivity check within the maintained iid Bernoulli assignment theory; it is not an extension to fixed-count or covariate-adaptive randomization.

\begin{table}[!htbp]\centering
\caption{Treatment-probability sensitivity at $n=1000$, $L=6$ (20,000 replications per row).}\label{tab:stressp}
\scriptsize\begin{tabular}{lrrrrrr}\toprule
Design & $p$ & IID reject. & Projected reject. & $V_n$ & $E[\widetilde V_{\mathrm{PD},n}]$ & Mean $\widehat r_n$\\\midrule
Positive MA(1) & 0.25 & 6.05\% & 4.96\% & 11.332 & 11.196 & 0.070 \\
Positive MA(1) & 0.75 & 6.60\% & 5.14\% & 7.999 & 7.853 & 0.102 \\
Cancellation MA(2) & 0.25 & 4.93\% & 4.69\% & 14.334 & 14.497 & 0.003 \\
Cancellation MA(2) & 0.75 & 4.86\% & 4.74\% & 8.334 & 8.487 & 0.010 \\
Negative MA(1) & 0.25 & 3.77\% & 4.50\% & 9.334 & 9.498 & -0.099 \\
Negative MA(1) & 0.75 & 3.48\% & 4.77\% & 6.001 & 6.158 & -0.147 \\
\bottomrule\end{tabular}\end{table}

\subsection{Dependence-strength sensitivity}
The final stress exercise varies the MA(1) coefficient $\theta$ in $b_i=\eta_i+\theta\eta_{i-1}$ while holding $n=1000$, $p=0.5$, and $\ell=6$ fixed. Table~\ref{tab:stresstheta} shows the phase diagram continuously rather than only at the benchmark $\theta=\pm0.5$: as surviving covariance becomes more positive, iid rejection rises above nominal size; as it becomes more negative, iid rejection becomes increasingly conservative. The projected correction remains close to $5\%$ across the grid. This is a dependence-strength sensitivity exercise, not an application-calibrated design; the latter would require an empirical setting with a substantively justified dependence model.

\begin{table}[!htbp]\centering
\caption{Dependence-strength sensitivity for MA(1) treatment effects $b_i=\eta_i+\theta\eta_{i-1}$ ($n=1000$, $p=0.5$, $L=6$, 20,000 replications per row).}\label{tab:stresstheta}
\scriptsize\begin{tabular}{rrrrrr}\toprule
$\theta$ & $R_{A,n}$ & IID reject. & Projected reject. & $V_n$ & $E[\widetilde V_{\mathrm{PD},n}]$\\\midrule
-0.75 & -0.266 & 2.69\% & 4.40\% & 5.627 & 5.853 \\
-0.50 & -0.182 & 3.39\% & 4.59\% & 5.501 & 5.646 \\
-0.25 & -0.089 & 3.98\% & 4.71\% & 5.625 & 5.697 \\
+0.00 & 0.000 & 5.11\% & 4.98\% & 6.000 & 6.004 \\
+0.25 & 0.075 & 5.73\% & 4.90\% & 6.625 & 6.554 \\
+0.50 & 0.133 & 6.91\% & 5.23\% & 7.499 & 7.360 \\
+0.75 & 0.174 & 7.75\% & 5.44\% & 8.623 & 8.400 \\
\bottomrule\end{tabular}\end{table}

The stress-test scripts \texttt{simulate\_projected\_stress.py} and \texttt{simulate\_projected\_strength.py} use fixed named random-number seeds recorded in \texttt{SEEDS.csv}, so every table above is exactly reproducible.

\subsection{Fixed-bandwidth calibration}\label{app:fixedbmc}
The fixed-bandwidth experiment uses the same ordered DGPs and the same projected Bartlett statistic, but replaces the standard-normal critical value with a simulated critical value for the fixed-bandwidth reference law in Theorem~\ref{thm:fixedb}. We use deterministic seeds and 20,000 DGP replications per reported cell. Table~\ref{tab:fixedbfull} reports the full selected grid. The correction is most useful when the Bartlett bandwidth fraction $b_n=(\ell+1)/n$ is not small: over the grid examined it substantially reduces the large-bandwidth under-rejection in the cancellation and negative designs, though it is not uniformly closer to $5\%$ in every cell. This is exactly the limitation emphasized by Theorem~\ref{thm:fixedb}: fixed-bandwidth asymptotics recalibrate the reference law but do not choose the dependence kernel.

\begin{table}[!htbp]
\centering
\caption{Fixed-bandwidth calibration sensitivity. Here $\ell$ is the largest included positive Bartlett lag and $b=(\ell+1)/n$. The projected statistic is identical across the last two rejection columns; only the reference critical value changes. Each DGP cell uses 20,000 replications. Parentheses report Monte Carlo standard errors in percentage points. Reference critical values use 60,000 Gaussian draws on a 1,000-point grid at the same kernel-bandwidth fraction.}
\label{tab:fixedbfull}
\scriptsize
\begin{tabular}{lrrrrrr}
\toprule
Design & $n$ & $\ell$ & $b$ & Fixed-$b$ cv & Normal reject. & Fixed-$b$ reject.\\
\midrule
Positive MA(1) & 250 & 2 & 0.012 & 1.941 & 5.27\% (0.16) & 5.51\% (0.16) \\
Positive MA(1) & 250 & 6 & 0.028 & 1.922 & 4.80\% (0.15) & 5.27\% (0.16) \\
Positive MA(1) & 250 & 16 & 0.068 & 1.864 & 4.01\% (0.14) & 5.38\% (0.16) \\
Cancellation MA(2) & 250 & 2 & 0.012 & 1.941 & 4.44\% (0.15) & 4.66\% (0.15) \\
Cancellation MA(2) & 250 & 6 & 0.028 & 1.922 & 4.34\% (0.14) & 4.78\% (0.15) \\
Cancellation MA(2) & 250 & 16 & 0.068 & 1.864 & 3.74\% (0.13) & 5.10\% (0.16) \\
Negative MA(1) & 250 & 2 & 0.012 & 1.941 & 4.20\% (0.14) & 4.40\% (0.14) \\
Negative MA(1) & 250 & 6 & 0.028 & 1.922 & 4.32\% (0.14) & 4.90\% (0.15) \\
Negative MA(1) & 250 & 16 & 0.068 & 1.864 & 3.77\% (0.13) & 5.11\% (0.16) \\
Positive MA(1) & 1000 & 2 & 0.003 & 1.953 & 5.29\% (0.16) & 5.38\% (0.16) \\
Positive MA(1) & 1000 & 6 & 0.007 & 1.944 & 4.96\% (0.15) & 5.16\% (0.16) \\
Positive MA(1) & 1000 & 25 & 0.026 & 1.926 & 4.66\% (0.15) & 5.12\% (0.16) \\
Cancellation MA(2) & 1000 & 2 & 0.003 & 1.953 & 4.35\% (0.14) & 4.40\% (0.15) \\
Cancellation MA(2) & 1000 & 6 & 0.007 & 1.944 & 4.57\% (0.15) & 4.76\% (0.15) \\
Cancellation MA(2) & 1000 & 25 & 0.026 & 1.926 & 4.39\% (0.14) & 4.86\% (0.15) \\
Negative MA(1) & 1000 & 2 & 0.003 & 1.953 & 4.23\% (0.14) & 4.33\% (0.14) \\
Negative MA(1) & 1000 & 6 & 0.007 & 1.944 & 4.47\% (0.15) & 4.66\% (0.15) \\
Negative MA(1) & 1000 & 25 & 0.026 & 1.926 & 4.48\% (0.15) & 4.86\% (0.15) \\
\bottomrule
\end{tabular}
\end{table}

\subsection{Misspecified grouping in the cluster correction}\label{app:groupmisspec}
To isolate dependence-model risk, fix the true treatment-effect cluster size at eight in the independent-cluster design and change only the grouping supplied to the projected variance estimator. The same 20,000 simulated samples are used for all working groupings. Table~\ref{tab:groupmisspec} shows that correct grouping restores nominal size. Splitting each true cluster into smaller groups omits part of the surviving covariance and leaves substantial over-rejection. Scrambling units into false size-eight groups is nearly as ineffective as singleton iid inference because the working groups no longer align with the latent effect modifier. Merging adjacent independent true clusters preserves all true within-cluster covariance and remains valid in this design, although such over-grouping can be less precise in other settings. The exercise illustrates a boundary of the projected approach: the theory reduces the covariance object to a scalar direction, but it does not identify the grouping structure for the researcher.

\begin{table}[!htbp]
\centering
\caption{Sensitivity to the working cluster structure when the true treatment-effect cluster size is eight. Assignment remains iid at the unit level. Parentheses report Monte Carlo standard errors in percentage points.}
\label{tab:groupmisspec}
\small
\begin{tabular}{lrr}
\toprule
Working grouping & Rejection & Mean projected variance\\
\midrule
Singleton (iid) & 18.39\% (0.27) & 5.991 \\
Split true clusters: size 2 & 15.00\% (0.25) & 6.990 \\
Split true clusters: size 4 & 10.38\% (0.22) & 8.981 \\
Correct: size 8 & 5.04\% (0.15) & 12.972 \\
Merge adjacent true clusters & 4.94\% (0.15) & 12.996 \\
Scrambled groups: size 8 & 18.14\% (0.27) & 6.050 \\
\bottomrule
\end{tabular}
\end{table}

\section{Supplementary implementation and simulation material}\label{app:supplementary}
\subsection{Independent-cluster treatment-effect geometry}
To complement the ordered time-series designs, consider $n=Gm$ observations partitioned into $G$ independent clusters of fixed size $m$. Treatment remains iid Bernoulli across individual units, exactly as in the maintained assignment theory. Let
\[
Y_{gc}(0)=u_{gc},\qquad Y_{gc}(1)=u_{gc}+\eta_g,
\]
where $u_{gc}$ and $\eta_g$ are independent standard normal variables and clusters are independent. Thus the individual treatment effect is $\tau_{gc}=\eta_g$. The one-observation variance contribution is unchanged by $m$, while the aggregate treatment-effect covariance is
\[
\Delta_{\tau,n}=m-1.
\]
At $p=1/2$, $V_n^\#=6$ and hence $V_n=6+(m-1)$. The projected cluster variance uses
\[
\widetilde V_{\mathrm{PD},n}
=\widehat Q_n^{-2}\frac1n\sum_{g=1}^{G}
\left(\sum_{i\in G_g}\widetilde\psi_i\right)^2,
\]
which is the known-cluster implementation recorded above. Main Table~\ref{tab:mccluster} reports the results from 20,000 replications per row. As cluster size grows from two to eight, the covariance omitted by iid weighting becomes increasingly important: iid rejection rises from about $7\%$ to about $19\%$, whereas the projected cluster correction remains close to $5\%$. This exercise changes the dependence geometry of heterogeneous treatment effects while preserving iid unit-level assignment; it is therefore an implementation check, not an extension to cluster-randomized treatment. The script \texttt{simulate\_cluster\_projected.py} reproduces the table.

\subsection{Additional Monte Carlo figures}
\begin{figure}[!htbp]\centering
\includegraphics[width=.76\textwidth]{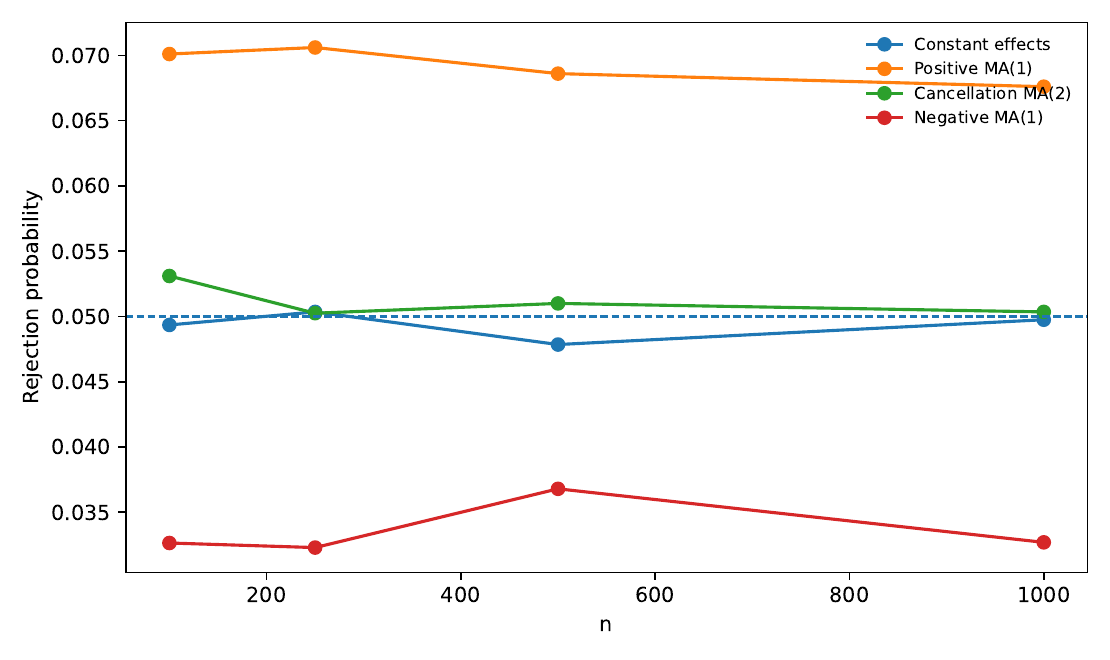}
\caption{Rejection probabilities of the null-imposed iid score test at $p=0.5$.}\label{fig:rejection-app}
\end{figure}
\begin{figure}[!htbp]\centering
\includegraphics[width=.76\textwidth]{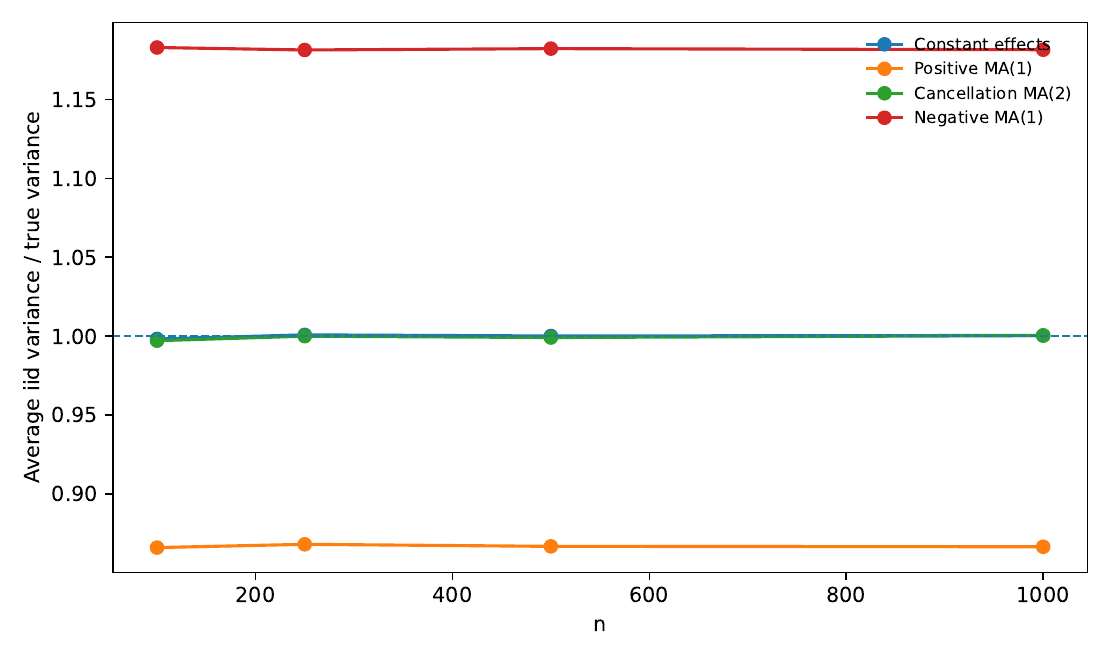}
\caption{Average HC0 variance divided by the true first-order variance at $p=0.5$.}\label{fig:variance-ratio-app}
\end{figure}

\end{document}